# Supporting Long-term Transactions in Smart Contracts Generated from Business Process Model and Notation (BPMN) Models

by

Christian Gang Liu

Submitted in partial fulfilment of the requirements
for the degree of Doctor of Philosophy

Dalhousie University
Halifax, Nova Scotia
January 2024

# TABLE OF CONTENTS

# LIST OF TABLES

# LIST OF FIGURES

# ABSTRACT

To alleviate difficulties in writing smart contracts for distributed blockchain applications, as other research, we propose transformation of Business Process Model and Notation (BPMN) models into blockchain smart contracts. Unlike other research, we use Discrete Event Hierarchical State Machine (DE-HSM) multi-modal modeling to identify collaborative trade transactions that need to be supported by the smart contract and describe how the trade transactions, that may be nested, are supported by a transaction mechanism. We describe algorithms to (i) identify the nested trade transactions and to (ii) transform the BPMN model into blockchains smart contracts that include a transaction mechanism to enforce the transactional properties for the identified trade transactions.

The developed proof of concept shows that our approach to automated transformation of BPMN models into smart contracts with the support of privacy and cross-chain interoperability is feasible. The thesis examines and evaluates automatically generated alternative transaction mechanisms to support such transactions using three use cases of varying degree of complexity, namely order processing, supply chain management, and a multi-faceted trade use case. The research enriches the academic dialogue on blockchain technology and smart contracts and proposes potential avenues for future research.

# LIST OF ABBREVIATIONS USED

| | |
|---|---|
| BPMN | Business Process Model and Notation … graphical representation for specifying business processes in a business process model |
| DB | Database |
| DBMS | Database Management System |
| EVM | Ethereum Virtual Machine |
| FSM | Finite State Machine |
| HSM | Hierarchical State Machine |
| DE-HSM | Discrete Event Hierarchical FSM Model |
| IPFS | Interplanetary File System |
| ACID | Transactional properties (Atomicity, Consistency, Isolation and Durability) |
| SESE | Single-entry and Single-exit Subgraph |
| LSI | Largest Smallest Independent (LSI) subgraph |
| LtB | Long-term Blockchain Transaction |
| PoC | Proof of Concepts |
| SC | Smart Contract |
| TABS | Tool for Automated Transformation of a BPMN Model into Smart Contracts |

# ACKNOWLEDGEMENTS

First and foremost, I am grateful to my supervisors, Drs. Peter Bodorik and Dawn Jutla for their invaluable advice and continuous support, and patience during my Ph.D. study. Their immense knowledge and plentiful experience have encouraged me in all the time of my academic research and daily life. I want to thank Drs. Srini Sampalli and Qiang Ye and for their time and effort in providing me with their reviews and guidance during my graduate studies. Thanks also to my external examiner, Dr. Wojciech Golab, for his thorough and thoughtful examination of my thesis. I also would also like to thank Dr. Michael McAllister for his administrative support during my Ph.D. study and research. It is their kind help and support that have made my study and life in the Dalhousie a wonderful experience. Finally, I would like to express my appreciation to my family and close friends for their unwavering support and patience, without which this journey would not have been possible.

# CHAPTER 1 INTRODUCTION

Introduction of the Bitcoin cryptocurrency (Nakamoto, 2008) and its subsequent rise (Marr, 2017) spurred great interest in cryptocurrencies and blockchains. Further blockchain platforms, such as Hyperledger fabric and Ethereum, followed the Bitcoin blockchain and many of them succeeded while many have failed (Shawdagor, 2023). The notion of a smart contract (Szabo, 1997, 2018; Buterin et al., 2015) has surfaced as an ensemble of methods inscribed in a Turing-complete high-level language. These methods are stored on the ledger, thereby inheriting the protection offered by blockchains. In addition, the blockchain infrastructure utilizes cryptographic concepts and methods to provide the blockchain's desirable properties of trust, immutability, availability, and transparency, amongst others.

The concept of a smart contract (Szabo, 1997, 2018; Buterin et al., 2015) emerged as a set of methods and stored on the ledger, thereby benefiting from the protection that blockchains offer. The blockchain infrastructure employs cryptographic concepts and methods to ensure desirable properties such as trust, immutability, availability, and transparency. However, as a new technology, blockchain and their smart contracts pose new challenges in both the blockchain infrastructure and in developing smart contracts and applications that use them. Thus, in addition to tackling the blockchain infrastructure issues of scalability, transaction throughput, and high costs, the development of smart contracts also received much attention by the research and development communities as can be seen by many literature surveys on the topic, such as (Taylor et al., 2019; Khan et al., 2021; Vacca et al., 2021; Belchior et al., 2021; Saito, 2016, Garcia-Garcia et al., 2020; Lauster et al., 2020; Levasseur et al., 2021).

Trade transactions, especially those encompassing high-value commodities, frequently engage multiple stakeholders and may extend over protracted durations. These long-term transactions pose difficulties when writing smart contracts due to their complexity and the need for multi-level transactions. An approach is needed that allows the construction of smart contracts as an interaction of independent modules, facilitating interaction across modules deployed in a multi-chain environment.

Belchior et al. (2021) conducted a comprehensive survey on the interoperability of cross-chain or multi-chain systems, identifying open issues concerning complex trade transactions. Solutions for building decentralized applications often lack interoperability, which hampers scalability (Besançon et al., 2019). Coordinating transactions across different blockchains to support multi-chain DApps is challenging due to differing blockchain properties (Abebe et al., 2019). Furthermore, privacy and security concerns, such as the right-to-forget and fine-grain access control, pose additional challenges (Tam Vo et al., 2018). Key open issues in blockchain interoperability include the gap between theory and practice (Thomas et al., 2021; Zhu et al., 2020), discoverability (Abebe et al., 2019; Liu et al., 2019), privacy and security (Tam Vo et al., 2018; Thomas & Schwartz, 2015; Wood, 2019; Zamyatin et al., 2019), and governance (Hardjono et al., 2019; Ilham et al., 2019; WEF, 2020).

Research on blockchain smart contracts, as categorized by Khan et al. (2021), falls into two categories: improvement and usage. The improvement category includes research and tools aimed at enhancing smart contracts either during their creation or after their existence. However, despite the existence of strategies that lead to smart contract programs with formally proven safety properties, their adoption has been limited due to their perceived complexity by software developers.

## 1.1 Goal And Objectives

The primary goal of this doctoral research is to

> *develop an approach for the automatic transformation of Business Process Model and Notation (BPMN) models into blockchain smart contracts that are intended to facilitate collaborative nested transactions that may span extensive periods and involve multiple participants, with activities emerging in unpredictable sequences.*

To achieve the goal, a number of technical issues need to be resolved. The resolution of these issues constitutes the specific objectives of this thesis, as outlined below:

- *Transformation of BPMN Models into Smart Contracts*: Develop algorithms and tools that facilitate the conversion of BPMN models into a cohesive set of smart contracts, specifically tailored for a cross-chain environment. This objective is designed to simplify and streamline the design and development process for smart contracts, ensuring both accuracy and efficiency in the transformation. By enhancing the accessibility and reliability of the process, it paves the way for new opportunities in implementing collaborative applications across various blockchains.

- *Support for Long-term Transactions*: Create mechanisms to manage transactions that may extend over prolonged durations, effectively accommodating the dynamics of extended processes spanning a long period of time. This includes handling participants' activities that may arise in unpredictable sequences, ensuring robustness and adaptability within the transaction process.
- *Interoperability of Off-chain Computations*: Design solutions to enable collaborative transactions to be executed across various blockchains. This approach aims to localize the computational components of collaborative transactions on different blockchains, ensuring access control and privacy for participants. By allowing participants to work on different blockchain, the design promotes compatibility and integration across various blockchains.
- *Blockchain Multi-Method Transactions*: Develop support for multi-method transactions within smart contracts, serving as independent computational components within collaborative transactions. The aim is to enhance access control and privacy of blockchain multi-method transactions.
- *Nested Transactions*: Recognizing that collaborative long-term transactions often encompass multiple nested transactions, which function as subprograms or sub-transactions within their parent transactions, the aim is to develop automated mechanisms to support such nested control structures.
- *Privacy Preservation*: Ensure that only participants involved in the processing of the transaction can access the operation and view the related data. This approach is aimed at maintaining the confidentiality and privacy integral to the transaction process.

- *Evaluation and Testing*: Develop a Proof of Concept (PoC) to demonstrate the feasibility of the proposed concepts and mechanisms developed in this thesis.

By successfully accomplishing these objectives, this research seeks to advance the research on creation of smart contracts using Model-Driven Engineering (MDE) approach. The intention is to create a synergy between conventional business process modeling and the still advancing field of blockchains. This integration not only enhances collaboration, security, and efficiency but also paves the way for collaborative blockchain applications across various industries.

## 1.2 CONTRIBUTIONS

The specific contributions of our research are:

- *Representation and Support of Collaborative Transactions*: Through an in-depth examination of use cases for collaborative transactions, we identified key properties that must be supported, including the long-term nature of transactions, unpredictability in the sequence of participant activities, the nesting of transactions with sub-transactions, access control, and privacy requirements. Based on this analysis, we introduced a concept for managing long-term transactions on blockchains that extend across multiple smart contract method executions. This approach draws insights from both database and blockchain transactions to ensure essential ACID, access control, and privacy properties. Furthermore, we proposed methods to (i) represent long-term transactions in a BPMN model and (ii) transform the BPMN models of applications with long-term transactions into smart contracts in order to significantly reduce the effort in implementing long-term transaction mechanisms for blockchain transactions.

- *Transformation of BPMN Models into Smart Contracts Using DE-HSM Modeling*: We proposed transformation of Business Process Model and Notation (BPMN) model into a Discrete Events Hierarchical Finite State Machines (DE-HSM) model and then into the methods of a smart contracts. The approach is aimed to support the development of cross-chain smart contracts for long-term transactions. By utilizing DE-HSM multi-modal modeling, which combines Finite State Machines (FSMs) with Discrete Event (DE) modeling to represent time, we bridged the gap between the high-level business process representation of BPMN and the technical demands of blockchain smart contracts. While BPMN excels in illustrating the overall flow of business processes, it lacks specific details necessary for handling complex system dynamics, concurrency, and hierarchical state management. DE-HSM enriches the model with these essential details, making it more suitable for direct transformation into hierarchical multi-method smart contracts. This innovative approach not only enhances the modeling of intricate smart contract processes but also fosters collaboration among actors in smart contract execution. Key features of this approach include:

- *Long-term Transaction*s: We integrated the concept of collaborative long-term transactions into BPMN graphical diagrams and developed algorithms to recognize and transform these transactions using DE-HSM models and then into the methods of a smart contract.
- *Interoperability for Off-chain Computations*: We designed solutions for communication between smart contracts on different blockchains, utilizing a bridge process to enable interaction between mainchain and off-chain components.
- *Nested Transactions*: Recognizing the complexity of collaborative long-term transactions, we developed mechanisms to support nested transactions, offering flexibility in recognizing and implementing nested structures.
- *Privacy Preservation*: We implemented sidechain processing to ensure that only participants involved in a transaction can access the operation and related data.

- *Definition and Support of Blockchain Multi-Method Transactions*: We introduced the concept of a multi-method (mm) transaction, defining it as a subset of the methods of a smart contract with independence properties. This approach preserves atomicity and isolation, and we developed mechanisms to support multi-method transactions with varying levels of privacy.

- *Development of the TABS+ Tool*: We created the Transformation of Automated Blockchain Smart Contracts (TABS+) tool as a proof of concept to demonstrate the feasibility of our approach. This tool was instrumental in performing initial cost analysis and performance evaluations, and it serves as a tangible representation of our research contributions.

By addressing these multifaceted challenges, our research aims at bridging the gap between the traditional business process modeling and the dynamic world of blockchain. Our research has led to publications (Bodorik et al., 2001; Bodorik et al., 2003; Liu et al., 2021a; Liu et al., 2021b; Liu et al., 2022a; Liu et al., 2022b; Liu et al., 2023a; Liu et al., 2023b) in highly respected venues, primarily conferences and journals published by Association for Computing Machinery (ACM), Institute of Electrical and Electronics Engineers (IEEE), Springer Nature (Springer), and Elsevier.

## 1.3 Related Work

Several researchers have explored approaches to the engineering of blockchain applications from BPMN models while specifically focusing on Ethereum for business processes and asset management (Weber et al., 2016; López-Pintado et al., 2019; Tran et al., 2018; Mendling et al., 2018; Loukil et al., 2021). Although our work also uses the same starting point of using a BPMN model to represent the distributed blockchain application, in contrast, our work has leveraged transformation of a BPMN model into the Discrete Events Hierarchical Finite State Machines (DE-HSM) multi-modal model in order to better model the application's collaborative requirements. This approach has enabled us to facilitate the automated transformation of a BPMN model into DE-HSM, and subsequently into smart contract methods. Our methodology extends beyond Ethereum, encompassing various blockchain platforms such as Hyperledger. By utilizing DE-HSM, we have modularized blockchain collaborative transactions, recognizing the distinct components of a blockchain application that involves multiple participants. Our proposed approaches to implement multi-method smart contracts in a cross-chain environment have been designed to address atomicity, isolation and privacy concerns. These innovative strategies aim to simplify the process of creating smart contracts and accommodate long-term transactions and to pave a promising direction for future research on automated creation of smart contracts.

During my Master research, we explored the automatic transformation of an application, represented using Finite State Machines (FSMs), into smart contract methods with support of sidechain processing (Bodorik et al., 2021; Liu et al., 2021a). This included detailing the transformation process and the system architecture, including a bridge between the mainchain and the sidechain. We also proposed algorithms to identify off-chain computation patterns and designed a proof-of-concept tool for automatic pattern recognition and off-chain execution.

Di Ciccio et al. (2019) compared Lorikeet and Caterpillar approaches, both of which transform a BPMN model into a smart contract, in terms of the following features.

- *Model Execution*: This feature identifies the ***M****odel* ***D****riven* ***E****ngineering* (MDE) model, which in all cases discussed here is BPMN. For the execution, it captures if the smart contract on the blockchain controls the flow of execution by generating the code to be executed or it introduces another layer of abstraction in that the workflow is controlled by an interpreter that invokes processes/tasks to be executed by or on behalf of actors.
- *Coverage of BPMN elements*: BPMN models are built using a set of BPMN elements as described in the background section. As BPMN contains a rich set of elements from which to build BPMN models, most, if not all approaches support a subset of BPMN elements and Coverage represents the size of the supported subset.
- *Discovery of Incorrect Behavior and Sequence Enforcement*: If inputs provided by actors are incorrect or appear in incorrect sequence, is it handled properly?
- *Participant Selection*: Are participant identities supported?
- *Access Control*: Is access control supported?
- *Asset Control*: Are assets controlled?

We compared our TABS+ approach (Liu et al., 2023b) with Lorikeet, Caterpillar, and CoBuP approaches using the above categories as a guidance. We note that the features of Access Control and Asset Control refer to fungible and non-fungible tokens supported directly by the Lorikeet approach through the transformation process from BPMN model to the smart contract. The access and asset control supports are hardcoded in the smart contract and are used to control access to the registry of tokens to determine if access to the asset is allowed (Access Control) or if changes to the asset are allowed (Access Control). Of course, Caterpillar, CoBuP and TABS+ approaches can also support both fungible and non-fungible tokens provided that BPMN models to support them are developed and transformed to smart contracts.

A feature that is missing in the comparison in (Di Ciccio et al., 2019) is privacy. Generally, a blockchain does not support privacy, in that any user having access to the blockchain can see anything stored on the blockchain; of course, only if the user knows where to look on the blockchain. From a practical point of view, privacy in a blockchain is the ability to keep a transaction state private so that actors/users who are not participating in the transaction are unable to view the details about its state. Consider a trade activity that involves several participants and has several transactions in which different actors participate, such as for the purposes of transport, insurance, or customs clearing. Clearly, if a transaction involves three participants, these participants should have access to the state of that transaction, but the other participants should not have access to state of that transaction. Thus, we introduce *transaction privacy*, or *privacy* for short, to represent such privacy provision and use it as another feature in the comparison to approaches.

Table 1 provides the comparison. However, our TABS+ approach and its tool support the following additional features that are not supported by the other approaches:

- Support of nested transactions.
- Deployment of smart contracts on either Ethereum or HLF blockchains. Other blockchains can be supported if the interpreter is written for that blockchain.
- Our approach supports Sidechain and Cross-chain processing.

- Synchronization of activities is achieved in a blockchain-agnostic way as it is represented by a DE-FSM model that is blockchain agnostic. However, the developer still needs to write code, but it is for the independent and isolated tasks with well-defined input and output parameters.

And, as the table indicates, our approach also supports privacy (as indicated in the table 1). Below we briefly overview how our previous work on blockchain smart contracts relates to this paper:

- In (Bodorik et al., 2021; Liu et al., 2021b) we describe how an application, described using an FSM, can be transformed automatically into the methods of smart contract such that sidechain processing is supported. We describe the transformation process to derive the methods of the smart contract and the system architecture that includes a bridge between the mainchain and the sidechain to support the sidechain processing.
- In (Liu et al., 2022a) we describe our early work on using DE-HSM models for multi-modal modeling when transforming a BPMN model to the methods of a smart contract.
- Bodorik 2023 et al. (2023) formalize the approach of the previous work to transforming an application expressed as a BPMN model into the methods of a smart contract. The results of (Liu et al., 2021b; Liu et al., 2022a) are used in formulating the transformations in design phase, while we use the results of (Bodorik et al., 2021) in supporting sidechain processing.
- Liu et al. (2022b) raises the issue of how to represent and support BPMN trade transactions.
- Finally, Liu et al. (2023a) describe a mechanism for a developer to specify blockchain transactions that span executions of multiple methods of a smart contract and thus extend the native blockchain mechanism that supports a transaction as a result of executing a single smart contract method. In addition, the paper also describes how pattern augmentation technique is used to automatically create a transactional mechanism for the multi-method transactions specified by the developer.

Table 1.1 Comparison of Approaches to Transform a BPMN (Adopted from (Liu et al., 2023b))

| Feature\Approach | TABS+ | Lorikeet | Caterpillar | CoBuP |
|---|---|---|---|---|
| Execution | Interpreter | Code generation | Code generation | Interpreter |
| BPMN coverage | High | Medium | High | Not Available |
| ID Inc. Behavior | Supported | Supported | Supported | Supported |
| Sequence enforcement | Supported | Supported | Supported | Supported |
| Participant Select | Identity based | Predefined | N/A | Predefined |
| Access Control | To SC* methods | Direct support | Via SC | Via SC |
| Asset Control | To SC Asset methods | Direct support | Via SC | Via SC |
| Privacy | transaction based | Registry access | Not supported | Not supported |

*SC … Smart Contract

## 1.4 Outline Of Further Sections

Chapter 2 provides background information that is relevant to the development of smart contract solutions for trade applications on multi-chain platforms, including topics such as Blockchains and Smart Contract creation.

Chapter 3 delves into our research, concentrating on defining the concepts of Long-term Blockchain transactions and multi-method smart contracts. It also presents mechanisms and strategies for supporting long-term transactional properties and implementing multi-method smart contracts.

Chapter 4 focuses on the transformation process from BPMN to multi-method smart contracts, aiming to facilitate collaborative long-term transactions on the blockchain.

Chapter 5 explores the supporting mechanisms for nested transactions within collaborative long-term transactions, providing insights into the complexities and solutions for handling these nested structures.

Chapter 6 introduces the TABS tool, specifically designed to assist long-term transactions through the automatic generation and evaluation of cross-chain smart contracts. This chapter also includes a demonstration and evaluation of our proposed approach on two selected blockchain platforms, Ethereum and Hyperledger, to assess the feasibility and effectiveness of our strategies.

Chapter 7 encapsulates the conclusions drawn from our work and outlines the milestones and completion dates associated with my Ph.D. research. This includes details of our academic publications and patent submissions, offering a comprehensive timeline of the intellectual journey and scholarly contributions made throughout my doctoral studies. Additionally, this chapter suggests potential avenues for future exploration and research.

The specific contributions made in each publication, along with other relevant details, are shown in appendices of this thesis, providing a complete overview of the work accomplished.

# CHAPTER 2 BACKGROUND

This chapter presents a context background to our research, commencing with a synopsis of blockchain technology and its applications in trade finance, including the inherent characteristics of applications in the domain of trade of goods and services. This is followed by an exploration of smart contracts and their associated development platforms. We then elaborate on the Business Process Model and Notation (BPMN) modeling, providing details of a straightforward BPMN use case commonly cited in academic literature with regards to the transformation of a BPMN model into a smart contract. Subsequently, we introduce Hierarchical State Machines (HSMs) and the proposed Discrete Events Hierarchical Finite State Machines (DE-HSM) multi-modal modeling. The latter integrates Discrete Event (DE) modeling for concurrent actions with HSM/Finite State Machine (FSM) modeling for functional aspects. In addition, we review the concept of transactions across various domains to comprehend and define the Long-term Blockchain (LtB) transaction, reflecting real-world collaborative trade transactions. This overview is aimed at establishing a foundation for understanding the context and technical nuances of our research and analysis.

## 2.1 BLOCKCHAIN TECHNOLOGY

A blockchain platform serves as a decentralized, unalterable, and replicated data repository autonomously governed via a peer-to-peer network and a distributed replicated virtual machine, employing distributed timestamping servers. As a secure, manageable, and distributed system, it has seen rapid adoption across various industries since its emergence in 2008. Blockchain networks come in two forms: public and private. Public blockchains welcome anonymous users, and their incentive mechanisms promote increased participation. Conversely, private blockchains employ predefined rules to regulate access to specific transactions, often incorporating sophisticated access control and organization management to maintain security and privacy. When contrasted with public blockchains, private ones deliver superior performance and throughput thanks to their expedited consensus mechanism and the integration of high-capacity NoSQL databases.

## 2.2 Smart Contracts

Smart contracts denote self-enforcing agreements, wherein the contractual stipulations are directly inscribed into code. These contracts autonomously execute preordained actions when the specified criteria are fulfilled, thereby ensuring enforcement of the contractual terms. The notion of smart contracts was extensively popularized by Ethereum through the introduction of a Turing-complete virtual machine (the Ethereum Virtual Machine or EVM). The EVM has the capability to execute varied code, thus enabling a high level of complexity and versatility in contractual functionality.

The multifarious applications of smart contracts transcend the boundaries of trade transactions, including financial transactions. Smart contracts play an essential role in various applications such as the construction of decentralized applications (dApps), automation of business procedures, orchestration of supply chains, and enabling of voting systems, to name a few. By dispensing with the need for intermediaries and facilitating automated contract execution, smart contracts contribute to cost reduction, enhancement of efficiency, and strengthening of transparency and trust within a wide array of industries.

However, the development of smart contracts, when juxtaposed with conventional software development, brings forth an array of challenges including but not limited to the diversity of development languages, absence of standard integrated development environments, inadequate smart contract modelling tools, intricacies in deployment and maintenance, and a steep learning curve. To surmount these challenges, design patterns such as "Fail Early and Fail Loud," "State Machine," "Upgradable Registry," "Transition Counter," and others are being given significant focus by blockchain researchers and practitioners (HeartBank, 2018; Mavridou & Laszka, 2018).

## 2.3 Trade Applications

Blockchains and their smart contracts are not suitable for all types of applications. Smart contracts are not suitable for applications that exhibit heavy computational requirements as execution of smart contact transactions requires validation involving many nodes of the blockchain network. As their name implies, smart contracts are particularly suitable for applications that involve collaboration of different actors, possibly of different organization, with varying requirements on information availability, security, and privacy, amongst others. And trade applications involve activities that exhibit those properties and thus should be suitable for blockchain applications - as is confirmed by a much higher adoption of blockchains in trade, which also includes finance in trade and hence also Decentralized finance (DeFi). Thus, trade and finance applications are our target area, which is a broad area involving third parties for the purposes of efficient processing of transactions by participants, such as banks, customs, shippers, and insurance companies.

## 2.4 Platforms for Smart Contract Development

Numerous platforms are available to facilitate the development and implementation of smart contracts, with Ethereum leading the pack due to its widespread use and support for smart contract development via its native language, Solidity. Ethereum offers a strong ecosystem for developers, complete with a large community and documentation.

Other noteworthy platforms encompass:

- *Ethereum Testnets*: Ethereum Testnets serve as experimental environments where developers can test their applications without spending real Ether (ETH). They mimic the Ethereum network and operate with the same rules. The various Ethereum Testnets include Ropsten, Rinkeby, and Kovan, each offering different consensus mechanisms or functionalities. Developers often use Testnets to test smart contracts and dApps, ensuring their smooth operation before they are deployed onto the Ethereum Mainnet, thus minimizing the risk of errors and loss of Ether (Ethereum Testnet, 2023).
- *Quorum*: A private, permissioned blockchain platform based on Ethereum, designed primarily for finance industry use cases. It supports the development of smart contracts using Solidity, like Ethereum, but also provides enhanced privacy and security features (Quorum, 2023).
- *Hyperledger*: An open-source, permissioned blockchain tailored for enterprise use. Hyperledger accommodates smart contract development in a variety of programming languages, including JavaScript, Go, and Java (Hyperledger Fabric, 2023).

- *Microfab*: Microfab, often associated with Hyperledger Fabric, is a permissioned, private blockchain infrastructure that serves as a development tool. Its primary utility is to create a lightweight and simplified development environment. This allows developers to experiment and debug smart contracts locally before deploying them to a production network. Microfab provides a quick and simple method to stand up a minimal Hyperledger Fabric network, accommodating the rapid prototyping of blockchain solutions (Microfab, 2023).

## 2.5 FSMs, Hierarchical State Machines, and Multi-Modal Modeling

Because FSM modeling has been used frequently in the design and implementation of software, the FSM modeling has been expanded with features, such as a guard along an FSM transition to specify a Boolean condition on the state's variables that must evaluate to true for the transition to take place. In the late 80's, FSMs were extended to address the issues of reuse of patterns with the concept of hierarchy, leading to Hierarchical State machines (HSMs) that may contain states that are themselves other FSMs. Harel (1987) showed that FSMs can be combined hierarchically: A single hierarchical state at one level can be considered to be in several states concurrently as represented by an FSM(s) in a lower level of the hierarchy, and FSMs may also be combined leading to concurrent FSMs.

An HSM can be defined using induction as follows (due to (Girault et al., 1999) and described in (Harel, 1987), (Yannakakis, 2000), and others): In the base case, an FSM is a hierarchical machine. Suppose that M is a set of HSMs. If F is an FSM with a set of states, S, and there is a mapping function f: S→M, then the triple (F, M, f) is an HSM. Each state, $s \in S$, that represents an HSM is replaced by its mapping (f(s)). HSMs recognize the same language as its corresponding flattened FSM. HSMs do not increase expressiveness of FSMs, only succinctness in representing them.

Girault et al. (1999) describe how HSM modeling can be combined with concurrency semantics of a number of several models, including communicating sequential processes (Hoare, 1987) and discrete events (Cassandras, 1993). Girault et al. (1999) describe how an HSM model can represent a module of a system under a concurrency model that is applicable only if the system is in that state. This enables representation of a subsystem using a particular concurrency model that may be nested within a hierarchical state of a higher-level FSM. This may be used in multi-modal modeling, in which different (hierarchical) states may be combined with different concurrency models that are best suitable for modelling of concurrent activities for that particular state. We exploit the concept of multi-modal modeling to allow the designer to model concurrent, but independent activities, by concurrent FSMs at the lower level of hierarchy. In (Liu et al., 2022a), we showed that a multi-modal model that combines Discrete Event (DE) modeling with FSM modeling may be used to model trade finance applications – such a combination of models is referred to as DE-HSM multi-modal modeling. We also showed that it is possible to automatically transform a DE-HSM model of such an application into methods of a smart contract deployed on a blockchain.

A DE-HSM has external inputs, and it produces outputs. The model represents how external inputs form inputs to the sub-models and how those sub-models are interconnected to produce the final output. However, if the sub-model's interconnection is such that there are no loops, then the model can be viewed as a zero-delay model (Yannakakis, 2000) in which the individual DE queues may be combined into one DE queue.

## 2.6 Business Process Model and Notation (BPMN)

BPMN was developed by the OMG organization (Business Process Model and Notation (BPMN), Version 2.0. 2023) with the objective of BPMN models to be understandable by all business users, from business analysts, through technical developers implementing the processes, to people managing those processes. It is viewed as a de-facto standard for describing business processes. That it has been adopted in practice is demonstrated by many software platforms available that provide for modeling of business applications with the objective to automatically create an executable application from the BPMN model. For instance, Oracle Corporation uses BPMN to describe an application and transform into a blueprint of processes described in an executable Business Process Execution Language (BPEL) (Dikmans et al., 2008): The blueprint represents the logic of the application in terms of concurrent processes and their interactions, while details of individual tasks are supplied by implementors. Another example is Camunda software platform that is also used to develop a BPMN model that is then transformed automatically into a Java application (Deehan, 2021).

BPMN standard categorizes the BPMN elements into five basic categories, wherein each category may have sub-categories. As there are many good introductions and overviews in the literature, and we overviewed BPMN in some detail in (Bodorik et al., 2023), we concentrate on the description of the BPMN use case, shown in Fig. 2.1, that has been used in several research reports on transforming BPMN models into smart contracts, for instance in (Weber et al., 2016; Tran et al., 2018; López-Pintado et al., 2019).

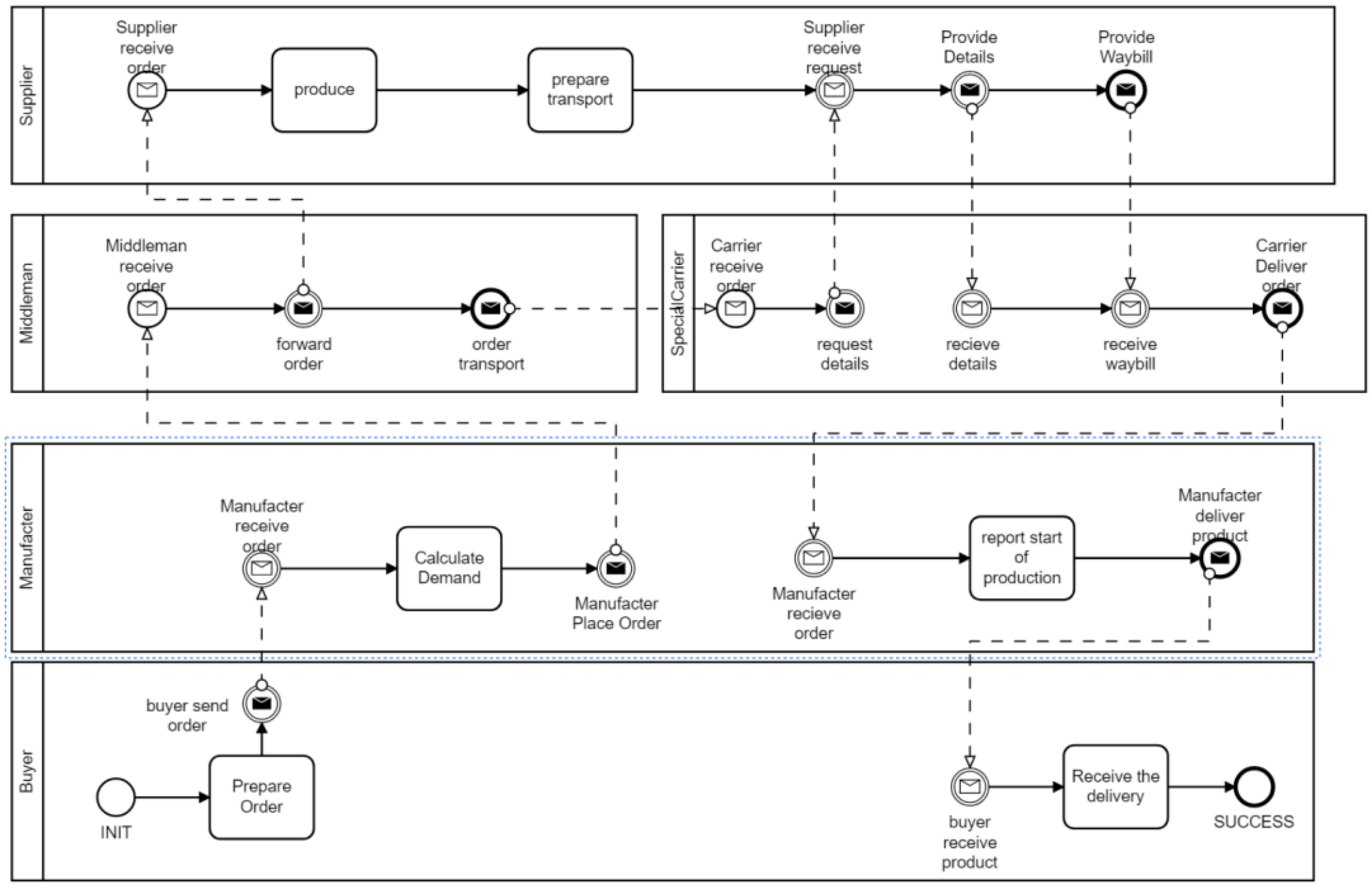

Fig. 2.1 A BPMN Model for A Supply-Chain Example (Adopted from (Weber et al., 2016))

This use case, depicted in Fig. 2.1, was adopted from (Weber et al., 2016) to be used as a sample use case. The supply-chain use case begins with the buyer issuing a new order. Once the manufacturer receives the order, she/he calculates demand and places an order with a middleman. Middleman concurrently sends the order to the supplier and orders transport from carrier. The producer fabricates the product and prepares it for transport. Carrier, upon receiving the request from the middleman, requests details from the supplier. Supplier provides the details to the carrier and then it prepares and provides the waybill to the carrier. Upon receiving details about the product and the waybill, both from the supplier, the carrier delivers the order to the manufacturer. Upon receiving the order, the manufacturer starts the production and when that is finished, it delivers the product to the buyer, who receives the order.

There are five actors (Buyer, Manufacturer, Middleman, Special Carrier, and Supplier), with each actor having a pool, or as a swimlane within a pool, that is represented by a large rectangle representing processing performed by the actor. Arrows, represent connecting objects, which are either sequence flows or message flows. Only message flows may cross pools. It should be noted that some of the BPMN elements represent tasks that appear in the diagram as smaller rectangles with rounded corners. A task is an activity, which is a script provided by the developer, that is executed when the flow of execution, as represented by arrows and other elements, reaches the task. For each task, the developer provides the code. The remaining BPMN elements are used to represent the coordination of the collaborative activities as performed by the actors. Furthermore, when two or more arrows exit out of an element, the element serves as a split/fork/diverging gateway, such that each arrow represents a concurrent stream of activities performed by the actor. When more than one arrow end at a BPMN element, that element serves as a join/merge/converge gateway representing joining of concurrent processing streams into one stream. Split gateways may be inclusive or exclusive, wherein an exclusive gateway allows only one stream to leave, while inclusive may allow many streams to proceed with execution concurrently. An inclusive joining gateway passes a token only if all incoming paths are enabled. Both splitting and merging gateways may have a condition specified on each path.

## 2.7 Bridge Process between Mainchain and Off-Chain

We have developed an approach for transforming the HSM model into smart contracts, enabling execution across multiple chains (Bodorik et al., 2021; Liu et al., 2021a). This transformation is essential for handling the interoperability of off-chain computations within the broader blockchain application.

To facilitate this, we included an interface within the smart contract, specifically designed for interaction with the pattern. This pattern refers to independent components of the blockchain application that are processed through off-chain computation. As the execution on the sidechain progresses, the pattern is automatically deployed there, ensuring seamless integration between the mainchain and sidechain. For the interaction and signaling between on-chain and off-chain computations, we utilized the InterPlanetary File System (IPFS) protocol, employing IPFS Message channels. IPFS serves as a distributed file system that enables data sharing across different nodes, providing a robust and efficient means of communication between the mainchain and sidechain (Bodorik et al., 2021; Liu et al., 2021a). This bridge process ensures that the off-chain computations are well-coordinated with the mainchain.

# CHAPTER 3 LONG-TERM BLOCKCHAIN TRANSACTIONS (LTB) AND MULTI-METHOD (MM) TRANSACTIONS

Understanding the concepts of ***L***ong-***t***erm ***B***lockchain Transactions (***LtB***) and ***m***ulti-***m***ethod (mm) transactions on the blockchain is crucial. It's essential to explore and define these concepts, as well as to address the specific transactional properties required to support collaborative transactions running on the blockchain. These issues are explored in subsections 3.1, 3.2, and 3.3.

- In 3.1 section, we compare the notions of transactions across different domains: databases (DB), real-world trade transactions, and blockchain. Through this comparison, it becomes apparent that there is no official, standardized definition of long-term transactions on the blockchain that adequately meets the requirements of real-world trade transactions.
- In 3.2 section, we delve into the details of Long-term Transactions (LtB), defining them and outlining the specific properties that must be supported. Long-term transactions are characterized by their extended duration and the potential complexity of interactions among multiple participants. Understanding these properties is essential for designing and implementing effective smart contract methods that can handle the intricacies of long-term transactions on the blockchain.
- In 3.3 section, we focus on the concept and properties of Blockchain multi-method Transactions. These transactions represent a technical approach to implementing LtB transactions. By breaking down the high-level concept of long-term transactions into specific methods and understanding their properties, we propose alternative mechanisms for supporting multi-method transactions properties through different access methods to private workspaces. This approach ensures that the implementation aligns with the unique requirements of enabling long-term transaction mechanisms on various deployment modes.

## 3.1 TRANSACTIONS IN DBS, TRADE, AND BLOCKCHAINS

To gain insight into trade transactions and ascertain the transactional properties, we provided an overview of the concept of transactions within the context of DBs, trade, and blockchains. We touch upon transactions within DB systems, as one could argue that the concept of transactions, concerning the storage, management, and manipulation of stored data, was initially comprehensively defined for DB systems. By understanding the differences in transactions among these domains, a better understanding of trade transactions can be achieved.

### 3.1.1 DB Transactions

Since the concept of transactions in databases is well-established and taught in DB courses at most post-secondary academic institutions, we provide only a brief overview of this topic. In summary, a transaction in a database is a sequence of read and write DB operations that are enclosed within transaction begin and end instructions. The database management system (DBMS) ensures that the transaction adheres to the ACID properties, which include *Atomicity* (all or nothing), *Consistency* (correctness while performing concurrent transactions), *Isolation* (no dirty reads), and *Durability*. When a transaction involves multiple users or participants, each DB operation must identify the user who issued the operation, which is necessary for the DBMS to enforce access control.

### 3.1.2 Trade Transactions

We already provided motivation for a trade transaction in the context of the trade of goods and services. A trade transaction often involves the collaboration of multiple participants in long-term activities. We define a trade transaction as a set of collaborative activities made by the transaction participants in unpredictable sequences, activities that may also be dependent on each other and span relatively long duration (Liu et al., 2022b). Currently, transactions in the exchange of goods and services lack a formal definition and explicit articulation of their properties.

### 3.1.3 Blockchain Transaction

In the realm of blockchain execution, the execution of a smart contract method results in the submission of any state changes for inclusion in a subsequent block. This block undergoes a verification process to confirm that changes induced by the smart contract method are in alignment with the current state of the ledger and all other transactions within the block. This process ensures that the writes initiated by a smart contract method are atomic, meaning that all the writes are either applied to the blockchain or not at all. These writes are also consistent and isolated, preventing the transaction's updates from being accessed by any other smart contract method or non-participant entities until their respective block is added to the blockchain. Once changes are disseminated to the replicated blockchain, they become permanent and are not subject to loss after the successful completion of a transaction.

Before delving into the intricacies of blockchain transactions, it is crucial to understand the developer's perspective of reading from and appending to the blockchain ledger. The immutability of the blockchain ledger allows only reads and appends at the ledger's end. However, creating smart contract methods using only reads and appends can be challenging and prone to errors. To simplify the process of writing smart contracts, blockchain infrastructure provides developers with a view of the ledger data as an associative data store, or database. Examples of such databases include Level DB, supported by Ethereum and Hyperledger, and Couch DB, supported by Hyperledger Fabric.

This associative database enables developers to use familiar database operations to read and write values. For instance:

- The *Write(key, value)* method takes two inputs - a key and a value - and stores the value on the ledger using that key.
- The *Read(key)* method retrieves the value associated with a given key from the ledger.

Each time a smart contract method writes to the ledger using a key, the blockchain infrastructure creates a new instance or version of that object if it already exists. The infrastructure then stores the new object instance on the ledger in a Merkle-tree data structure and keeps track of the order of the versions of any object. As a result, developers view the ledger as an associative data store, while the blockchain infrastructure transforms the reads and writes on the associative data store to reads and appends of new object versions on the immutable ledger itself.

The documentation for different blockchains often lacks precision when describing the concept of a transaction, especially when a blockchain has its native currency. In such cases, the documentation may refer to transferring funds as a transaction, or it may refer to changes in the state of the ledger. To address this ambiguity, we adopt the following definition of a transaction: *A blockchain transaction is a set of updates to the ledger state that result from the execution of a single smart contract method.* According to this definition, a transaction encompasses a series of reads and writes to the ledger that are performed by an execution of any smart contract method.

## 3.2 Transactional Properties for Long-term Blockchain Transactions (LtB) and Multi-Method Transactions

From comparing transaction concepts across various domains, it became evident that there is not an official or standardized definition for long-term transactions on the blockchain that mirrors real-world collaborative trade transactions. Current blockchain infrastructure lacks support for such transactions. Thus, this subsection focuses on defining Long-term Transactions (LtB), detailing the necessary transactional properties for blockchain support, and creating a transactional mechanism in an automated fashion to enforce the transactional properties.

We introduced collaborative trade transactions and defined long-term transactions for real-world trade applications using blockchain (Liu et al., 2022b). Trade transactions can be lengthy, involving multiple participants over varying periods, and may necessitate collaboration to achieve a shared objective. A subset of participants may perform activities specific to their group, requiring atomic collaboration, meaning all activities must be completed successfully or not happen at all.

We proposed modeling trade transactions with BPMN and then transforming them into multi-chain smart contract methods. We identified BPMN limitations in modeling long-term trade transactions, and we introduced the concept of BPMN collaborative long-term transactions (Bodorik et al., 2023). According to the BPMN specification (BPMN 2.0

Introduction, 2022), a BPMN transaction is confined to a subprocess within a swimlane, limiting it to local execution by a single actor. However, trade transactions need collaboration among multiple actors, prompting the authors to extend the BPMN transaction subprocess concept across swimlanes.

Another BPMN limitation is that transaction process boundaries are determined by an actor's entry and exit from the subprocess through incoming and outgoing sequence flows. In trade transactions, a subprocess represents a collaboration of actors/participants in performing the operations of the transaction. We defined a collaborative BPMN transaction that has the following characteristics (Bodorik et al., 2023):

- One incoming and one outgoing sequence flow, situated in its pool.
- Transactional activity scope defined by specific semantics, including transaction-status and activity duration.
- Atomicity, consistency, privacy, isolation, and durability are enforced.

Additionally, we introduced the concept of a Long-term Blockchain (LtB) transaction to meet the requirements of supporting the collaborative long-term transactions of trade of goods and services using multi-method transactions that may be deployed in multi-chain environment (Liu et al., 2023a).

Definition: An LtB transaction is a subset of smart contract methods with the following features:

- Transaction methods only share transaction state and don't refer to external data/objects or invoke external methods.
- Two special methods: LtB-transaction-begin (initiating the transaction) and LtB-transaction-end (completing the transaction).
- Transaction state represented by an object with attributes, persisting only for the transaction duration.

The necessity to define certain properties to facilitate enforcement through long-term transaction support mechanisms prompted us to review database and blockchain transaction properties. This review aimed to discern which properties are required and are desirable for long-term blockchain transaction mechanisms. Properties shared by both database and blockchain systems are considered essential, while those supported by only one system are viewed as desirable. When it comes to ensuring atomicity, consistency, and isolation, the blockchain system has a unique approach. Whenever a smart contract method is executed, the resulting updates are bundled into a block, which is then added to the blockchain. The process of block verification ensures that these updates align with the existing ledger state and all other transactions within that block. This process ensures that all changes are atomic, consistent, and isolated. Furthermore, the consensus mechanism in place ensures that all updates within a block are consistent. Since smart contract methods can only access and modify transaction updates after their respective block has been added to the chain, there's no risk of "dirty reads", ensuring isolation. Durability is a given in the blockchain world; once updates are in a confirmed block, they're there to stay, replicated across the network. This ensures no loss of data post-transaction completion. Thus, while both traditional databases and blockchains adhere to the ACID principles, the way they handle transactions and the scope of application access differ between the two systems.

Therefore, the LtB transactions must adhere to the ACID properties, as both database and ledger operations require atomicity, consistency, isolation, and durability within transactional operations. The main difference between databases and blockchains is the scope of transactions. In databases, a transaction may involve multiple CRUD operations on the database, with each operation individually submitted by an application. In contrast, a blockchain transaction consists of a set of ledger reads and writes executed by a smart contract method. Both database and blockchain systems follow the ACID properties, but to support LtB transactions, the constraint that a transaction is limited to ledger updates made by an execution of a smart contract method must be overcome to allow long-term transactions across multiple smart contract method invocations.

Access control also differs between database and blockchain systems. In relational databases, access control manages access to relations with different permissions assigned to various CRUD operations. Blockchains provide access control to smart contracts rather than ledger data, specifically they provide access control to smart contract methods. The type of access control depends on whether the blockchain is permissioned or public. In a public blockchain, anyone can create an account and develop a smart contract that accesses ledger data. Access control to smart contract methods is not enforced by the public blockchain infrastructure. However, control over funds stored in a contract is restricted: only the smart contract owner with the associated private key can transfer or "spend" funds within the contract. In a permissioned blockchain, access control for smart contract methods is based on authenticating the actor initiating the smart contract method invocation. As a result, access control for LtB transaction methods is essential and should be supported.

Privacy, as previously mentioned, ensures that only specific users can access an object written to the ledger. In databases, privacy is achieved by controlling access to relations, allowing users to view data only if they have read access. In blockchains, there are no limitations on read/write access to the ledger—if a user can create smart contracts and execute their methods, no restrictions apply to the ledger data accessed by these methods as long as they have information on where on the blockchain the data is stored. However, an LtB mechanism should support privacy by permitting only LtB transaction participants to access the associated ledger data.

In summary, to facilitate long-term blockchain transactions:

- An LtB transaction must include ledger reads and writes executed by a specified subset of a smart contract's methods.
- The ACID properties must be upheld.
- Access control for users interacting with LtB transaction methods needs to be supported.
- Privacy should be maintained, ensuring that only actors with access to transaction operations can view the results of those operations within the transaction, preventing non-participants from accessing such information.

In (Liu et al., 2022b), we investigated the extent to which blockchains natively support the ACID properties of consistency, durability, and isolation, which are supported by all major relational database systems. We discussed how these properties are supported in blockchains and also highlighted additional properties that distinguish blockchain transactions from database transactions.

Smart contracts must support privacy, especially when coordinating activities among users of different organizations or departments. Public blockchains like Bitcoin or Ethereum do not offer privacy as they rely on the anonymity of account owners, while private blockchains like Hyperledger provide identity verification and mechanisms like channels for confidential information sharing among selected groups of users. However, the initial setup and use of channels can be complex, potentially introducing errors by developers.

For scalability, blockchain-based applications have limitations due to the characteristics of the blockchain network. For example, Bitcoin faces scalability issues due to its limited capacity to handle large amounts of transaction data, while Ethereum costs approximately $17,100 USD for 1MB of data (Miner US, 2023). The high cost of storing data using Bitcoin has been discussed by many researchers, making blockchain systems not feasible for storing distributed application data due to the expensive storage cost. Additionally, computation frequency is also a challenge for scalability, currently with Bitcoin only allowing 5 transactions per second and Ethereum only allowing about 15 transactions per second due to its network consensus.

As a current transaction on blockchains, transaction that is the result of execution of any smart contract method, supports ACID properties, an LtB transaction on a blockchain that involves executions of more than one smart contract method must also support the ACID properties. And (i) how to define a mm transaction mechanism on a blockchain and (ii) how to enforce the ACID properties of LtB transactions are the objectives of this section.

## 3.3 Transactional Mechanism for Blockchain Multi-method Transactions

As discussed in previous subsections, blockchains do not inherently provide the concept of a ***m***ulti-***m***ethod (***mm***) transaction. In (Liu et al., 2023a) we describe how to define an mm-transaction as a subset of the methods of a smart contract while ensuring that the transaction's methods are independent. The independence property is vital as it prevents the transaction methods from referencing objects or invoking methods that are declared externally to the methods of the transaction. Such external references would violate the atomicity and isolation properties of the transaction. For example, if a transaction invokes an external method X, and the transaction is then aborted, the activities of method X would also have to be aborted or negated. This would violate the isolation property, as the effects of the transaction would not be confined solely to the transaction itself.

This section emphasizes the importance of understanding and carefully managing the relationships between methods within a multi-method transaction. By ensuring that the methods are independent and that the atomicity and isolation properties are maintained, it is feasible to create robust and reliable multi-method transactions on the blockchain.

As expounded in (Liu et al., 2023b), the implications of providing a transactional mechanism to support a blockchain mm-transaction are as follows:

1) Blockchain is immutable. Thus, to support an mm-transaction, optimistic methods need to be used in support of the transaction commitment and the transactional properties. An mm-transaction method must perform ledger writes in some private workspace and only when the transaction is committed are the writes written from the private workspace to the ledger.
2) A private workspace is required to facilitate transaction methods that involve deferred modification. This workspace stores reads and writes to the ledger until the transaction is committed. The cached data is then applied to the ledger only upon transaction completion. Furthermore, the private workspace must be shared across executions of the transaction methods.

In (Liu et al., 2023a), it is demonstrated that to share information across the executions of the transaction method(s) of a smart contract, such information needs to be stored on the ledger or in a special data structure that persists across executions of methods of a smart contract, which is only possible if a blockchain provides such a data structure. An example is private data provided by the Hyperledger Fabric blockchain in the form of a private blockchain shared only by specified actors. If such a data structure is not provided, then there is no alternative but to store the shared information on a blockchain, which may be further supported by off-chain storage techniques.

To automatically generate a transaction mechanism, we use pattern augmentation techniques (Liu et al., 2023a). The developer first creates the smart contract, including the methods intended to constitute the transaction, wherein the transaction methods must satisfy the independence property. Once the smart contract methods are created, the developer identifies/marks the transaction methods so that they are known to the preprocessor as methods belonging to the mm transaction. Before the smart contract methods are compiled, they are processed by a preprocessor that amends each of the transaction methods in the way described below. In addition to the smart contract methods, the preprocessor is provided with additional information on who the actors are that participate in the transaction and information on how the private workspace is provided, which is discussed later. In brief, the preprocessor amends the smart contract methods with patterns as follows:

- Begin transaction method is inserted. It initializes the private workspace that must persist across executions of the transaction methods and be accessible to all methods of the transaction. The private workspace acts as a cache to store:
  - The state of the transaction that is managed only by the patterns augmented by the preprocessor. In other words, the transaction methods prepared by the developer are unaware of both the cache and the object that is used to store the state of the transaction – these are augmented by the preprocessor.
  - To store in the cache the ledger reads and writes performed by the transaction methods.
- Each read or write to the ledger made by any of the transaction methods must be replaced with a pattern to make that read or write using the cache instead.
- End of the transaction method is inserted. It propagates all transaction reads and writes, made using the cache, to the ledger itself.

The begin and end transaction methods and the declaration and manipulation of the object representing the state of the transaction are facilitated by the preprocessor using pattern augmentation techniques.

We support the following options for the private workspace/cache (Liu et al., 2023b):

1) Cache is hosted in the private data structure if it is provided by the native blockchain.

2) The cache is hosted on the ledger in locations that are known only to the methods of the smart contract and hence the cache is inaccessible by other non-transaction methods.
3) Slave smart contract is used to host the cache and the methods of the mm-transaction. The methods are invoked by the main smart contract that contains all non-transaction methods that invoke the transaction methods of the slave smart contract. There are two variations:
    a. The slave smart contract is deployed on the same chain as the main smart contract.
    b. The slave smart contract is deployed on a sidechain, while the main smart contract is deployed on the mainchain.

It is shown in (Liu et al., 2023a) that all the above options support the ACID properties. How augmentation is used to support the properties of access control and privacy is also described. Access control assures that only those actors who are participating in the transaction can invoke the transaction methods. This is attained by augmenting each transaction method with a pattern that checks that the actor, causing the invocation of the transaction method, is in the list of the actors that may participate in the transaction.

There are difficulties with supporting the privacy property that states that only the actors that participate in the transaction have access to the cache. Although access control to the methods is supported, the issue is that any actor that has access to deploy smart contracts on the blockchain has access to any data that is on the blockchain ledger. For privacy, (Liu et al., 2023b) examines architectures available to support the transaction as described above. For instance, if the transaction methods are deployed in a separate smart contract on a sidechain, privacy is supported as long as the actors who do not participate in the transaction do not have privileges to access the sidechain. Another option for privacy is to use encryption on the private workspace; however, the cost estimates indicate that this may be expensive if public-key cryptography is used to encrypt all data stored in the private workspace. As an alternative, in order to decrease the cost of cryptography, instead of encrypting all data stored in the cache, which is on the ledger, only the location of the cache on the ledger is encrypted: If an attacker does not know where on the ledger the data is stored, the attacker cannot view the stored data.

Cost estimates for each of the alternative approaches is also provided by comparing the costs. Further details may be found in Chapter 6.

# CHAPTER 4 TRANSFORMATION OF A BPMN MODEL INTO SMART CONTRACT METHODS

This chapter delves into the process of transforming a Business Process Model and Notation (BPMN) model into the methods of a smart contract, as outlined in (Bodorik et al., 2023). The initial subsection provides a comprehensive review of the significant aspects of this transformation process. Subsequently, the focus shifts to the analysis of the graph representation of the multi-modal model to identify Single-Entry, Single-Exit (SESE) subgraphs. It is posited that a collaborative transaction in a BPMN model should be such that the transaction's activities do not export any information outside the transaction. This localization property is inherent in SESE subgraphs, making them suitable candidates for specifying a transaction at a BPMN level as a collaboration of activities by different actors. It is noteworthy that the BPMN specification does not include such a transaction construct, as in BPMN, a transaction may only be defined on a sub-process executed within the context of a single actor, thereby precluding its use in supporting collaborations of several actors.

Our approach identifies all SESE subgraphs within the DE-FSM model and subsequently transforms the subgraphs back to the BPMN model representation. This transformed model is presented to the developer, who decides which of the SESE subgraphs should be transformed into transactions satisfying the Atomicity, Consistency, Isolation, Durability (ACID) properties. Moreover, when the BPMN model is transformed into the methods of a smart contract, the activities for each of the SESE subgraphs selected by the developer as a transaction are transformed into a separate smart contract method. This transformation process satisfies the independence property as defined in (Liu et al., 2023a). As a result, the transactional mechanisms described in (Liu et al., 2023a) for multi-method transactions support are applied to the methods generated from the SESE subgraphs selected by the developer.

The subsequent subsections elucidate the overall process of transforming a BPMN model into the methods of a smart contract under the developer's guidance. This includes support for specifying nested collaborative BPMN transactions and a description of how transactional support for the BPMN collaborative transactions is achieved, even when nested to support sub-transactions.

## 4.1 TABS Transformations of a BPMN Model into Smart Contract Methods

The process of converting a Business Process Model and Notation (BPMN) model into the methods of a smart contract, as outlined in the description of TABS(+) tool detailed in (Bodorik et al., 2023; Liu et al., 2023b), is succinctly recapped in this section, albeit with some specifics omitted for brevity. The sub-sections of 4.1 elucidate the following facets:

- Section 4.1.1 elucidates the primary assumptions imposed on the BPMN model.
- Section 4.1.2 portrays the overarching architecture of the transformation process, which converts a BPMN into smart contracts deployed in a multi-chain environment.
- Section 4.1.3 introduces the pre-processing steps and the formation of a Directed Acyclic Graph (DAG) representation of the BPMN model. It further explains how a DAG model is transformed into the multi-modal Discrete Event Hierarchical State Machine (DE-HSM) and subsequently into Discrete Event Finite State Machine (DE-FSM) models.
- Section 4.1.4 clarifies the concept of long-term transactions, the transactional properties that need to be supported, and how the proposed DE-HSM supports long-term transaction execution and transactional properties such as Atomicity, Consistency, Isolation, Durability (ACID), access control, and privacy concerns.
- Section 4.1.5 delineates the transformation of the DE-FSM model into the methods of a smart contract(s).

### 4.1.1 Assumptions

This subsection briefly outlines the assumptions made in (Bodorik et al., 2023) regarding a BPMN model and its transformation into a smart contract:

- *Assumptions on BPMN looping and parallel construct*: As in other work on transforming the BPMN model into a smart contract, we limit the semantics of BPMN looping and parallel constructs to ensure model determinism. These constructs are treated as special cases in the final phase of generating smart contract methods. The elimination of loops from BPMN models facilitates the use of multi-modal modeling, where Discrete Event (DE) modeling represents concurrent execution, and functionality is expressed using Hierarchical State Machine (HSM) or Finite State Machine (FSM) modeling.

- *Assumptions on BPMN converging gateways*: We make a simplification for an inclusive converging gateway in that we simply pass the token - we check the other pathways neither for enablement nor for a token arrival.
- *Coverage of BPMN*: Not all BPMN symbols are supported. The list of currently supported symbols can be found in (Bodorik et al., 2023). Fig. 4.1 shows the symbol that our TABS tool supports, albeit as noted in assumptions above, we impose limits on looping and parallel subprocesses and arbitrary loops in BPMN diagrams.

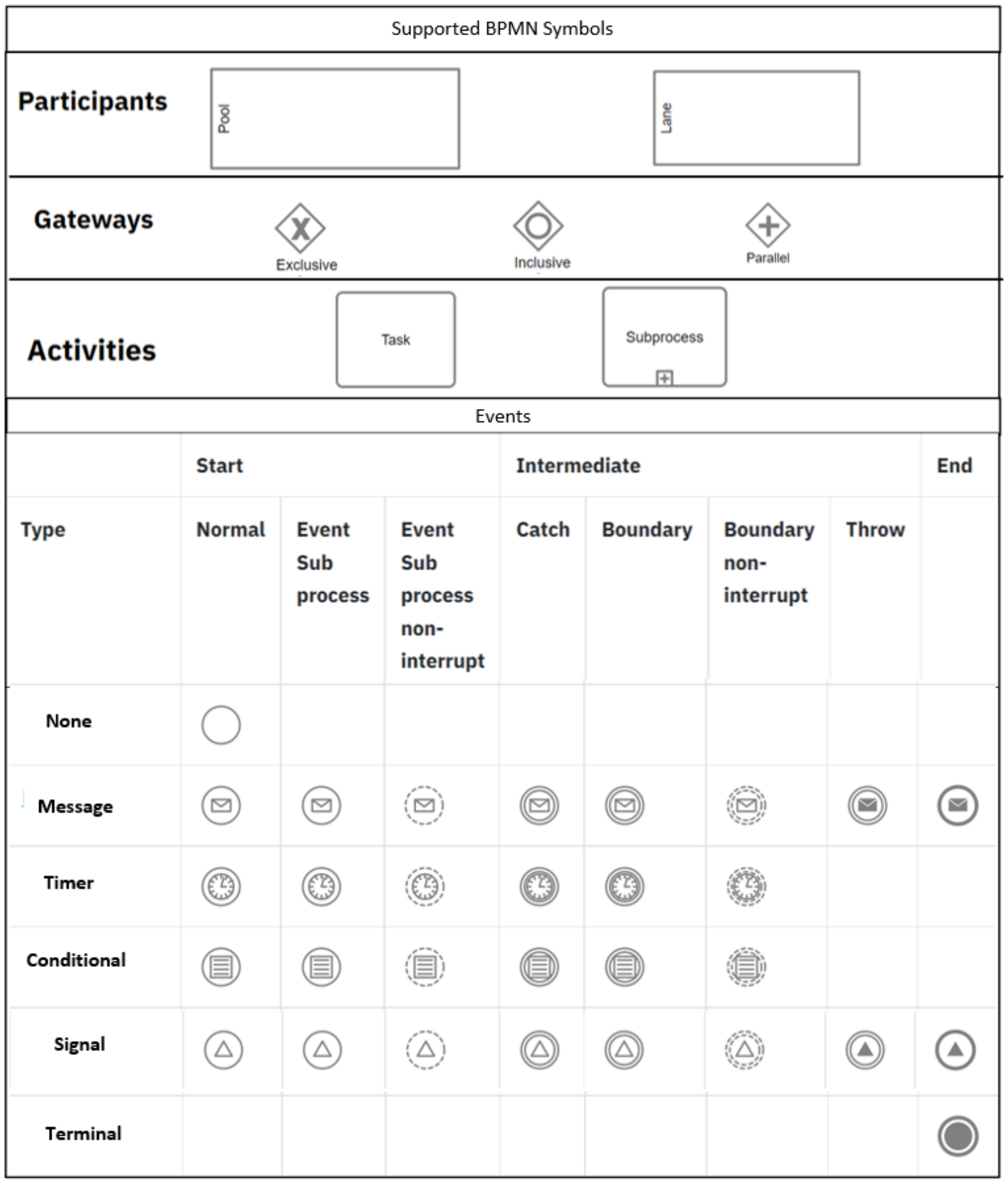

Fig. 4.1 BPMN Symbols Supported by TABS+ (Adopted from (Bodorik et al., 2023))

- *Task element*: The BPMN task element represents a self-contained task. The transformations prepare the method skeleton for each task element, and a task is invoked according to the BPMN model description. However, the code/script for the task must be supplied by the developer. This approach is standard in creating applications from a BPMN model. For instance, Oracle Corporation uses the same approach in its Oracle Business Process Analysis Suite when transforming a BPMN diagram into a blueprint for executable Business Process Execution Language (BPEL) processes (Dijkman et al., 2008) and a similar statement also applies to the Camunda platform that also asks the modeler to provide the script for each of the PBMN model's tasks.
- *Off-chain storage*: We adopt the standard practice to store large objects/data off-chain and store on the mainchain only the hash-code of the data stored off-chain. Any time the object is retrieved by a smart contract, to ensure the immutability property of an object stored off-chain, its hash-code is recalculated and checked against the hash code stored on the blockchain. We facilitate off-chain storage by using IPFS, which is a distributed system for storing and accessing files, websites, applications, and data (Steichen et al., 2018). IPFS storage is content addressable using the hash-code of the object. Storage in IPFS may be replicated and by controlling replication desirable resilience to storage failures may be achieved (Obe, 2022; Nabben, 2022).

### 4.1.2 Architecture Overview

We provide a high-level overview of our proposed system architecture, that is designed to facilitate the transformation of Business Process Model and Notation (BPMN) into multi-method smart contracts within a cross-chain environment (Bodorik et al., 2023). The system architecture for the transformation of BPMN into multi-method smart contracts is depicted in Figure 4.2, which delineates two primary phases: the Design and Deployment/Installation phase, and the Execution/Run-time phase. In the initial phase, the BPMN diagram undergoes transformation first into the DE-HSM model, subsequently into the methods of a smart contract. However, the transformation process not only generates the smart contract methods along with its supporting run-time management information, but also prepares API methods for utilization by a Distributed Application (dApp) to interact with the methods. Post the preparation of the smart contract, it is deployed during the Design and Deployment/Installation phase, which involves the deployment of the smart contract on the blockchain and the installation of the API and associated methods on a network node, thereby enabling a dApp to communicate with the smart contract. The smart contract incorporates a monitoring module with the run-time management information stored in the data structures prepared during the design phase. These include the DE queue of events and FSM definitions corresponding to the DE-FSM sub-models and FSMs to support concurrent processing within individual DE-FSM sub-models.

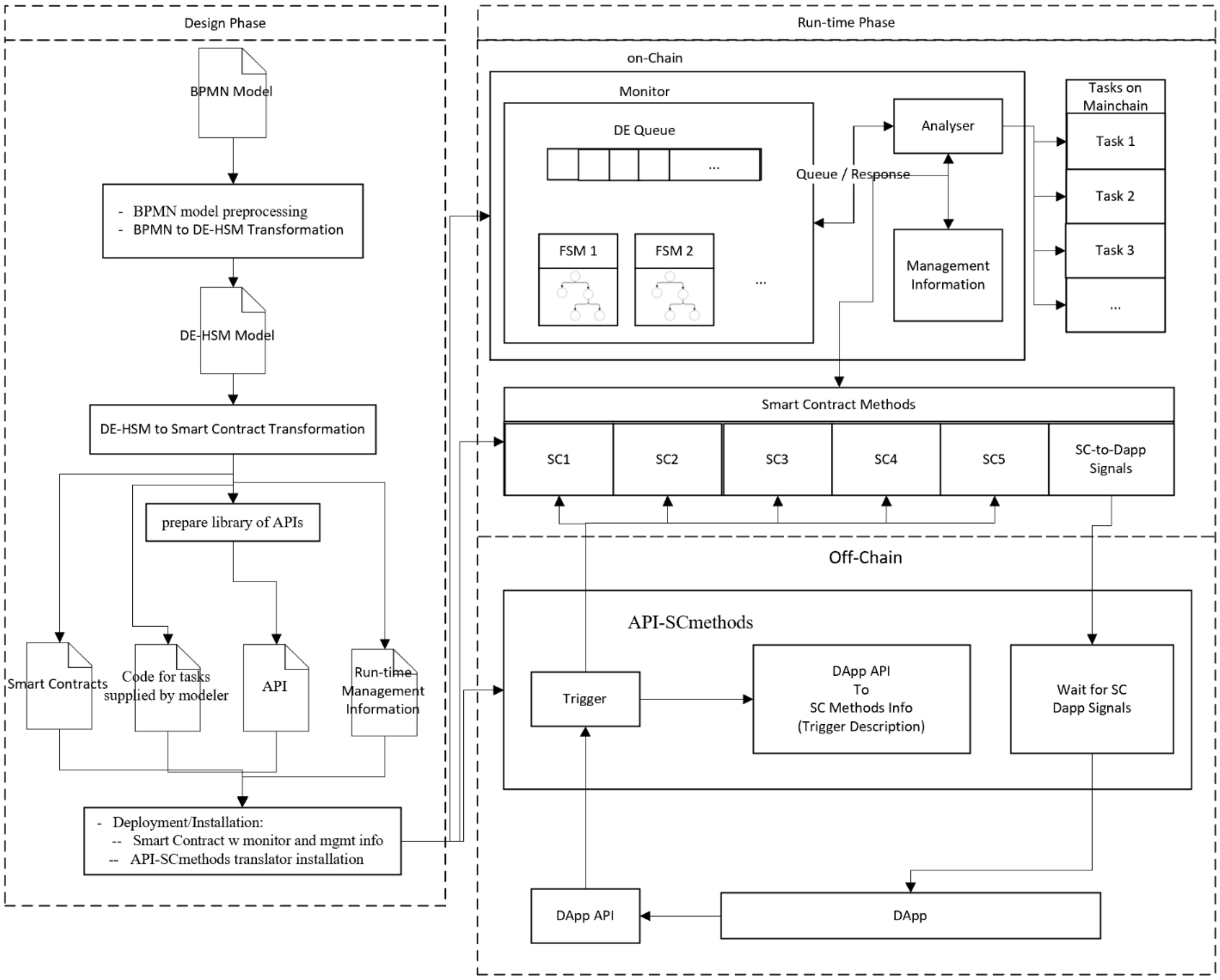

Fig. 4.2 Architecture Overview (Adopted from (Bodorik et al., 2023))

Following the deployment of the smart contract and installation of API methods, run-time processing involves user interaction with the dApp, which in turn invokes appropriate

API methods in response to user input. This input is scrutinized by the *API-SCmethods* module, which translates it into an invocation of a suitable smart contract method.

The Monitor module within a smart contract performs the following functions:

- It accepts inputs from a smart contract method, analyzes them, and queues an appropriate event into the DE queue in response to the input.
- It sequentially removes events from the DE queue, and for each event:
    - Analyzes the event's information and invokes an appropriate FSM, passing it input derived from the event's information formed by input to a smart contract method.
    - The invoked FSM processes the input, examines its state, and triggers a transition while generating output that is returned back to the monitor.
    - The Monitor analyzes the output produced by the FSM's firing and may perform one or more of the following actions:
        - Invoke one of the tasks.
        - Queue an event into the DE queue.
        - Respond to the smart contract invocation.
        - Trigger a signal to be delivered to the dApp.

#### *4.1.2.1 Design and Deployment/Installation Phase*

The Design and Deployment/Installation Phase encompasses the following primary steps:

1. Initialization and Transformation of the BPMN model into the DE-HSM multi-modal model.
2. Transformation of the DE-FSM sub-model into a collection of smart contract methods.
3. Deployment and installation: This involves deploying the smart contract(s) and installing the API-SCmethods and the dApp API.

As previously mentioned, our architecture bears similarities to those presented in (Weber et al., 2016; Tran et al., 2018; López-Pintado et al., 2019), and for comparative purposes, we adopt the same terminology as in (Weber et al., 2016), which has also been utilized in other works. Figure 4.3 illustrates a BPMN model that is transformed by the Design phase's Translator into a DE-HSM model, followed by the generation of smart contract methods. Correspondingly, the run-time phase involves a mediator/monitor that processes external events/triggers, which the dApp delivers to the run-time methods as parameters when calling them. The response is delivered to the dApp via output parameters upon return.

The initialization of the design phase involves the preparation of data structures for subsequent design steps, which will incorporate information used during the execution phase. Post initialization, the transformation from the BPMN model into the DE-HSM model involves determining which parts of the BPMN model should be represented by which of the DE-FSM sub-models. Each DE-FSM sub-model includes a DE model to represent concurrency and an FSM model to represent functionality. Choreography (not a BPMN term), which orchestrates the execution of activities in potentially concurrent processes, is represented by a DE-HSM model that represents the control flow amongst the DE-FHM sub-models. Furthermore, each DE-FSM model may also have internal parallel activities if the BPMN collaboration it represents has parallel or inclusive gateways.

In addition to the smart contract methods and information for the run-time Monitor, the design phase also prepares API methods to be invoked by the dApp. The API methods presented to the dApp are focused on activities amongst actors and thus roughly correspond to the flows of objects between either pools or swimlanes. Consequently, when input is received via a message from one actor to another, receiving a message will result in a call to the API with input being the content of the received message. The Design phase prepares appropriate information for the API-to-SCmethods translator that assists in determining which of the smart contract methods should be invoked and which input parameters to the API should be included as input to the smart contract method invocation.

The final steps are relatively straightforward as they include the deployment of the smart contract methods and the installation of the dApp API and API-to-SCmethods modules on network nodes.

#### *4.1.2.2 Execution Phase*

The Execution Phase comprises two stages:

- Initialization
- Execution

The Initialization stage addresses aspects such as the authentication of actors, also referred to as participants. Once the smart contract is deployed, the Execution stage is managed by the Monitor module.

It is important to note that the activities delineated above must be performed by any approach that generates smart contract methods from a BPMN model. As a result, parts of the architecture for the run-time phase bear similarities to other architectures, such as those presented in (Weber et al., 2016; Tran et al., 2018; López-Pintado et al., 2019). Consequently, we employ the same or similar terminology, including terms such as monitor/mediator, and triggers. Figure 3.2 depicts an off-chain dApp API module that provides an API to the dApp for the run-time phase. The module also features an API-to-SCmethods component that transforms an API call into a call to a smart contract method. In other research, this component is referred to as "triggers", but we believe that the term API-to-SCmethods is more apt.

The dApp API methods correspond to information flow across pools or swimlanes. When a connecting object crosses a pool or a lane, information about the origin and destination of the connecting object is included in the input parameters to the *API-to-SCmethods* component to determine which smart contract method should be invoked.

At times, a smart contract needs to inform external actors of certain events to which a user should respond. For instance, a contract may undergo computation along different paths that eventually converge, at which point a user(s) may need to be informed of the results before the computation proceeds. Our systems facilitate this by having an off-chain API process that invokes a special API method (in Fig. 4.2, this process is depicted as the *Wait for SC to dApp signals* process) which in turn invokes a smart contract method that waits for a "wakeup" event that may be generated by another smart contract method. When a smart contract needs to signal the external dApp, it raises a "wakeup" event that releases the method waiting for a wakeup event, and it can thus return to the dApp information about an event that has occurred within the execution of the smart contract.

When a dApp makes a call to the dApp API, the call information with its parameters is provided to the API-to-SCmethods module that determines which smart contract method is to be invoked, and then it invokes the smart contract method and waits for its response.

The smart contract method provides its information to the Monitor that uses its Analyzer sub-module to examine the input parameters and use them to queue an appropriate event into the DE queue. When that event reaches the head of the queue, it is dequeued and passed to the appropriate FSM for firing. The FSM receives its input, checks the current state, and fires, that is, it makes a transition while generating output. The transition's output indicates if a task or a subprocess, associated with the new FSM state, needs to be performed. For instance, if the transition was to a state that represents a task to calculate demand, then a task created for that purpose is invoked. As a part of the FSM transition, the smart contract analyzes the returned output from a transition and consults information on tasks and subprocesses and determines which of the following should be performed: (i) schedule another event by queueing it into the DE queue; (ii) prepare output parameters and return it to the smart contract method that invoked it; (iii) produce output and cause a dApp signal to be generated; (iv) invoke another smart contract method.

#### *4.1.2.3 Example*

Consider the BPMN diagram depicted in Fig. 2.1. Assume that the processing has advanced successfully such that (i) the manufacturer has dispatched a message to the middleman with an order for parts; (ii) the message containing the order has just been received by the dApp, and the dApp has just invoked the API with information about the received message. The API method takes the information from its input parameters and packages it as input to the API-to-SCmethods component, along with information on the source and destination of the message flow-object. The component consults its information, invokes an appropriate smart contract method while passing it input parameters, and then waits for the response from the smart contract method.

The smart contract method conveys its information to the Analyzer component of the Monitor. The Analyzer checks its management information and transforms the smart contract input information into an event, which includes parameters that are stored into the DE queue. The component then removes elements from the DE queue, processes them, and repeats until the queue is empty. In this example case, there is only one element that was dequeued from the DE queue, an element representing the reception of the message containing the order.

When the event, which represents the reception of an order, is dequeued from the DE queue, it is processed by the Analyzer by: (i) determining which of the FSMs is to be fired to process the dequeued event; (ii) preparing the input parameters for the FSM; and then firing the FSM, i.e., causing the FSM's transition. The FSM takes the input, the current FSM state, and determines the transition while producing output that contains the message content and also information about the new state that was reached, the state that corresponds to a task ProcessOrder. Since that state has been reached and it represents the ProcessOrder task, the corresponding task is invoked that produces output from the transition that includes the processed order. As the output from the task, the processed order, is passed to the next element, it is returned to the Analyzer as transition's output. The Analyzer then examines the output from the transition and checks the new state, which is sending a message to the Manufacturer while it is still a part of the same FSM. Consequently, the Analyzer creates a new event with input containing the processed order and queues it into the event queue – and the process repeats.

### 4.1.3 Transforming A BPMN Model into A DE-HSM Multi-Modal Model and Its DE-FSM Sub-Models

To better understand the transformation process, we briefly review the key properties of the DE-HSM modeling (Bodorik et al., 2023). As our system transforms a BPMN model into a DE-HSM model, we need to identify the DE-HSM sub-models and their interconnection. However, before we proceed, we need to be aware of the model semantics and its implications on modeling. Furthermore, we also explain some of the more subtle points on how certain BPMN elements are represented in the transformation process. Only then we describe the transformation process itself.

#### 4.1.3.1 DE-HSM Modeling Semantics and Loops

We use DE modeling for concurrency combined with HSM modeling for functionality, wherein the modeling is applied hierarchically. As an example, Fig. 4.3 shows a system that contains two sub-models.

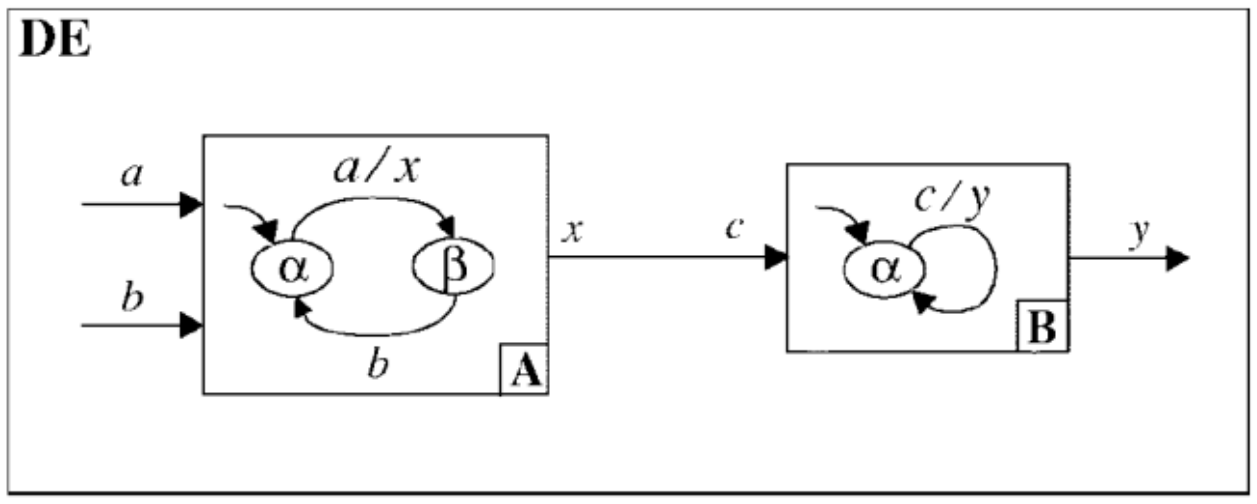

(a) Two child modules refining a parent

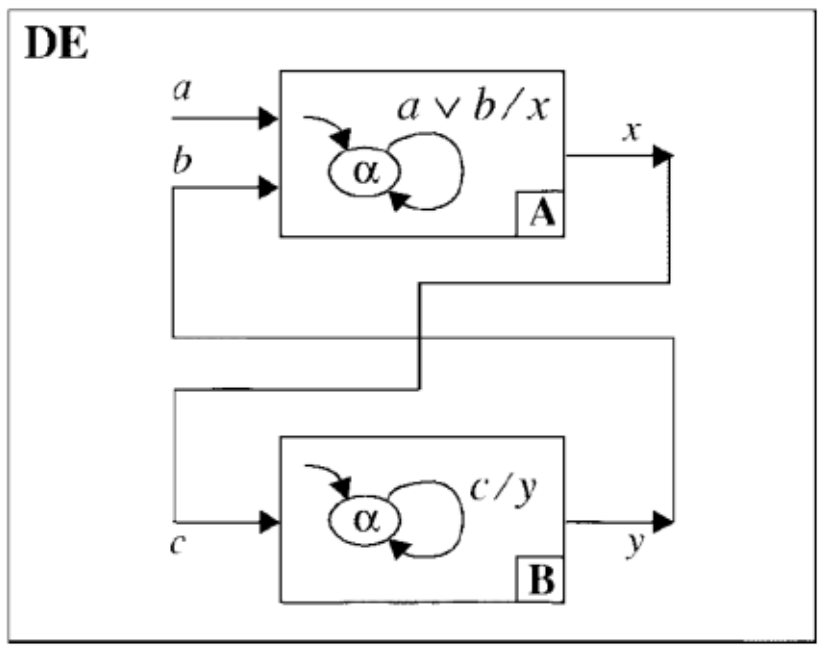

(b) Two child modules with loops refining a parent

Fig. 4.3 Two Examples of DE-HSM Models (Adopted from (Yannakakis, 2000))

Fig. 4.3 (a) shows a DE-FSM master model containing two DE-FSM sub-models, A and B, and their interconnection. Each of the master and the two sub-models has its own queue of events for DE modeling and an FSM for functionality modeling. It should be noted that we referred to the model as a DE-FSM model as the subsystems have FSMs and not HSMs. In semantics for DE-HSM models, adopted from (Yannakakis, 2000), an FSM system appears to the DE system as a zero-delay participant/actor. Under the assumption of zero-delay semantics, a system with sub-models is such that the FSM of a child subsystem is the first to fire, that is it performs the transition on inputs and produces output that is timestamped with the same timestamp as input (hence zero-delay). Only after this does the parent system react and initiate a transition, leading to its output.

Furthermore, without loops, we can use one global queue of events, instead of each subsystem having a separate queue of events (Yannakakis, 2000).

However, loops cause difficulties under the assumed zero-delay semantics. Consider Fig. 4.3 (b) and the two subsystems A and B. There is a loop because output from A forms input to B, and there is another loop in which B's output is input to A. Such loops cause difficulties, and we cannot apply zero-delay semantics in such a case, as execution may never terminate. Consequently, representing a BPMN model using DE-HSM modeling, we wish to avoid such feedback loops due to the semantic difficulties described above. In the subsection 4.1.1, we assumed that a BPMN model does not include any loop that cannot be represented by a looping subprocess.

#### *4.1.3.2 Initialization (in Design Phase)*

In the initialization phase, the translator first transforms a given BPMN model into an equivalent model that is well-formed as described in Section 4.1.3.1.

In general, a BPMN specification does not require the modeler to provide description on all information, which is associated with the BPMN model, that is required for execution. However, since our objective is to produce executable smart contracts, the TABS system must obtain such information from the modeller. Thus, in the initialization phase, the software interacts with the modeler to ensure that transformation has information about all data flowing along the edges of a BPMN graph to ensure that transformation has all necessary information about guards on gateways and about the content of messages. During initialization, we ask the modeler to annotate the BPMN graph accordingly using data associations. The issue is that BPMN specification does not insist on the modeler to provide such information and hence all approaches that strive to create an executable smart contract from a BPMN model need to seek information from the modeler if it is missing in a BPMN model (BPMN 2.0 Introduction, 2022).

Once the modeler supplies information on the data flowing along the connecting objects, the system asks the modeler to provide the code to be executed by each individual task. Each task has a single input and a single output, wherein each may be a complex object. The tasks are incorporated by the system into the smart contract methods so that they can be executed according to the control flow as specified by the BPMN model.

#### *4.1.3.3 Representing BPMN elements for DE-HSM Modeling*

Transformation of a BPMN model into a DE-HSM model is based on a graph representation of a BPMN model given its BPMN diagram description using XML. For the DE-HSM modeling and the transformation process, we use a Directed Acyclic Graph (DAG) representation of the BPMN diagram. In the DAG representation, each BPMN flow element/object is represented using a graph vertex, while connecting objects, that is sequence and message flows, are represented by edges in which the flow control is indicated by the edge direction.

For most of the BPMN flow object elements, the flow-control graph representation is straight-forward in that it can be represented by a graph vertex, with one or more input flows and with one or more output flows. For instance, the task and subprocess BPMN elements belong to the category of activities of BPMN symbols that are represented as graph nodes, each with one incoming and one outgoing flow object. A gateway can also be represented by a graph node with one or more incoming or outgoing edges. Events, which also belong to the category of activities, are not as obvious and shall be discussed here in more detail in terms of how they are represented in a DAG used for DE-HSM modeling.

There are three types of events: start, intermediate, and end. The Start and End events are special in that they occur at the very beginning and the very end of a process, respectively – hence they are specialized cases of the flow of control. Events have a Dimension attribute that describes the type of an event, which may be one of: message, timer, cancel, compensation, conditional, link, signal, terminate, multiple and parallel. Out of this list, we currently support (as is shown in Fig. 4.1) the following events: message, timer, conditional, and signal. We are in the process of incorporating the support for the rest of the BPMN elements. We first discuss how intermediate events are represented and then discuss representation of the begin and end events.

Intermediate events that are not boundary have explicit connections, represented by connecting objects between the corresponding throwing and catching events. Thus, they can be represented by vertices connected by edges that represent the connecting objects. A boundary event does not have an explicit connection via a connecting object to its counterpart, such as which signal in a task/subprocess is handled by which of the boundary catch events. An explicit connection between the rise of an event and its consequence, however, needs to be established, which is done at the conclusion of the modeling process when the modeler is asked by the transformation process to provide details on the flow of inputs and outputs along the edges of the graph. For each boundary catch event, the modeler is asked to identify, by using labels, which of the signal event caused the catch event by matching the labels of the throw event firing and its catching it as a boundary event. Thus, each event does have at least one incoming and one outgoing edge. A boundary catch event, such as catching a signal event, is represented by a graph node that is connected to its corresponding throw signal event, within the activity to which it is attached (as a boundary event), by an incoming edge from the vertex representing throwing the corresponding event. If the boundary activity is also interrupting, the state of execution for the interrupted activity is ended. However, if it is not interrupting, in essence two concurrent activities will proceed, one to continue with the execution of the activity with the boundary condition and one due to the catching of the boundary non-interrupting event and processing it.

A boundary begin event is a throwing event. If a boundary begin event is on a boundary of an activity, i.e., a subprocess or a task, that is a graph node B, the boundary event is represented on the DAG representation of the BPMN model by inserting an inclusive gateway on the incoming edge to the node B as is identified via label numbers between the boundary start event and the label of its corresponding catch event as provided by the modeler. In other words, boundary throwing events are fired/thrown at the beginning of an activity and, if it is non-interrupting, then two processes proceed concurrently during execution, one for the boundary start-event activity and one for the uninterrupted activity represented by the vertex B to which the boundary start event is attached.

A boundary end event is a catching event, and it is represented in the DE-HSM model in the usual way by a graph node with the incoming edge that is also an outgoing from the corresponding throwing event as represented by the labeling of events.

#### *4.1.3.4 Transformation of a BPMN Model into a DE-HSM Multi-modal Model*

The DE-HSM model, which we are pursuing and that symbolizes a BPMN graph, is an interconnection of DE-HSM sub-models, where each sub-model is a child model for the parent model. Moreover, each of the sub-models may itself be a DE-HSM sub-model, i.e., it may be modeled as an interconnection of its child-models. As described in a previous sub-section, the TABS system represents a BPMN model using a DAG, and then it determines the DE-HSM model as an interconnection of the sub-models, wherein each sub-model is represented without further sub-models: The topic of this section. However, once we do find the sub-models, we also need to describe how to build each sub-model, which is the topic of the next section.

Commencing with a BPMN diagram, we first describe how we represent a BPMN model as a DAG and then how we approach defining the problem of finding a DE-HSM model for a given DAG. To find a DE-HSM model, we use the concept of independent subgraphs and their variations, which are described next. We then present an algorithm to find a DE-HSM model that is an interconnection of DE-FSM sub-models, i.e., sub-models that do not have hierarchical states.

##### DAG representation of a given BPMN graph

At execution time, a DE-FSM sub-model contains a DE queue of events and an FSM. At execution time, processing within a sub-model proceeds by dequeuing an event from the DE queue and then using it to form an input that is given to the FSM. Upon input, the FSM fires and produces output. The main property of the DE-FSM sub-model, however, is that it accepts inputs that are fed to an FSM and outputs produced by the FMSs firing produce the output out of the sub-model. Thus, a sub-model consumes one set of inputs and produces one set of outputs.

We represent a BPMN model as a DAG G = (S, E), where S is the set of vertices and E is a set of edges. For a given BPMN diagram, vertices in S represent the BPMN non-connecting objects, while edges represent the connecting objects, that is sequence flows or message flows. We use a DAG G^ = (S^, E^) to represent a DE-HSM model. To transform this BPMN model into a DE-HSM representation, we decompose it into distinct subgraphs. Each of these subgraphs comprises a subset of vertices from the original BPMN model, and the edges that interconnect them. It is imperative to note that these subgraphs are mutually exclusive, meaning each vertex from the original BPMN model is allocated to one and only one subgraph. Within each subgraph, the original vertices and their interconnecting edges are retained, effectively rendering each subgraph as a self-contained fragment of the original BPMN model. When aggregated, these subgraphs reconstruct the entirety of the original BPMN model. The edges in our DE-HSM representation can either interconnect vertices within a singular subgraph or bridge multiple subgraphs, but they will not simultaneously serve both functions. This refined representation facilitates a more structured and systematic analysis of the BPMN model. Subsequent to this overview, we delineate the transformation methodology through a formal algorithmic exposition. We find a DE-HSM model in which:

i. For each element $S^{\wedge}_i \in S^{\wedge} = \{S^{\wedge}_i, i = 1, 2, \ldots n^{\wedge}\}$, $S^{\wedge}_i$ is a subset of vertices in S, $S^{\wedge}_i \subseteq S$.
ii. For each element $e^{\wedge}_k \in E^{\wedge} = \{e^{\wedge}_k, k = 1,2, \ldots, m^{\wedge}\}$, $e^{\wedge}_k$ is an edge in E, such that each edge $e^{\wedge}_k = (s_x, s_y)$ corresponds to an edge $(s_x, s_y) \in E$, wherein $s_x \in S^{\wedge}_i$, $s_y \in S^{\wedge}_k$, $i \neq k$, $i = 1, 2, \ldots, n^{\wedge}$ and $k = 1, 2, \ldots, m^{\wedge}$.
iii. $S^{\wedge}_i \cap S^{\wedge}_k = \emptyset$, i.e., $S^{\wedge}_i$ and $S^{\wedge}_k$ are mutually exclusive.
iv. For each $S^{\wedge}_i$, $i = 1, 2, \ldots, n^{\wedge}$, vertices in $S^{\wedge}_i$ are vertices of a subgraph $G'_i = \{S'_i, E'_i\}$, which is a subgraph of G, $G'_i \subseteq G$, such that $S'_i = S^{\wedge}_i$, i.e., vertices of $G^{\wedge}_i$ and $G'_i$ are the same, and the set of edges $E'_i$ includes all those edges $e = (s_x, s_y) \in E$, where $s_x, s_y \in S^{\wedge}_i$ and $S^{\wedge}_i = S'_i$; i.e., $s_x, s_y \in S^{\wedge}_i$ and the edges are internal to $G'_i$.
v. $(\cup S^{\wedge}_i, i = 1, 2, \ldots, n^{\wedge}) = S$: Union of all $S^{\wedge}_i$ is S, i.e., each vertex in S appears in exactly one of the $S^{\wedge}_i$.
vi. The following properties apply to the sets of edges in E:
    - $E^{\wedge} \cap (\cup E^{\wedge}_i, i = 1, 2, \ldots, m^{\wedge}) = \emptyset$; i.e., $E^{\wedge}$ intersect (union of all edges $E'_i$) is empty. Edges in $E^{\wedge}$ form interconnections amongst the subgraphs $G'_i$, where each subgraph $G'_i$ represents a DE-FSM sub-model.

- Each edge $e \in E$ is either in $E^\wedge$ or in the union of all edges in $E'_k$, $k = 1, 2, \ldots, m^\wedge$.

In other words, each $S^\wedge_i$ is a set of vertices of a subgraph $G'_i = (S'_i, E'_i)$, where $S'_i = S^\wedge_i$ forms a set of vertices of a sub-model of the DE-HSM model, and edges in $E'_i$ are all those edges $(s_x, s_y) \in E$ that are internal edges of the subgraph $G'_i$. Thus, each $S^\wedge_i \in S^\wedge$ identifies a DE-HSM sub-model and edges in $E^\wedge$ represent the interconnection of the DE-HSM sub-models. Furthermore:

- $G'_i$ must be a connected graph.
- As a sub-model must have only one entry and one exit node to be represented by an HSM or an FSM, its subgraph can only have one entry and one exit vertex.

### Finding a DE-HSM multi-modal model

Given a BPMN diagram, we represent it by a DAG G= (S, E) for which we wish to find its equivalent DAG $G^\wedge = (S^\wedge, E^\wedge)$ in which each node $S^\wedge_i$, $S^\wedge_i \in S^\wedge$, represents a subset of nodes of S that are members of a subgraph $G'_i = (S'_i, E'_i)$, such that:

i. Each $G'_i$ represents a sub-model $G'_i = (S'_i, E'_i)$ of $G^\wedge$ and hence $G'_1, G'_2, \ldots, G^\wedge_n{}'$ are such that
   - $G'_i \subset G$, $S'_i \subset S$
   - $S'_i \cap S'_k$, $i \neq k$, i.e., the sets $S'_i$, $i = 1, 2, \ldots, n^\wedge$, are mutually exclusive
   - $S = \cup\ S'_i$, $i = 1, 2, \ldots, n^\wedge$; vertices in subgraphs cover the whole graph S
   - For each edge $e = (s_x, s_y) \in E$:
     - $e \in E^\wedge$, i.e., it is one of the interconnections of the sub-models, or
     - $e \in E'_i$, for some i and where $E'_i \cap E^\wedge = \emptyset$, $i = 1, 2, \ldots, n^\wedge$, i.e., intersection of the set of edges that interconnects the sub-models with any of edges of the subgraphs is empty

ii. $G'_1, G'_2, \ldots, G'_n$ are such that for $G'_i$ and $G'_j$, where $G'_i$ and $G'_j$ are both subgraphs G', *are either*

- disconnected, i.e., not connected to each other directly (but connected indirectly via transitivity through some other $G_k \in G'$),

  *or are*
- connected by one edge $e = (s_x, s_y)$, where $s_x \in S_i'$ and $s_y \in S'_k$, $i \neq k$

iii. Each sub-model is characterized by having a single entry and exit point, with two notable exceptions:
- The subgraph encompassing the start event node of the original BPMN graph lacks an entry node due to the absence of an incoming edge from the subgraph's external nodes to any of its internal nodes.
- The subgraph containing the end event node of the original BPMN graph lacks an exit node as there is no outgoing edge from any of the subgraph's nodes to any of the subgraph's external nodes.

The conditions stipulated in (i) ensure that the subgraphs are mutually exclusive and collectively encompass the entire BPMN graph. Condition (ii) asserts that each subgraph is connected to another subgraph by precisely one edge, an edge that does not appear in any of the subgraphs. The final condition, (iii), arises from the fact that each DE-HSM sub-model accepts input fed to its FSM, which generates output upon firing. In other words, a sub-model possesses an FSM, and the sub-model's input forms the input to its FSM. Following the firing of the FSM, its output constitutes the sub-model's output. Consequently, each subgraph $G_i'$, which represents a sub-model in G^, must have one vertex (the subgraph's entry node/vertex) that receives the FSM's input and one node/vertex (the subgraph's exit point) that forwards the FSM's output for subsequent processing after exiting the subgraph.

### Independent subgraphs

The concept of *independent* subgraphs is now introduced.

<u>Definition</u>: Given a graph $G = (S, E)$, a subgraph $G' = (S', E')$, $G' \subset G$, is termed an *independent subgraph* if it fulfills the following criteria:

i. All nodes within the subgraph G', barring its entry and exit nodes, possess only internal edges (x, y), where (x, y), $x, y \in S'$.

ii. There exists precisely one node/vertex, denoted as $s_{en}$, within the subgraph that, in addition to its internal edges, has only one vertex with incoming edges from external nodes. This node is referred to as an entry node. (An *entry node* does not have any outgoing edges to external nodes.) The exception to this is the node of G' representing the Start event BPMN element, which does not have any incoming edges from external nodes.

iii. There exists precisely one node/vertex, denoted as $s_{ex}$, within the subgraph that, in addition to its internal edges, has any outgoing edges to any external nodes. This node is referred to as an exit node. (An *exit* node does not have any incoming edges from external nodes.) The exception to this is the node of G' representing the END event BPMN element, which does not have any outgoing edges to external nodes.

In the above definition, all internal nodes of an independent subgraph, excluding the entry and exit nodes, have edges connected solely to other internal nodes. As the graph S signifies the control flow, and the subgraph S' has only one entry and one exit node, this implies that once the initial state of the subgraph is reached during execution, the execution will proceed locally within the subgraph S' until execution reaches the subgraph's exit vertex, $s_{ex}$. This is precisely the property required for a DE-FSM sub-model. The properties of independent subgraphs are now demonstrated using Theorem 1, as presented in (Liu, 2021; Bodorik et al., 2021).

*Theorem 1*: Consider a BPMN model, which is represented as a DAG G = (S, E), and that has two *independent subgraphs* $G_1 = (S_1, E_1)$ and $G2 = (S_2, E_2)$, where $S_1 \subset S$ and $S_2 \subset S$ and where $|S_1| > 1$ and $|S_2| > 1$. Then one of the following holds:

a. $S1 \subset S2$ or $S2 \subset S1$
b. S1 and S2 are such that they either
    - share nodes that form an independent subgraph or
    - they share at most one edge and, furthermore, that edge is such that it is an outgoing edge from an exit node of one of the subgraphs and, furthermore, that edge is also an incoming edge to the entry node of the other subgraph.

Proof: The proof can be found in pages 39-41 of (Liu, 2021).

The concept of an independent subgraph will be utilized in the determination of the DE-HSM model. However, it is possible that a DE-HSM model may encompass several other DE-HSM models, which are also independent subgraphs. The objective is to identify independent subgraphs such that each one can be represented by a DE-FSM sub-model, i.e., a subgraph that does not contain any further internal sub-models. However, it is not desirable to perform unnecessary decomposition.

Definition: Let G = (S, E) be a DAG. A subgraph G', G' ⊂ G, is called a *Smallest Independent (SI) subgraph* if it has more than one vertex and it *does not have any proper subset that is also an independent subgraph*.

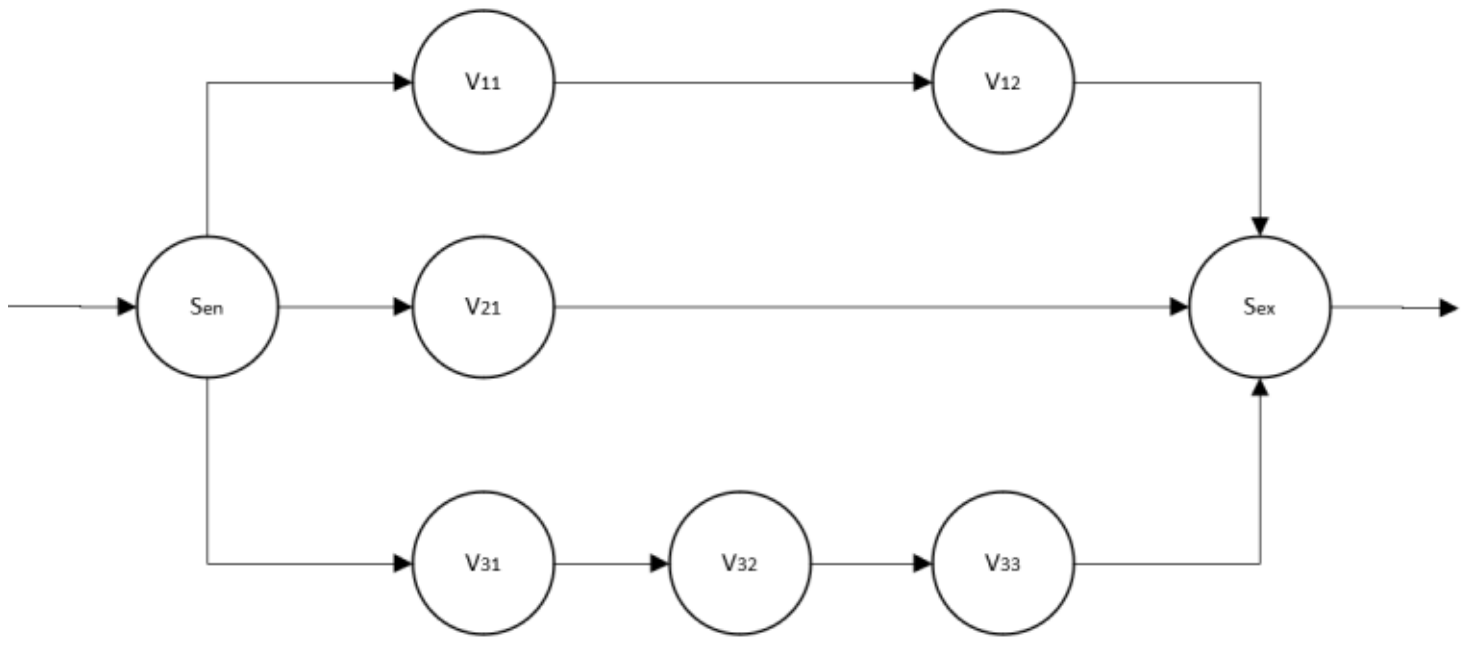

Fig. 4.4 Independent Subgraph with Proper Subsets That Are Also Independent Subgraphs (Adopted from (Bodorik et al., 2023))

We can use SI subgraphs to form the sub-models, however, such a decomposition breaks up sub-models when there is no need for further decompositions. Consider Fig. 4.4, which shows an independent subgraph G', of a part of a BPMN graph of G = (S, E). G' = (S', E'), where S' = { $s_{en}$, $v_{11}$, $v_{12}$, $v_{21}$, $v_{31}$, $v_{32}$, $v_{33}$, $s_{ex}$ }, E' = { ($s_{en}$, $v_{11}$), ($s_{en}$, $v_{11}$), ($v_{11}$, $v_{12}$), ($v_{12}$, $s_{ex}$), ($s_{en}$, $v_{21}$), ( $v_{21}$, $s_{ex}$), ($s_{en}$, $v_{31}$), ($v_{31}$, $v_{32}$), ($v_{32}$, $v_{33}$), ($v_{33}$, $s_{ex}$)}. However, G' has two independent subgraphs $G_1$ = ($S_1$, $E_1$) and $G_2$ = ($S_2$, $E_2$), where:

i. $G_1$ = {$v_{11}$, $v_{12}$, $v_{21}$, $s_{ex}$} and $E_1$ = {($s_{en}$, $v_{11}$), ($v_{11}$, $v_{12}$), ($v_{12}$, $s_{ex}$), ($s_{en}$, $v_{21}$), ( $v_{21}$, $s_{ex}$)}

ii. $G_2$ = {$v_{21}$, $v_{31}$, $v_{32}$, $v_{33}$, $s_{ex}$} and $E_2$ = {($s_{en}$, $v_{21}$), ($v_{21}$, $s_{ex}$), ($s_{en}$, $v_{31}$), ($v_{31}$, $v_{32}$), ($v_{32}$, $v_{33}$), ($v_{33}$, $s_{ex}$)}

Clearly, it is desirable to have only one DE-HSM model representing the subgraph G' as opposed to the two sub-models, one for $G_1$ and one for $G_2$. Further decomposition of G' into $G_1$ and $G_2$ is not only unnecessary, but it would also result in more complex interconnection of the sub-models then is necessary. Another example is an independent subgraph G'' = ({$v_{31}$, $v_{32}$}, {($v_{31}$, $v_{32}$)}) or an independent subgraph G''' = ({$v_{31}$, $v_{32}$, $v_{33}$}, {($v_{31}$, $v_{32}$), ($v_{32}$, $v_{33}$)}). They are both independent subgraphs, but clearly, there is no need to have separate sub-model to represent them.

We note that for any smallest independent subgraph G' = (S', E'), any outgoing edge from the entry vertex, $s_{en}$, identifies a path of connected nodes, such that the path terminates at the exit node, $s_{ex}$. Furthermore, any node on that path has only two edges that are on the path from $s_{en}$ to $s_{ex}$ and no other edges. If a node, say a node $s_1 \in S'$, which is on a path from $s_{en}$ to $s_{ex}$, did have another edge, say to a node $s_2 \in S'$, that is not on the same path from $s_{en}$ to $s_{ex}$, it would violate the properties of the smallest independent subgraphs as either: (i) If $s_2 \notin G'$, then an edge from the node s1 to $s_2$ within the subgraph G' leads to the node $S_2$ that is outside the subgraph G', which means that the subgraph is not an independent subgraph as it has two nodes with edges outgoing to nodes that are external to the subgraph. Or (ii) $s_2 \in G'$, but then the edge ($s_1$, $s_2$) would create an independent subgraph within the smallest independent subgraph G', which is a contradiction.

### Largest Smallest Independent (LSI) subgraph

Consequently, we wish sub-models to represent independent subgraphs such that each subgraph does not have any independent subgraph itself unless that independent subgraph (i) has the same entry and exit nodes as its parent graph or (ii) that independent subgraph is fully on a path from the subgraph's entry node to its exit node. We introduce the concept of the Largest Smallest Independent (LSI) subgraph.

Definition: Let G be a graph, G = (S, E). A subgraph G' = (S', E'), G' ⊂ G, is called a *Largest Smallest Independent (LSI) subgraph* of G, if it is an *independent subgraph* and, furthermore, any vertex s ∈ S', which is neither an entry nor an exit node of an independent subgraph, has exactly one incoming and one outgoing edge.

We use the properties of the LSI subgraphs to construct, from the given graph G, an equivalent representation G^ = (S^, E^), where each $S^\wedge_i \in S^\wedge$ is a set of vertices of a subgraph $G'_i$, i = 1, 2, …, n^. Each subgraph $G'_i$ is an LSI subgraph that represents a DE-FSM sub-model, and E^ is a set of edges in E that represents interconnections of the sub-models represented by the subgraphs.

## Algorithm to find LSI subgraphs

The previous sections dealt with formalities and abstractions suitable for presentation of the concepts of independent and LSI subgraphs. We now provide a pseudocode of an algorithm to find LSI subgraphs for a given DAG G that represents a BPMN model. The algorithm, called *findLSIsubgraphs,* has as its input a global DAG G = (S, E), which represents a BPMN diagram. The method results in a DAG G^ = (S^, E^), such that E^ represents edges and each $S^\wedge_i \in S^\wedge$ represents a subset of nodes $S'_i$ of the graph $G'_i$, where the nodes of $S_i^\wedge$, i = 1, 2, …, n^, form a subgraph $G_i' = (S'_i, E'_i)$, where $S'_i = S^\wedge_i$ is a set of nodes representing an LSI subgraph that is an FSM sub-model. Furthermore, $E_i'$ represents interconnections of the nodes in $S'_i$, wherein each edge e = $(s_x, s_y)$ is an internal edge to $S'_i$, i.e., where $s_x \in S'_i$ and $s_y \in S'_i$. Thus, each $S^\wedge_i$ defines an LSI subgraph G' = $(S'_i, E'_i)$, while the set of edges E^ of G^ form interconnections amongst the LSI subgraphs.

The method uses a global variable G = (S, E) to mark the nodes with information on vertices incoming and outgoing degrees, where outgoing degree of a vertex s is the number of directed edges of the form $(s, s_x)$, for some vertex $s_x \in G$, and an incoming degree is the number of edges of the form $(s_x, s)$, where $s_x \in G$.

Recall from the section 4.1.2, which dealt with preprocessing of the BPMN graph, that preprocessing replaces any gate that is both a join gate and a fork gate with an equivalent representation by one merge gate and one fork gate, wherein the merge gate has one or more incoming edges but only one outgoing edge that is also the only incoming edge to a fork gate that has only one has more than one outgoing edge.

Before we proceed with the presentation of the algorithm to find LSI subgraphs, we make a few observations about LSI subgraphs:

a) With the exception of the entry of exit nodes, there is no vertex that has both incoming and outgoing degrees being greater than one.
b) If a vertex has an incoming degree of more than one, then it is a merge/join gate with one outgoing edge. In such a case, the vertex may only be an exit node from an LSI subgraph.
c) If a vertex has an outgoing degree of more than one, then it is fork/split gate, with an incoming degree of one. In such a case, the vertex may only be an entry node to an LSI subgraph.
d) An internal node of an LSI subgraph has an incoming degree and outgoing degree being exactly one.

**Algorithm 1: findLSIsubgraphs**

```
Input: DAG G = (S, E);
Output: setOfLSI subgraphs G* = { Gi' = (Si',Ei'), i = 1, 2, ... }
            where each Si' corresponds to the Si^ an element of S^, where G^ = (S^, E^),
            and E^ is the interconnection of the LSI subgraphs Gi', i = 1, 2. ...
BEGIN
Initialization;
for each node s in S do
    Let s.inDegree = the number of incoming edges to s ;
    Let s.outDegree = the number of outgoing edges from s ;
    Let s.root = s ;
end
Let G* be empty;
var vertex s, sx, sy;   var edge e = {sx, sy};   var LSIsubgraph G' = (S', E') ;

 // invocation of the breadth-first search traversal of the DAG G ;
 while ( ( s = nextInLevelOrder (G)) ≠ nil ) do
     // Check if vertex s represents the START BPMN element
     if (s.inDegree = 0) then
        let s.root = s  ; //Assumption: BMPN diagram has more than the START and END element
        exit while loop;
     end
     // s may be an internal LSI subgraph node or an exit node or may represent END element;
     if (s.inDegree > 1)  then   // s node is an exit node from an LSI subgraph
          G'i = isLSIsubgraph (s.root, s); // Should return an LSI subgraph
          add G'i to G*;  // G'i should not be null
          exit while loop; // exit the current iteration of the while loop and check for the next iteration
      if (s.inDegree = 1)  then
          if (s.outDegree = 0) then //s represents the END BPMN element => it is exit node
               G'i = isLSIsubgraph (s.root, s); // Should return an LSI subgraph
               add G'i to G*;                // G'i should not be null
               exit while loop; // exit the current iteration of the while loop and check for the next iteration
          elseif (s.outDegree = 1) then   // there is only one edge from s and s may or may not be an exit node
               Let e =( s, sy); // Find the destination vertex sy of the only outgoing edge from s
               if ( sy.inDegree > 1) ) then  //  s is an exit node from an LSI subgraph as sy is an entry node
                    G'i = isLSIsubgraph (s.root, s); // Should return an LSI subgraph
                    add G'i to G*;  // G'i should not be null
                    exit while loop; // exit the current iteration of the while loop and check for the next iteration
               else
                    if ( sy.inDegree = 1) then // s is neither exit nor entry node
                         s.root = (pointedToBy(s)).root=s ;
                    endif // if ( sy.inDegree = 0) then * The node sy represents the END BPMN that will be
                                                       recognized as such and processed on the next
     iteration */
               endif
          endif
     endif
 endwhile;
 // Post processing:
 Combine any subgraphs G' in G* that have the same entry and exit nodes into one subgraph;
 END;
```

Fig. 4.5 Independent Subgraph with Proper Subsets That Are Also Independent Subgraphs (Adopted from (Bodorik et al., 2023))

The algorithm/method *findLSIsubgraphs*, shown in Fig. 4.5, has a DAG G = (S, E) as its input. Recall that the start event has only one edge, which is an outgoing edge. Each node s ∈ S is assumed to have the following attributes:

- ***s.inDegree*** … the number of edges that are incoming into the vertex s
- ***s.outDegree*** … the number of edges outgoing out of the vertex s
- ***s.root*** … reference to a node that may be an entry of an LSI subgraph to which the node s may belong

The method *findLSIsubgraphs()* uses the following auxiliary methods:

- *vertex x = nextInSearchOrder (G)* … the method facilitates a breadth-first search traversal of the connected DAG G, in which that vertex s ∈ S that represents the BPMN START event is the start of the traversal. Each of the other nodes has at least one incoming edge and one or more outgoing edges, with the exception of the vertex that represents the BPMN End event, which does not have any outgoing edge. Calling the method *nextInSearchOrder(G)* repeatedly results in the method returning a reference to the next node in the breadth-first search traversal of G and returning null once all vertices have been traversed.
- *LSIsubgraph G'i = isLSIsubgraph (s, d)* … given vertices s and d, the method determines whether the vertices s and d identify an LSI subgraph by checking the cases that are discussed below. If it does discover an LSI subgraph, it is returned by the method; otherwise, the method returns null. We first discuss the special cases that exist for the vertices that represent the *START* and *END* BPMN elements.
    - If the in-degree of the vertex, say s, is zero, then the vertex represents a *START* BPMN element and hence it cannot be an exit node. The s.root attribute is set to s as it will be the entry node of an LSI subgraph, which is also the root node of the subgraph.
    - If the out-degree of vertex is zero, then the vertex represents the *END* BPMN element. It is treated as an exit vertex of an LSI subgraph.
- *vertex sx = pointedToBy(s)* … The method returns null if the vertex s does not have any incoming edges or if it has more than one incoming edge. Otherwise, it returns that vertex $s_x$ that is the source for the only edge leading to s, i.e., it returns that vertex $s_x$ or for which there exists an edge e = ($s_x$, s).

The algorithm uses a breadth-first search traversal of the DAG G, starting with that graph node s that represents the BPMN element that is the Start event, while its output has been described already. We now discuss the details of the *findLSIsubgraphs()* algorithm that is shown in Fig. 4.5.

The first part deals with initialization that sets the attributes of each node of the DAG G, namely attributes *s.inDegree*, *s.outDegree*, and s.root. Following that there is a while loop that is used to determine whether the node s, returned by the *nextInSearchOrder(G)*, is an exit node of an LSI subgraph and whether that node identifies an LSI subgraph, which has been found and should be included in the set of LSI subgraphs output by the method, or whether it is some other case and how it should be handled. These cases are now discussed by making specific references to the lines of the algorithm presented in Fig. 4.5.

The line 18 within the while loop checks the indegree of the vertex s. If the indegree of the vertex s is zero, then the vertex s represents a *START* BPMN element and hence it is an entry node of an LSI subgraph. If it is, then the s.root attribute is set to s as it will become a root node of an LSI subgraph. The while iteration is exit and another iteration starts (lines 24-26). Note that for the vertex representing the *START* BPMN element, its outdegree cannot be greater than one as only a fork gate can have an out-degree greater than one.

- If the *s.inDegree* > 1 (line 23), then the vertex s is an exit node and an LSI subgraph has been discovered and added to the set of LSI subgraphs G^, and the loop is exited (lines 24-26)
- If the *s.inDegree* is equal to 1 (line 27), then the node's outdegree is checked.
    - If the outdegree of the vertex s is zero (line 28), then the vertex s represents the *END* BPMN node and, consequently, the LSI subgraph identified by the entry and exit nodes s.root and s, respectively, is returned as an LSI subgraph G' that will be added to the set of LSI subgraphs G^ (lines 29-31).

- If the outdegree of the vertex s is equal to one (line 32), then it cannot be determined as yet whether the node s is or is not an exit node of an LSI subgraph. The vertex s could be an exit node such that there is only a single path to the node s, while the out-going edge from s, edge (s, sy), leads to a node that has an out-degree of greater than one and hence is the entry node of another LSI subgraph, in which case the node s is an exit node from an LSI subgraph. If the outdegree of the node sy is one, then the node s is not an exit node from an LSI subgraph as the node sy could be safely added to the LSI subgraph identified by the vertices s.root and s without violating the LSI subgraph properties. Thus, to determine whether or not s is an exit node, we must examine the out-degree of the node sy, where sy is the destination node of the outgoing edge from s (line 33):
  - If the outdegree of the vertex $s_y$ is greater than zero (line 34), then the vertex $s_y$, where $s_y$ is the destination vertex of the out-going edge from s, i.e., the edge (s, $s_y$), is the entry node of the next LSI subgraph and hence the node s is the exit node of the LSI subgraph. The found LSI subgraph is added to the set of LSI subgraphs that is output (lines 35-37).
  - If the outdegree of the vertex $s_y$ is equal to 1 (line 39), then the vertex s is neither exit nor entry node and the s.root attribute is set to refer to the s.root attribute of the vertex that has an outgoing edge to points top s.
  - As a note, if the outdegree of the vertex $s_y$ is equal to 0, then $s_y$ represents the *END* BPMN element, in which case nothing needs to be done as the node $s_y$ will be processed on a subsequent iteration of the while loop.

Once all nodes are processed, the algorithm performs a post-processing step that ensures that any LSI subgraphs that were discovered and added to the set G^ and that have the same entry and exit nodes are combined into one LSI subgraph. For instance, if the algorithm outputs two separate independent subgraphs $G_1$ and $G_2$, where $G_1$ and $G_2$ are:

i. $G_1 = \{v_{11}, v_{12}, v_{21}, s_{ex}\}$ and $E_1 = \{(s_{en}, v_{11}), (v_{11}, v_{12}), (v_{12}, s_{ex}), (s_{en}, v_{21}), (v_{21}, s_{ex})\}$

ii. $G_2 = \{v_{21}, v_{31}, v_{32}, v_{33}, s_{ex}\}$ and $E_2 = \{(s_{en}, v_{21}), (v_{21}, s_{ex}), (s_{en}, v_{31}), (v_{31}, v_{32}), (v_{32}, v_{33}), (v_{33}, s_{ex})\}$

Then the post-processing step combines all independent subgraphs that have the same entry and exit nodes into one subgraph. For the example of $G_1$ and $G_2$, the post-processing step combines the two sub-graphs into one as is shown Fig. 4.4.

The importance of decomposing the Business Process Model and Notation (BPMN) into Discrete Event Hierarchical State Machine (DE-HSM) sub-models, which adhere to the characteristics of Least Significant Independent (LSI) subgraphs, lies in the fact that these LSI subgraphs result in sub-models that can be directly represented by a Discrete Event Finite State Machine (DE-FSM) model. This implies that these sub-models do not necessitate any further decomposition.

### Example

Consider the experimental supply-chain BPMN model depicted in Fig. 2.1, which has also been utilized in (Weber et al., 2016; Tran et al., 2018; López-Pintado et al., 2019). The application of the algorithm *findLSIsubgraphs()* yields the LSI subgraphs, enumerated by the method *foundSubgraph*, as illustrated in Fig. 4.6 and delineated below.

- *S1* … This subgraph encompasses *INIT, Buyer sends offer, Manufacturer receives order, calculate demand, Manufacturer places order,* and *Middleman receives order*.
- *S2* … This subgraph comprises *forward order, order transport, supplier receives order, produce, order transport, prepare transport, carrier receive order, supplier receive request,* and *request details*.
- *S3* … This subgraph includes *provide details, provide waybill, receive details*, and *receive waybill*.
- *S4* … This subgraph incorporates *carrier deliver order, manufacture receives order, report start of production, manufacturer deliver product, buyer receives product,* and *SUCCESS*.

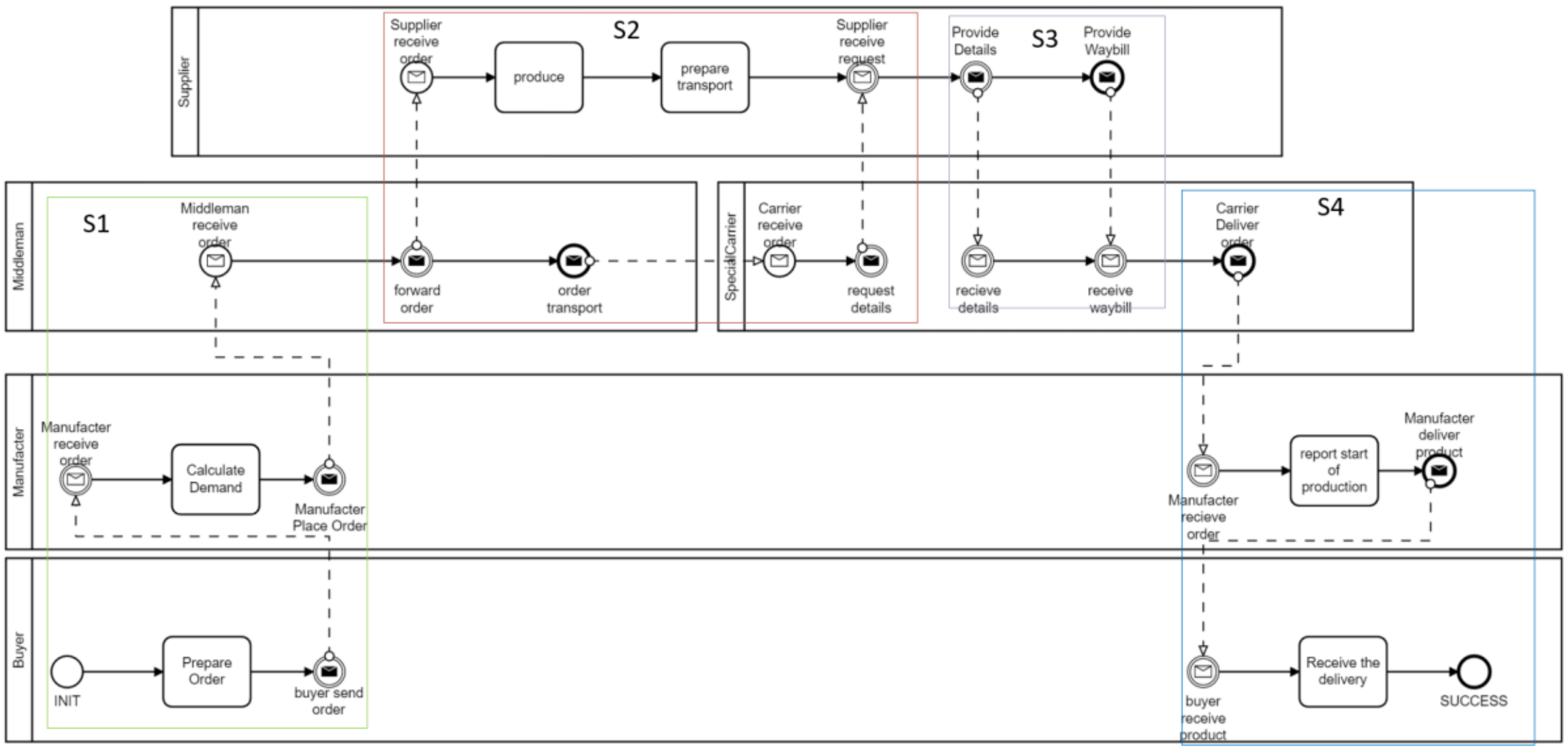

Fig. 4.6 BPMN Model and Found Subgraphs (Adopted from (Liu et al., 2022b))

### 4.1.4 Building DE-FSM Sub-models and Their Interconnection

Up to this point, we have discussed the identification of a DE-HSM model, specifically, the process of determining a DE-HSM model in which none of its sub-models contain additional sub-models. We elucidated the construction of each DE-FSM sub-model and their subsequent integration in (Bodorik et al., 2023).

#### *4.1.4.1 Building a DE-FSM submodel*

The previous section described how to find a DE-HSM model for a given graph G that represents the BPMN model. The DE-HSM model is represented by a graph $G^\wedge = (S^\wedge, E^\wedge)$, such that each $S^\wedge_i \subset S^\wedge$ represents a subset of nodes of the graph G, where the nodes of $S^\wedge_i$, $i = 1, 2, \ldots, n^\wedge$, are also the nodes of a subgraph $G'_i = (S'_i, E'_i)$ of G, where $S'_i = S^\wedge_i$ is a set of nodes representing an LSI subgraph that is an FSM sub-model.

Given an LSI subgraph $G' = (S', E')$ that represents a DE-FSM sub-model, we build its model using a breadth-first search traversal of its graph $G'_i = ( S'_i, E'_i )$, starting at its entry node $s_{en}$. Recall that each sub-model represents a DE-FSM model that uses one DE queue, which is shared by all sub-models, and an FSM that represents the functionality. Also recall from the previous section that described how to find the DE-HSM model, that a fork gate denotes an entry node of an LSI subgraph, while a merge gate is an exit node/vertex from an LSI subgraph. However, an LSI subgraph is such that each of its vertices that is neither the entry nor the exit vertex of the subgraph has exactly one incoming and one outgoing edge. Thus, with the exception of its entry or exit nodes, there are no other vertices that represent a gate. As a consequence, the translator creates the sub-model in two phases.

The initial phase corresponds to the flow control entering the sub-model during the execution phase, that is when the flow control reaches the subgraph’s entry node. Depending on the type of the entry node, further execution may proceed as one processing stream for an exclusive fork gate, or more than one processing streams in case of an inclusive fork gate. It's worth noting that an exclusive or inclusive fork gate may have a guard along each outgoing edge, wherein a guard along an outgoing edge represents the constraint that the FSM’s input must satisfy for the execution to proceed along that edge. Of course, for an exclusive gate, guard constraints must be such that only one evaluates to true at any one time. Thus, an inclusive fork gate also covers a parallel fork gate when all guards are set to true.

If the entry node of the LSI subgraph represents an exclusive fork gate BPMN element, then in the initial phase, the root of the FSM is created. For an exclusive fork gate, in addition to the root of the FSM, additional states are appended for a binary search that will result in identifying a path that represents the control flow along each of the paths from the entry node to the exit node. For instance, if there are four outgoing edges from an exclusive fork gate with the guards being A, B, C, and D, then Fig. 4.7 shows the vertices added to the FSM in order to identify the path for each of the outgoing edges to the exit node. In the second phase, on each iteration of the search, a new state is added on a path for which the input satisfies the guards. In short, there is one FMS for an entry node that represents an exclusive fork gate, such that the FSM’s root is followed by transitions that result in n paths, one path for each of the outgoing streams from the gate. In the subsequent phase, state transitions are built along the outgoing paths, each one representing functionality for one of the guards on the exclusive fork gate.

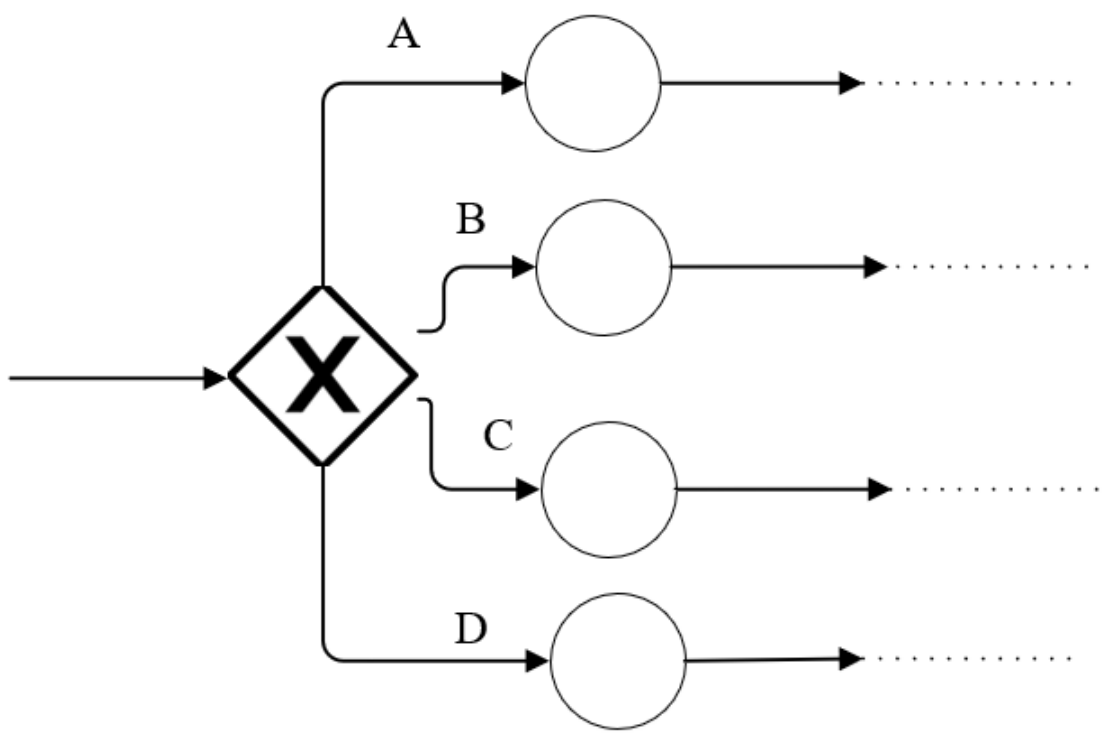

(a) Exclusive fork gate with guards A, B, C, and D

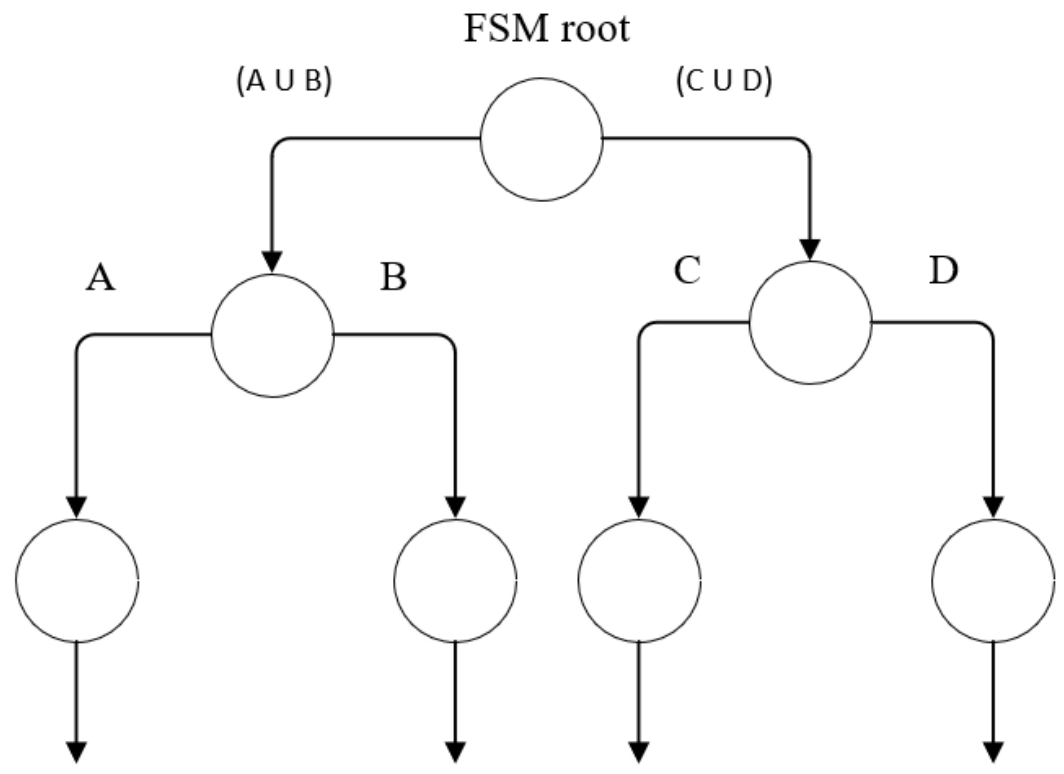

(b) FSM nodes to identify outgoing paths

Fig. 4.7 Entry Node that Represents An Exclusive Fork Gate and FSM Transitions as Paths of FSM (Adopted from (Bodorik et al., 2023))

If the entry node represents an inclusive fork gate BPMN element, then the translator creates n+1 FSMs if there are n outgoing streams out of the inclusive fork gate. In addition, one FSM, referred to as a *control FSM*, is created to determine at the execution time, to which of the outgoing streams from the fork gate does the input to the FSM belong when an FSM fires, i.e., when it receives input to which it needs to react. The controlling FSM is used in the execution phase to determine to which of the FSMs should the input be directed depending on the evaluation of the guards – the identification of the FSMs for which guards evaluate to true is the output of the controlling FSM's firing. In addition, when the exit from the firings of the controlling FSM is reached, the flow of control is directed towards each of the concurrent FSMs for which the guard evaluated to true. This is achieved the monitor component at run-time by creating a DE for each of the guards that evaluates to true. For each one, the monitor inserts a DE event that causes it to fire and hence causes execution of the firing of the corresponding concurrent FSM when the event reaches the head of the queue. Fig. 4.8 shows the inclusive fork gate and the resulting control FSM and the concurrent FSM.

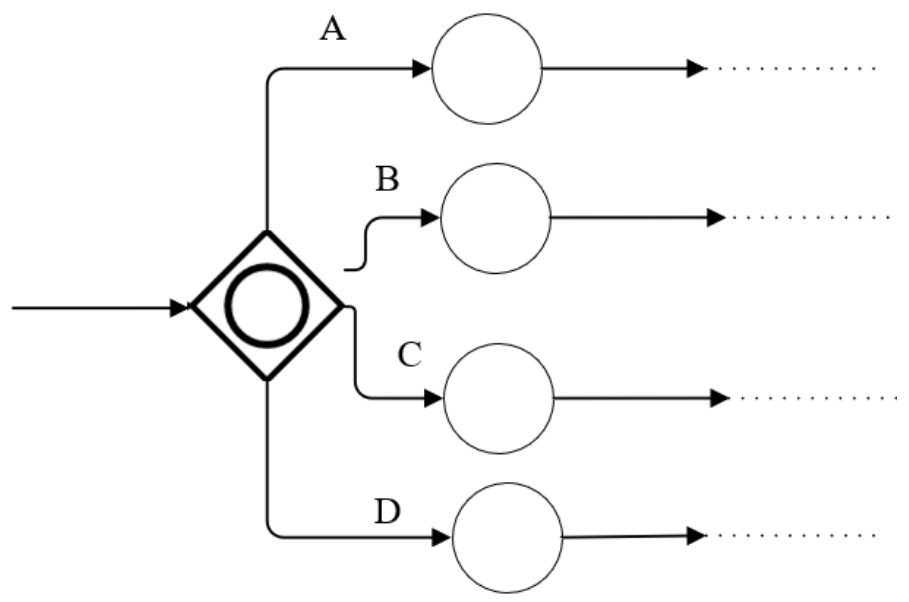

(a) Inclusive fork gate with guards A, B, C, and

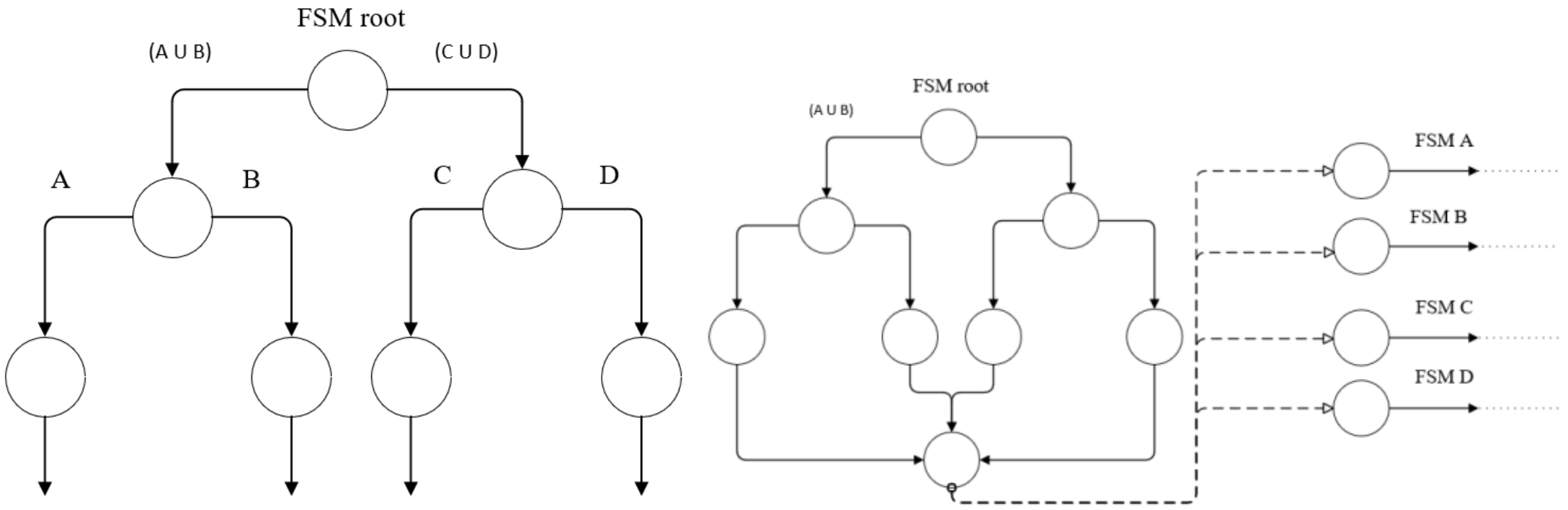

b) Controlling FSM and roots of n concurrent FSMs, one for each concurrent stream

Fig. 4.8 Entry Node That Represents Inclusive Fork Gate and Control FSM and One FSM for Each of The Concurrent Process Streams (Adopted from (Bodorik et al., 2023))

Once the FSMs are prepared, a breadth-first search of the nodes of the subgraphs is used to build the transitions states of the concurrent FSMs. For an exclusive fork gate, all transitions are in the one FSM along the paths from the root to the exit node, wherein the initial vertices are created by the translator in the initial phase to guide the input to the appropriate path using the guard conditions. For an inclusive fork gate, for each node of the breadth-first search, the translator first determines to which of the FSMs the node belongs and then the translator augments the corresponding FSM with the state transition for the node. Both phases are now described.

### Initial Phase – Entry into a Sub-model

The translator scrutinizes the BPMN element that the entry vertex of the sub-model represents. The transformation into a sub-model is contingent on whether the BPMN element is not a gate, is an exclusive fork gate, or is an inclusive fork gate:

- If the entry node is not a gate, the processing within the sub-model is represented by a single stream with the functionality represented by an FSM. The translator constructs the root of the FSM, with the remainder of the FSM states created by examining each of the nodes of the subgraph, as described below in the main phase that constructs the FSM transitions.
- If the entry node is an exclusive fork gate, the processing within the sub-model is represented by a single stream with the functionality represented by one FSM. The translator constructs the root of the FSM, and then it inserts states that examine the input and perform a binary search using guards of the exclusive gateway to determine on which path, from the entry node to the root, a new transition is inserted (as depicted in Fig. 4.7).
- If the entry node is an inclusive fork gateway, the processing within the sub-model is represented by multiple concurrent streams by creating one concurrent FSM for each of the outgoing concurrent streams from the entry node that represent the inclusive fork gate BPMN element. Additionally, the translator creates the control FSM that is used to direct input coming into the sub-model into an appropriate FSM for an outgoing flow-control stream, determined by examining the input to the FSM using constraints represented by the guards on outgoing edges of the BPMN gateway. As in the previous case of an exclusive gateway, the translator inserts into the controlling FSM states that examine the input using Boolean expressions appearing in the guards of the gate that the node represents, as shown in Fig. 4.7.

### Main Phase

In the main phase, subgraph vertices are processed, vertices that have exactly one incoming and one outgoing edge. Such a vertex results in a new FSM state and transition to be added to an FSM: For an inclusive fork gateway, the new state and transition are added to the appropriate concurrent FSM, while for the exclusive fork gateway, the new state and its transition are added to the appropriate path of the one FSM that has one path, from the subgraph's entry node to its exit node, for each of the outgoing edges out of an inclusive-fork gateway.

#### *4.1.4.2 Integration of Sub-models*

Once the sub-models are constructed, the overall model is built by identifying the subgraph that contains the node representing the BPMN START event – referred to as the start subgraph. The overall model is then constructed by integrating the sub-models. The exit node of the start subgraph is connected to entry nodes of LSI subgraphs i, for which there is an outgoing edge from the exit node of the start subgraph and to the entry node of the LSI subgraph i. For each such interconnection, the translator takes the output from the FSM of the exit node and creates a DE event that is queued into the DE queue, with the event being targeted for processing by the FSM that represents the functionality of the LSI subgraph i. This process is repeated for each of the LSI subgraphs.

### 4.1.5 Transforming DE-FSM Model into the Methods of a Smart Contract(s)

We proposed a mechanism for transforming DE-HSM models into the methods of smart contracts within a multi-chain environment in (Bodorik et al., 2023). In the ensuing subsections, we provide an overview of the transformation process of the DE-HSM model into the methods of a smart contract. A comprehensive explanation is not provided here as our approach has been previously detailed in (Bodorik et al., 2021; Liu et al., 2021a) and shares similarities with other works. However, our unique contribution lies in equipping developers with the functionality to deploy portions of the smart contract for processing on a sidechain, with the aim of reducing costs and/or enhancing privacy. These aspects are highlighted in greater detail.

#### *4.1.5.1 Generation and Deployment of Smart Contract Methods*

During the contract execution phase, user participants and their systems interact with a Decentralized Application (dApp), which in turn makes calls to the dApp Application Programming Interface (API) provided by our system. At runtime, when a smart contract method is invoked, its input is analyzed and used to create a corresponding event that is queued into the Discrete Event (DE) queue. The DE processing component continuously dequeues and processes events. The processing of a DE event includes: (i) analysis of the DE event to determine which Finite State Machine (FSM) should fire and preparation of input for that FSM; (ii) provision of input to the FSM; (iii) firing of the FSM, resulting in output; and (iv) transformation of the FSM output into a response to the smart contract method invocation. This entire process is managed by the monitor/mediator using information prepared during the design phase.

The runtime architecture of the system is similar to other approaches in terms of system interaction with the dApp and the mapping of dApp interactions with the API into calls to the methods of smart contracts, as well as how the system responds to dApp requests. However, our approach diverges in terms of flow control management.

The internal representation of the system state is based on the DE-HSM model, which determines the flow control between the FSM sub-models that represent the computational state of individual DE-FSM sub-models. We accurately model the BPMN graph using a DE-HSM model, which is used to direct input into the appropriate sub-model. Here, the input is provided to the sub-model's FSM, which fires and produces output. Depending on the DE-HSM model, the output produced by the FSM may be used as input to another sub-model and the process repeats, or the output is analyzed and provided as output for the call to the smart contract method, which in turn provides its returned parameters to the dApp method that invoked the dApp API. The channeling of outputs produced by sub-models is according to the interaction between the sub-models as defined by the DE-HSM model, while the function of each sub-model is represented by an FSM, or a group of FSMs if the sub-model has concurrent processing indicated by the inclusive fork gates, that also accurately models a part of the BPMN graph that is a BPMN LSI subgraph.

Our approach also differs from others in that we facilitate off-chain processing on a sidechain, which is described in the following section.

#### *4.1.5.2 Sidechain Processing for Privacy and Cost Reduction*

Upon the transformation of the BPMN model into a DE-HSM model and the development of individual sub-models, but prior to the transformation into smart contract methods, the developer is presented with a design choice. This choice involves deploying a portion of the smart contract for processing on a sidechain, which can enhance privacy, reduce processing costs, or both. To facilitate this decision, sub-models are presented to the developer for selection of those to be deployed on a sidechain.

In fact, as part of the testing process and before final deployment, two versions of the smart contract are developed and deployed. In one version, the developer-selected sub-models are deployed and executed on a sidechain. In the second version, all smart contract methods are deployed and processed on the mainchain. Delays are measured for both versions, and the results are provided to the developer. This allows the developer to compare the cost of smart contract execution on the mainchain only versus on the mainchain with sidechain processing of selected sub-models. The developer is provided with information on the delays incurred in both cases, and if the mainchain or sidechains use Ethereum EVM, such as Quorum, cost estimates in ETH gas are also provided to the developer.

We first outline our system architecture for the run-time phase with sidechain processing. We then explain how we assist the developer in determining which parts of the smart contract are suitable for sidechain processing, either for privacy or for cost reduction. Examples of sidechain usage are provided in a subsequent section that describes our Proof of Concept (PoC), the *TABS* tool.

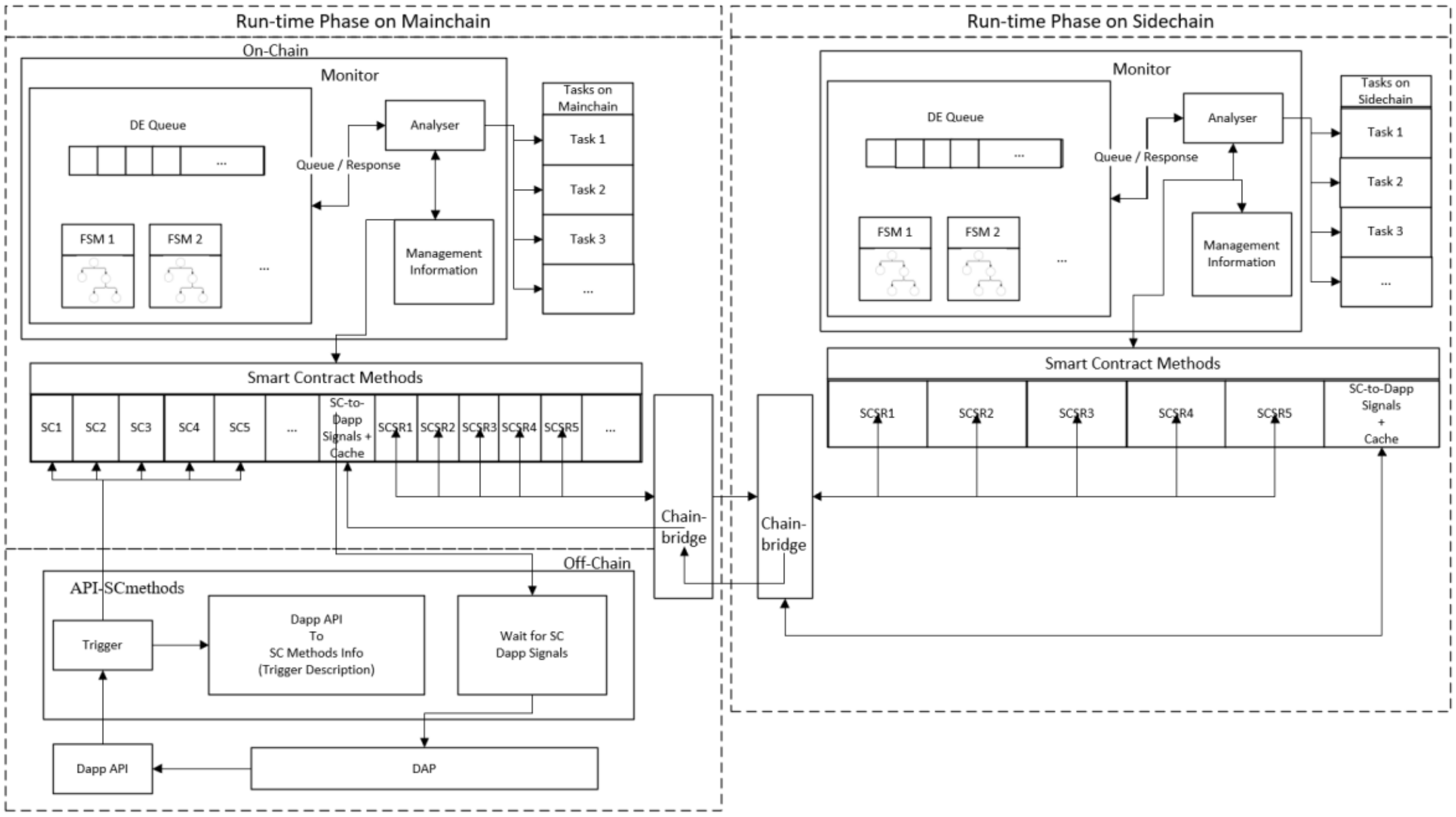

Fig. 4.9 System Architecture with Mainchain and Sidechain Processing (Adopted from (Bodorik et al., 2023))

### System Architecture with Sidechain Processing

Fig. 4.9 illustrates the system architecture at run-time for both the mainchain and the sidechain. The internal architecture for sidechain processing largely mirrors that for the mainchain, as shown in Fig. 4.9. Interoperability between the mainchain and the sidechain is partially achieved by a mainchain contract method calling a smart contract method on a sidechain. However, additional complexities arise, as our sidechain contract is essentially an integral part of the mainchain contract, but it is executed on a sidechain.

When smart contract methods are moved from the mainchain to a sidechain, they must be invoked from the mainchain. When a call is made on a mainchain to a smart contract that is deployed on a sidechain, the mainchain method must redirect the call to the sidechain. However, challenges arise when state variables are accessed during a method executed on a sidechain – a method deployed on a sidechain does not have access to any state variables stored on the mainchain. For this reason, they need to be fetched from the mainchain and stored on a sidechain in a cache. Consequently, on the first call to a method deployed on a sidechain, in addition to the call parameters, blockchain data that will be accessed by the method is also included. At the design time, static analysis of the sidechain code is used to determine which data is provided on the first call to a method so that it would populate the cache on a sidechain. Of course, there will still be cache misses, due to the sidechain method referencing objects dynamically.

#### *4.1.5.3 Sidechain Processing: What, When, How*

In order to facilitate off-chain processing of a portion of the smart contract, several decisions need to be made:

- Identification of the components (referred to as patterns) of the smart contract that should be processed off-chain.
- Determination of the conditions under which sidechain processing should be utilized.
- Establishment of the methodology for achieving sidechain processing.

1) Identifying Patterns Suitable for Sidechain Processing

The concept of independent subgraphs, utilized in determining the DE-HSM model from a BPMN model, proves useful in identifying which patterns are suitable for off-chain processing. Off-chain processing incurs overhead, partly due to the sidechain's lack of access to the mainchain. Therefore, if a pattern interacts with mainchain methods, the overhead cost of communication between the mainchain and sidechain arises. Consequently, a pattern moved to the sidechain should be independent from the other methods of the smart contract executing on the mainchain. Independent LSI subgraphs possess this property, as once the flow of control enters the LSI subgraph, it does not exit until the computation within the subgraph is completed. Thus, if the LSI-subgraph processing is deployed on a sidechain, apart from the mainchain contract passing the application's input as parameters to the sidechain, there is no processing on the mainchain until the execution exits from the independent subgraph executed on a sidechain. Therefore, deploying parts of the smart contract to a sidechain is equivalent to deploying DE-HSM sub-models on the sidechain. The developer is thus provided with the choice of which DE-HSM sub-models can be selected for processing on a sidechain.

2) Deciding When to Process on a Sidechain

Sidechain processing may be more cost-effective and could be used to decrease the cost of processing a pattern. However, processing a pattern on a sidechain incurs overhead due to interaction between the mainchain and the sidechain. Thus, from a cost perspective, sidechain processing should only be used if its cost, including the overhead cost, is less than processing the pattern on the mainchain. We determine the cost of processing on a sidechain, including the overhead cost, by conducting a live test of running the dApp when (a) the smart contract is deployed on the mainchain only and (b) the pattern is deployed on the sidechain, thus measuring all delays, including delays due to overhead. If processing on a mainchain or a sidechain involves EVM machine, the cost is also estimated in Eth for each of the smart contract methods and is provided to the modeler.

However, another reason to process a pattern on a sidechain may be privacy. A smart contract deployed on the mainchain is such that all participants can explore all data stored on the blockchain by the smart contract – hence there is no privacy that can be provided so that some collaboration be private, that is confidential to other actors. For instance, a subset of actors performs a local collaboration represented by a pattern and that collaboration should be private in that its details (its blockchain data) should not be visible to the other participants to the smart contract who do not participate in that collaboration. In such a case, we should process the pattern on a sidechain, which will provide privacy and the other actors will not be privy to the pattern's processing details. Of course, all smart contract participants will be able to see any data returned by the sidechain computation that is stored on the mainchain.

3) How to Process Off-Chain

Once it is determined which pattern is to be processed off-chain, the next question is how to facilitate it. A pattern is a set of smart contract methods, such that they are invoked either due to a signal from the dApp API or because they collaborate and invoke each other, but "within" the pattern. A smart contract method executing on a sidechain is likely to need to know the state of execution and other information stored on the blockchain – however, a method on the sidechain does not have access to the mainchain and its data. Consequently, when transitioning to off-chain execution of a smart contract method, in addition to the parameters passed to the method, the system also needs to find, retrieve, and deliver to the off-chain processing any mainchain data that the methods on a sidechain read. Similarly, upon execution returning back to the mainchain, any writes to the blockchain that were performed by the off-chain execution of the pattern need to be collected and handed-off, together with transition outputs, for recording on the blockchain. Finally, upon completion of a method executed off-chain, before the blockchain is updated with the new state and data values written by the off-chain method, the results of off-chain execution must be reviewed and approved/attested by each of the participants affected by the off-chain computation. As a consequence of the above, we provide interfaces for the following three phases when methods of a pattern are deployed on a sidechain:

Pattern Start: Upon the first invocation of an off-chain pattern method, our software tool prepares appropriate data structures to support on-chain / off-chain interaction. Most important is a cache for mainchain data accessed by the pattern executed off-chain. Off-chain code uses a local cache for reading and writing blockchain data – its data structure needs to be prepared. On a cache miss, data is retrieved from the main blockchain and then stored in the off-chain cache using a getter method prepared just for that purpose to retrieve the data from the mainchain. For the cache, we use IPFS as a distributed system for storing and accessing temporary files and data. When Blockchain needs to access or examine the data in the cache, TABS allocates a data resource stored on IPFS by tracking the content address of corresponding data stored on IPFS.

Pattern Middle: After the first invocation, interaction between the off-chain and on-chain communication is concerned with provision and return of appropriate parameters and data for methods executed off-chain. Also, on a cache miss to the off-chain cache, data needs to be retrieved from the main chain using a getter method in order to service the cache fault.

Pattern End: Upon completion of the last method that is executed off-chain, which occurs upon a transition from the off-chain pattern execution to the on mainchain execution, in addition to the returned parameters, the results produced by the off-chain computation that were saved in a cache are collected and are provided to the attestation procedure executed on the mainchain. Furthermore, once the results are approved by attestation, they are recorded on the mainchain. Our default attestation is to deliver the results of sidechain processing to each of the participants collaborating on the sidechain processing, including the hash of the results, and receive from each actor the digitally signed hash-code of the result (signed by the private key) and thus signifying that the actor accepts the results. If the results are not accepted by an actor, an exception is raised. Of course, the default attestation method may be replaced by an attestation method provided by the developer.

#### *4.1.5.4 Sidechain Processing Control*

As the sidechain processing may be used for either cost reduction or for privacy, our tool, described in the next section, delegates the responsibility, of determining which patterns to process on a sidechain, to the modeler/developer. We display the list of LSI patterns of the smart contract, DE-HSM sub-models, to the developer who determines which pattern is or is not processed on a sidechain. For informed decision-making regarding the cost on mainchain vs sidechain, we deploy the smart contract with and without the pattern executing on a sidechain and present both costs to the developer to assist in the decision-making. An example will be provided in the chapter 6 that describes the TABS tool.

In summary, the decision to process on a sidechain is multifaceted, involving considerations of cost, privacy, and the specific requirements of the smart contract. Our tool provides the developer with the necessary information and flexibility to make informed decisions about which parts of the smart contract should be processed on a sidechain, and under what conditions. This approach allows for a more efficient and effective use of blockchain resources, potentially reducing costs and improving privacy where necessary.

## 4.2 BPMN Collaborative Transactions: Properties and Specifying

Prior to delving into the automatic creation of the transaction mechanism, it is imperative to address the transactional properties that need support and how developers specify the BPMN pattern that constitutes a collaborative transaction. These issues are explored in subsections 4.2.1 and 4.2.2, respectively. Subsection 4.2.3 subsequently elucidates the automated creation of the transactional mechanism.

### 4.2.1 Transactional Properties for BPMN Collaborative Transactions

Given that Business Process Model and Notation (BPMN) serves as the modeling framework for collaborative transactions, and that it's used to generate multi-method smart contracts to support long-term transactions on the blockchain, it's essential to adopt the blockchain mm-transaction properties. These properties must be enforced by the BPMN collaborative transaction mechanism to ensure proper functionality and security. Specifically, the following transactional properties must be supported:

- ACID properties: The fundamental properties of Atomicity, Consistency, Isolation, and Durability (ACID) must be maintained. These properties ensure that the transaction is processed reliably, with all-or-nothing execution, consistent state maintenance, isolation from other concurrent transactions, and permanent recording of the transaction.
- Access control and privacy: The mechanism must include provisions for controlling access to the transaction and maintaining the privacy of the data involved. This includes restricting access to authorized participants and ensuring that sensitive information is not exposed to unauthorized entities.

By adhering to these properties, the BPMN collaborative transaction mechanism can effectively support the complex requirements of long-term transactions on the blockchain. This alignment between the modeling framework and the underlying blockchain technology ensures that the generated smart contracts are both robust and secure, facilitating the implementation of collaborative applications across blockchains.

### 4.2.2 BPMN Collaborative Transaction: How to Specify

BPMN modeling specification (Business Process Model and Notation (BPMN), 2023) provides the concept of an all-or-nothing transaction only for the subprocess element, which is a BPMN sub-model with one start and one end element. However, as the sub-process is executed within the context of a single actor, it cannot be used to represent the collaborative activities of more than one actor. Here we address the issues of how to determine which BPMN patterns are suitable for representing a collaborative transaction and how the developer indicates to the system which BPMN pattern is to be treated by the system as a collaborative transaction.

Clearly, selecting any subset of BPMN elements from a BPMN model is not appropriate as the pattern selected as a transaction must have the all-or-nothing property. Consider the sample use case, shown in Fig. 2.1, and the activities of the Middleman actor, which are: Middleman receive order, forward order, and order transport. It might appear that the three activities might be a suitable pattern to be declared as a BPMN transaction; however, difficulties arise with the atomic commitment. If the first two activities are successfully performed but the third one fails, then the transaction is aborted, and the question arises: What to do about the effects of the forward activity that has already been executed, which means that the order was already sent to the Supplier actor? The Supplier actor might have produced the product that is now prepared for transport – and these activities would also have to be aborted and further cascaded aborts might be required. The problem arises because the activities selected as a transaction were not localized in that they affected other activities not belonging to the transaction. This is similar to the isolation property of the transactions in the DB system: No activities external to the transaction should view updates made by a transaction that has not been completed.

In terms of the graph representation, the three activities that were selected as a transaction are such that there is a flow of execution, from the selected BPMN pattern of the three activities, to an activity outside the pattern. To avoid such a situation, nodes that are selected to be the pattern representing a transaction must be such that there is no outgoing edge out of any vertices in the pattern to vertices outside the pattern. The exception is one vertex that may have an outgoing edge(s) corresponding to the outgoing flow of execution in the BPMN model when the transaction is completed. And the above requirements are satisfied when a pattern chosen to be a transaction is a SESE subgraph. Once the flow of computation enters a SESE subgraph via its entry node, it remains within that subgraph until the computation exits via the subgraph's exit node – we refer to this as the localization property of SESE subgraphs and we exploit it, together with additional properties exhibited by the SESE subgraphs described below, in defining the BPMN collaborative transactions and in creating the transactional mechanisms to support the BPMN collaborative transactions.

For the developer to determine which of the BPMN patterns are to be supported as collaborative transactions, the DAG representation of the BPMN model is analyzed in order to find all SESE subgraphs. Once all such subgraphs are identified, they are mapped back to the BPMN model and presented to the developer who determines which of the subgraphs should be transformed into a BPMN collaborative transactions. Once the developer decides which of the SESE subgraphs are to be processed as collaborative transactions, the BPMN model is further processed to create a transactional mechanism to enforce the transactional properties.

Consider the sample use case shown in Fig. 2.1 that was also used in (Weber et al., 2016; Tran et al., 2018; López-Pintado et al., 2019). The graph is analyzed to find all SESE subgraphs in the DAG representation of the BPMN model. Fig. 4.6 shows the SESE subgraphs when mapped backed to the BPMN model shown in Fig. 2.1. It should be noted that that there are many other SESE subgraphs that exists but are not shown in the Fig. 4.6, as the figure would become too busy. How SESE subgraphs are further classified and how they are used by the TABS tool will be discussed later.

The developer chooses which of the subgraphs should be used to form a collaborative transaction. In the next subsection we shall discuss the SESE subgraphs property that, if any two SESE subgraphs share vertices, then one of the subgraphs is a proper subset of the other subgraph. This property is used to define nested collaborative transaction that will be discussed in the next section. In this subsection 4.2.2, we described how a BPMN collaborative transaction is identified and specified by the developer, while how a transactional mechanism is created for a BPMN collaborative transaction is described in the next subsection 4.2.3. How the nested collaborative transactions are supported is described in the chapter 5.

### 4.2.3 Transaction Mechanism to Enforce Collaborative Transaction Properties

In this subsection, we describe how the transformations are amended to support the BPMN collaborative transactions. The main effect is on the modeling at the DE-HSM level that is used to determine which BPMN patterns are suitable for defining as collaborative transactions. Thus, BPMN model preprocessing and transformation of the BPMN model into the DAG representation are not affected, and we describe here the effects of supporting the collaborative transactions on DE-HSM modelling and subsequent transformations.

#### *4.2.3.1 DAG Transformation into DE-HSM and DE-FMS Sub-models and Properties of SESE Subgraphs*

As was described above, the developer uses the TABS tool to find and show all SESE subgraphs from which the developer chooses which of the subgraphs are to be treated as collaborative transactions. As was shown in (Bodorik et al., 2023), SESE subgraphs satisfy the following property:

- As the BPMN graph representation is a DAG, each of the SESE subgraphs is a DAG.
- As a SESE subgraph is acyclic, the subgraph may be analyzed using multi-modal modeling in which concurrency is modeled using DE modeling and functionality is expressed using HSMs.

There are two consequences due to the above properties:

- In DE modeling, during execution, each subsystem represented by a subgraph contains a queue of events ordered by their timestamps. As none of the SESE subgraphs have feedback loops (they are acyclic), all DE-HSMs models may use just one DE queue of events ordered by the event timestamps.
- More importantly, recall that the edges in the DAG represent the flow of computation. As SESE subgraphs have only one entry node, once computation enters the subgraph via the entry node, it proceeds with the execution within the subgraph until the execution exists via its exit node – hence we refer to a SESE subgraph as representing localized computation.

It is the localization property of the SESE subgraphs that we exploited in (Bodorik et al., 2023) for automated generation of smart contracts with sidechain processing, wherein the functionality of a selected SESE subgraph is transformed into the methods of a smart contract that is deployed and executed on a sidechain. It is a “slave” contract of a master smart contract that is executed on the mainchain and that invokes the methods of the slave smart contract. As the slave contract executed on a sidechain has localized computation, once the flow of execution enters the subgraph via its entry node, it remains within the subgraph until the flow of computation exits via the subgraphs exit node. It should be noted that if a transaction is aborted, then its abort compensation activities are performed within the scope of the transaction and exit from the transaction is via the subgraph's exit node.

In section 4.1.3.4, we discussed two additional SESE subgraph properties for any two SESE subgraphs found to exist in a DAG, they are either

1) mutually exclusive in that they do not share any vertices, or
2) one of the subgraphs is a proper subgraph of the other one.

However, the concept of the SESE subgraphs is further refined to avoid arriving at a model that has too many subgraphs. Consider the graph G of Fig. 4.4, wherein G = (S, E), S = { $s_{en}$, $v_{11}$, $v_{12}$, $v_{21}$, $v_{31}$, $v_{32}$, $v_{33}$, $s_{ex}$ }, and E = { ($s_{en}$, $v_{11}$), ($s_{en}$, $v_{11}$), ($v_{11}$, $v_{12}$), ($v_{12}$, $s_{ex}$), ($s_{en}$, $v_{21}$), ( $v_{21}$, $s_{ex}$), ($s_{en}$, $v_{31}$), ($v_{31}$, $v_{32}$), ($v_{32}$, $v_{33}$), ($v_{33}$, $s_{ex}$) }. G has two independent subgraphs $G_1$ = ($S_1$, $E_1$) and $G_2$ = ($S_2$, $E_2$), where:

i. $S_1$ = { $s_{en}$, $v_{11}$, $v_{12}$, $v_{21}$, $s_{ex}$} and $E_1$ = { ($s_{en}$, $v_{11}$), ($v_{11}$, $v_{12}$), ($v_{12}$, $s_{ex}$), ($s_{en}$, $v_{21}$), ( $v_{21}$, $s_{ex}$) }
ii. $S_2$ = { $s_{en}$, $v_{31}$, $v_{32}$, $v_{33}$, $s_{ex}$} and $E_2$ = { ($s_{en}$, $v_{31}$), ($v_{31}$, $v_{32}$), ($v_{32}$, $v_{33}$), ($v_{33}$, $s_{ex}$) }

Clearly, it is desirable to have only one DE-HSM model representing the subgraph G as opposed to the two sub-models, one for $G_1$ and one for $G_2$. Further decomposition of G into $G_1$ and $G_2$ is not only unnecessary, but it would also result in more complex interconnection of the sub-models then is necessary. Another example is an independent subgraph G' = ( S', E'), where S' = { $s_{en}$, $v_{31}$} and E' = { ($s_n$, $v_{31}$)} and an independent subgraph G'' = (S'', E''), where S'' = {$v_{32}$, $v_{33}$, $s_{ex}$ } and E''={ ($v_{32}$, $v_{33}$), ($v_{33}$, $s_{ex}$}). They are both mutually exclusive SESE subgraphs, but clearly, there is no need to have separate sub-models to represent them – hence, in (Bodorik et al., 2023), the concept of ***L**argest **S**mallest **I**ntendent (**LSI**) subgraphs* was introduced:

*LSI subgraph:* An LSI subgraph G = (S, I) is a SESE subgraph, which has an *entry vertex* $s_{en}$ and an *exit vertex* $s_{ex}$, such that for any vertex s ∈ S, which is neither an *entry* nor an *exit vertex* of the SESE subgraph, has exactly one incoming and one outgoing edge.

We note that an LSI subgraph is such that it does not contain any proper subgraph that is also an LSI SESE subgraph. Furthermore, when we make a reference to a SESE subgraph in this paper, we mean a reference to either an LSI SESE subgraph or a SESE subgraph that contains one or more SESE or LSI subgraphs. See (Bodorik et al., 2023) for the full description of the SESE subgraph properties.

#### *4.2.3.2 Transformation of the DE-HSM Sub-models into Smart Contract Methods*

Recall from the subsection 4.2 that when there is no sidechain processing, then activities of each actor are represented by a smart contract method representing the flow of execution for that actor. The flow of execution for one actor's model is represented using an FSM, if there is one flow of execution, or using concurrent FSMs if there are concurrent executions within the actor context due to the BPMN flow for that actor having fork gates. However, if the developer chooses a transaction representing a collaboration of several actors, i.e., if the developer chooses a SESE subgraph as a collaborative transaction, our approach represents such a transaction by using a separate flow of execution. Thus, for each SESE subgraph chosen by the developer to be treated as a transaction, a separate smart contract method is created. The flow of execution within that method is represented by the concurrent FSMs that represent the flow of execution within that SESE subgraph chosen by the developer as a transaction. Thus, the smart contract is represented by (n+m) smart contract methods, where n is the number of actors and m is the number of BPMN collaborative unnested transactions as chosen by the developer.

A smart contract method representing a collaborative transaction does not invoke any of the other smart contract methods. Thus, although the smart contract method for a collaborative transaction is invoked multiple times by an application or monitor, the method does satisfy the property that once computation enters the transaction as represented by a SESE subgraph, it stays within that computation until the computation exits the subgraph, which satisfies the independence property of an mm-transaction as described in (Liu et al., 2023b). Consequently, the methods for automated creation of an mm-blockchain transaction mechanisms are applicable also for the transactional mechanism for collaborative transactions as is described in section 3.3.

# CHAPTER 5 NESTED BPMN COLLABORATIVE TRANSACTIONS

Collaborative long-term transactions often encompass multiple sub-transactions or sub-programs, referred to as nested transactions. These nested structures add complexity and require specific handling within the blockchain environment. In this chapter, we will define and explore the concept of nested transactions based on the BPMN model. We will also discuss methodologies and strategies to support these nested transactions, ensuring their seamless integration within the broader context of long-term collaborative transactions on the blockchain.

## 5.1 NESTED BPMN COLLABORATIVE TRANSACTIONS

Consider Fig. 4.6, which illustrates our BPMN model with four SESE subgraphs identified by our proposed *findLSIsubgraphs* algorithm depicted in Fig. 4.5. It's worth noting that there are additional SESE subgraphs containing other LSIs or SESE subgraphs. For example, Fig. 5.1 reveals an extra SESE subgraph, S5, encompassing proper subgraphs S1 and S2. Both S1 and S2 are LSI SESE subgraphs, as they do not contain any proper subgraphs that are also SESE subgraphs. When it comes to selecting the collaborative transaction, the developer has several options, including:

1. Select S1 and/or S2 as separate collaborative transactions.
2. Select S5 as a collaborative transaction (but without selecting S1 or S2 as a sub-transaction).
3. Select S5 as a collaborative transaction and select one or both of S1 and S2 as sub-transactions.

The last option results in nested transactions, which is an issue that we address in this sub-section. We shall concentrate on describing a simple 2-level nesting of a transaction with two sub-transaction and then we comment on how the approach is generalized to an arbitrary number of nesting levels and sub-transactions.

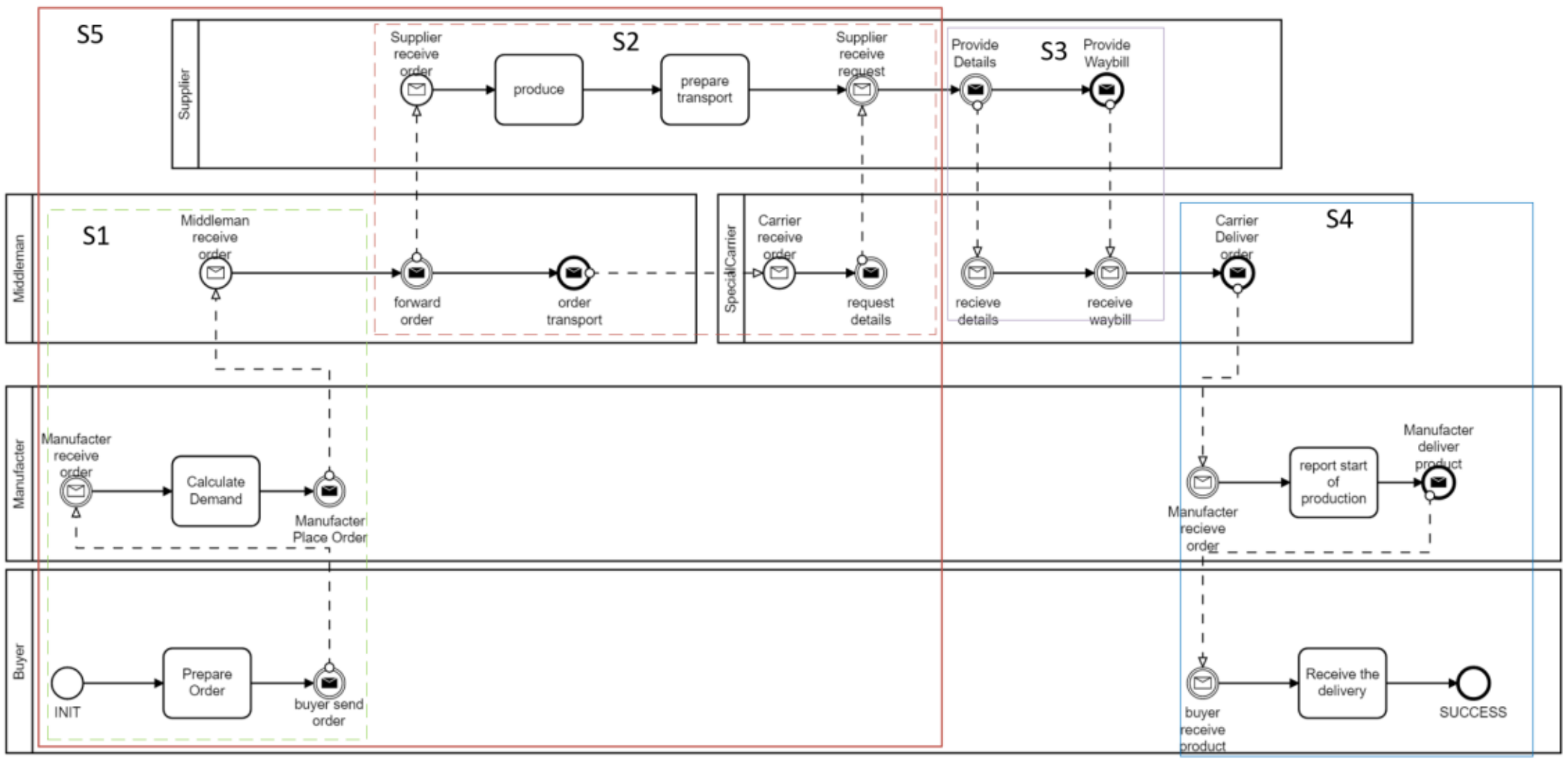

Fig. 5.1 SESE subgraph S5 contains proper LSI SESE subgraphs S1 and S2 (Adopted from (Liu et al., 2022c)).

If the user selects only S1 and S2 as two separate collaborative transactions, then the smart contract will have (n+2) smart contract methods: One method for each of the n actors of the contract, plus one method for each of S1 and S2. For each of the collaborative transactions S1 and S2, pattern augmentation techniques described in the previous section are used to amend the smart contract methods with patterns to create a transactional mechanism for each of S1 and S2.

When a developer selects a transaction with nested sub-transaction, the construction of the transactional mechanism for the nested transaction proceeds bottoms up, starting with creating the mechanism for the innermost transaction. First, the smart contract methods for the innermost transactions are constructed in the usual manner, one method for each of the innermost collaborative transactions as described above. If there is a collaborative transaction with m transactions nested within it, then there will be (n+m+1) smart contract methods: One for each of the actors of the smart contract, one for the outer (the parent) transaction, and one smart contract method for each of the child sub-transactions.

However, atomic commitment needs to be supported for the parent and its child sub-transactions; that is, the commitment of the parent and child transitions together must be atomic. To support the atomicity, we use the 2-Phase-Commit (2PC) protocol, wherein the parent-transaction smart-contract method is augmented with a pattern (Liu et al., 2022b) for the 2PC coordinator of the 2PC protocol, while each of the sub-transactions is augmented with the pattern (Liu et al., 2022b) for the 2PC participant. Thus, before the parent transaction may commit, each of its participating sub-transactions needs to be prepared to commit. Once the parent and children of the 2PC are ready to commit, the parent commits and then child sub-transactions commit (Gray & Lamport, 2006).

Recursive application of the above approach yields transactional support for nested transaction to an arbitrary level. As the 2PC protocol is a standard technique that has been used extensively in distributed systems, we shall not elaborate on details, but rather estimate the overhead caused by supporting nested transactions in the next section.

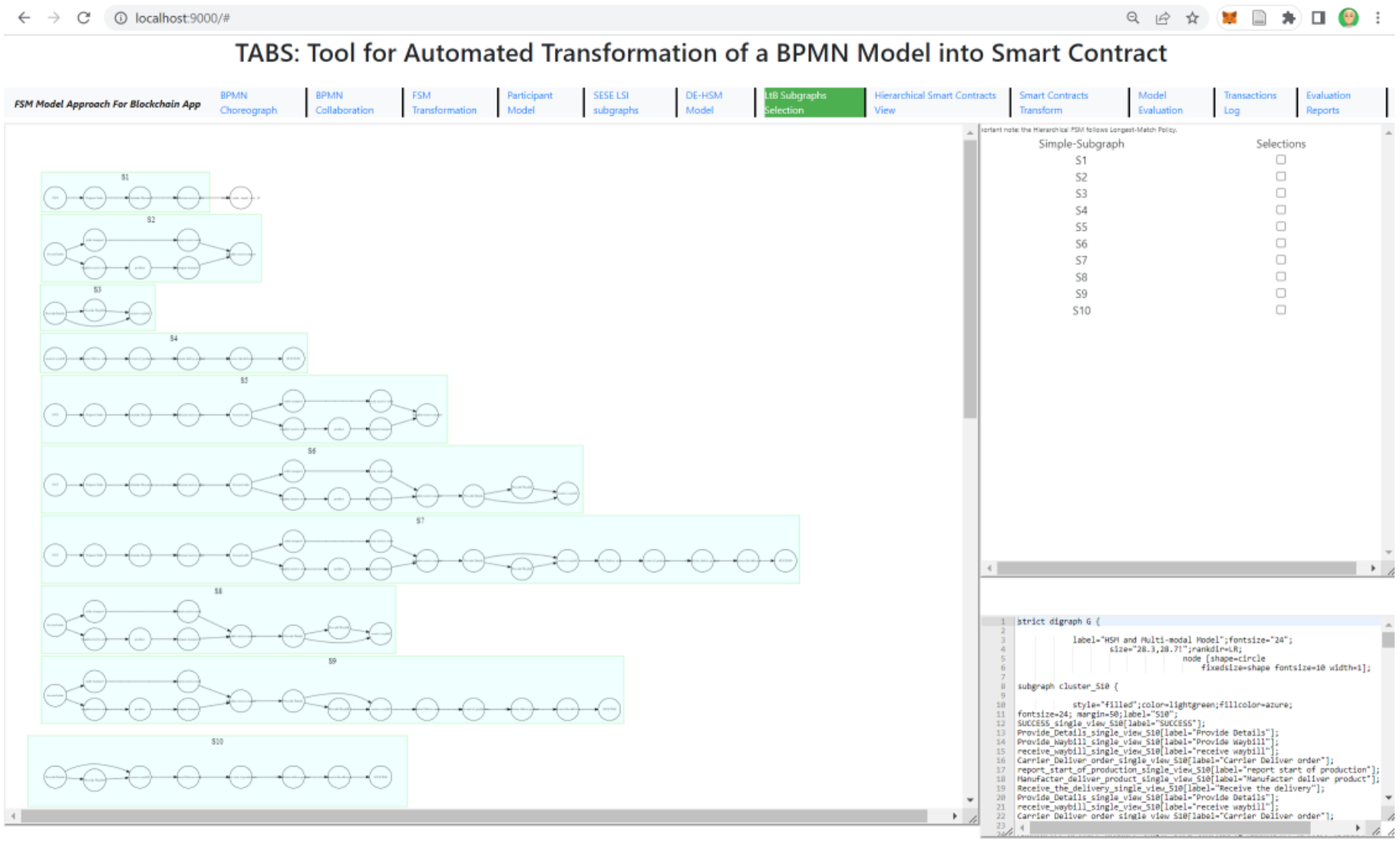

Fig. 5.2 Listing and selecting SESE subgraphs (adopted from (Liu et al., 2023b))

## 5.2 Supporting Collaborative Transactions with Nesting

It's imperative to note that the initial steps in the transformation of a BPMN model, leading to the identification of SESE subgraphs, remain consistent regardless of whether the objective is to support sidechain processing or to facilitate BPMN collaborative transactions. However, once the SESE subgraphs are shown to the user, as in Fig. 5.2, further transformation and processing differ from the case when the TABS tool is used to support sidechain processing. The developer is asked to select those SESE subgraphs that should be treated as a BPMN collaborative transaction. For instance, the developer may select the subgraphs S3, S4 and S5 as collaborative transactions. S3 and S4 are LSI subgraphs and hence do not contain other SESE subgraphs. However, the subgraph S5 contains the LSI subgraphs S1 and S2 as its subgraphs and hence they are shown to the user, as is shown in Fig. 5.3, as they lead to nested transactions in that S5 contains sub-transactions S1 and S2. The developer may then choose any one of S1 and S2, or both, or neither as sub-transactions. Following this, the developer is provided with options for the selection of the type of the collaborative transaction mechanism to be used to support the collaborative nested transactions as discussed below.

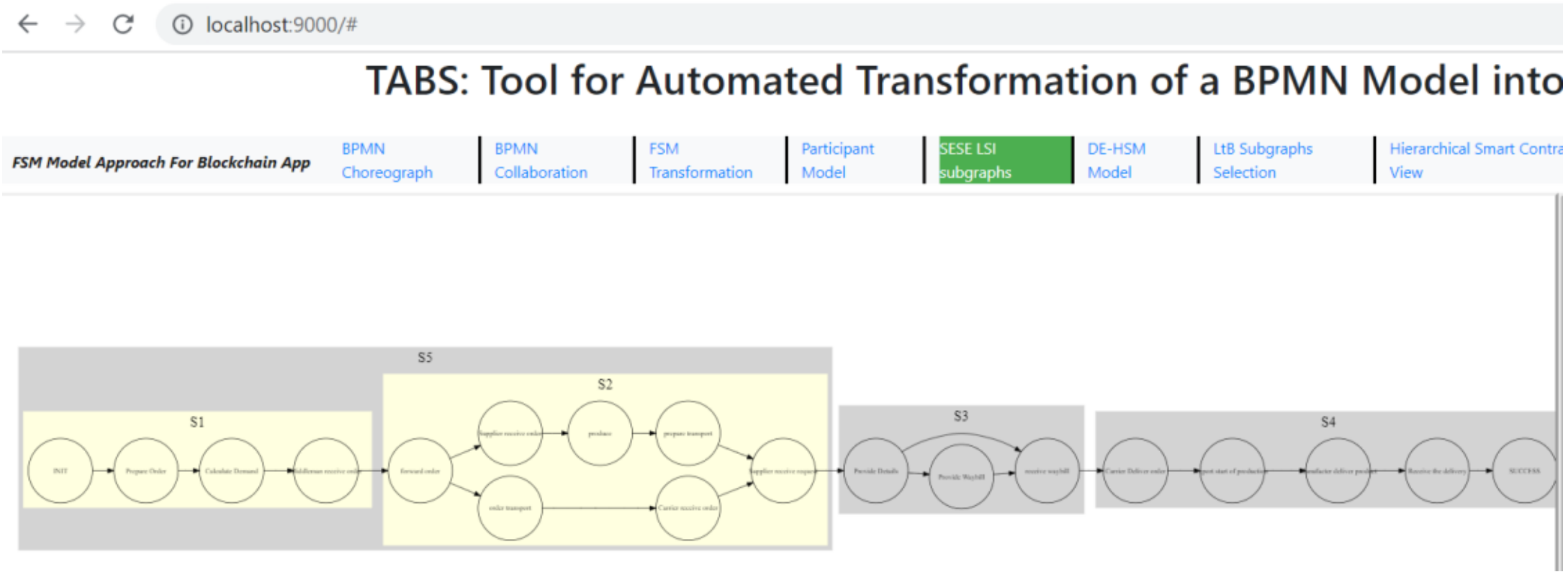

Fig. 5.3 User Selecting Subgraphs S3, S4, and S5, wherein S5 Contains LSI Subgraphs S1 and S2 (Adopted from (Liu et al., 2023b)).

The graphical representation of the methods for the collaborative smart contracts are shown in Fig. 5.4. For each of the collaborative transactions S3, S4, and S5, there is a method that is denoted by three icons labelled *method_start_SX*, *tx_SX*, and *method_end_SX*, where X is either 3, 4, or 5 depending on for which transaction it is. However, as S5 has two sub-transactions S1 and S2, it further contains two methods, one each for the sub-transactions S1 and S2, respectively. The figure was generated using (Node.js, 2023; Graphviz, 2023; Bostock, 2023) software.

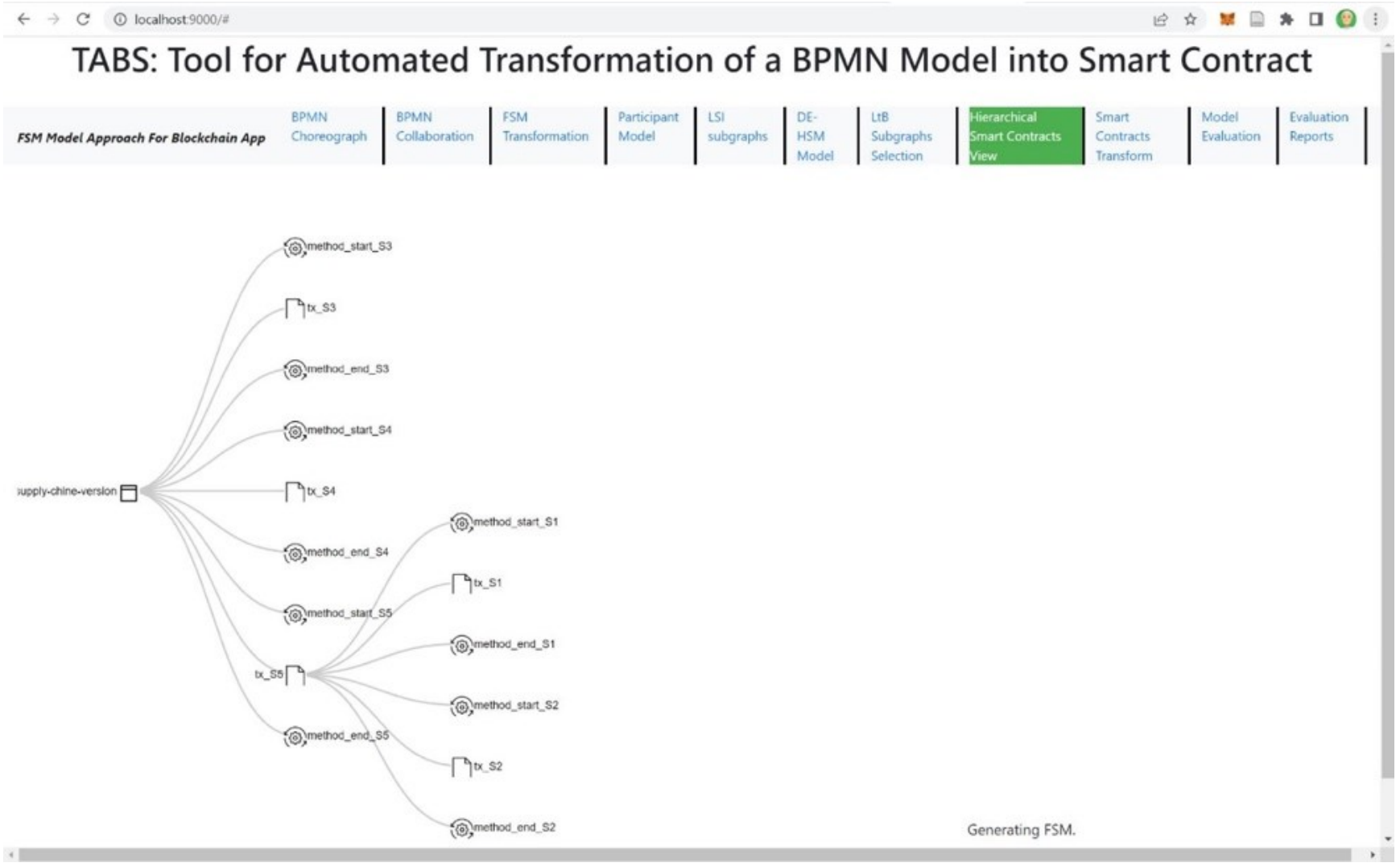

Fig. 5.4 Methods for Collaborative Transaction S4, S4, and S5, wherein S5 Has Sub-Transactions S1 and S2 (Adopted from (Liu et al., 2023b)).

To ensure the atomicity of the primary transaction in conjunction with its sub-transactions, a Two-Phase Commit (2PC) protocol is employed. The general workflow of our 2PC solution is delineated as a sequence diagram in Fig. 5.5.

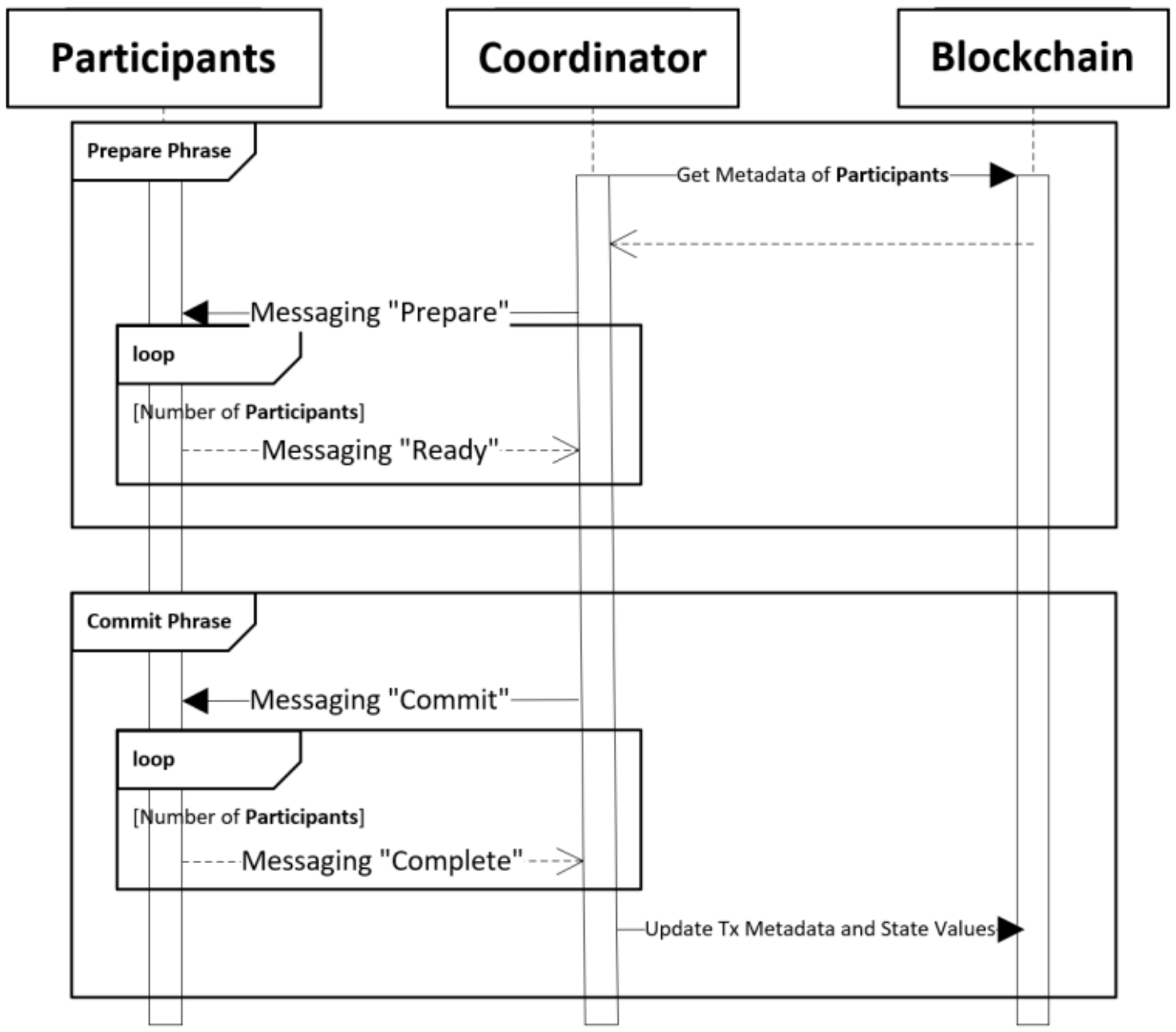

Fig. 5.5 Sequence Diagram Illustrating the 2-Phase Commit Protocol for Nested Transactions (Adopted from (Liu et al., 2023b)).

During the commitment stage of nested transactions within long-term transactions, the coordinator initiates the 2-phase commit by triggering the 'event' of the smart contracts, which sends a 'ready' message to all its 2-PC participants for the subordinate nested transactions. The execution of this event-emitting function incurs an additional cost. Subsequently, the coordinator awaits responses from the participants through its 'listener' interface. Each subordinate responds by sending back a 'yes' or 'no' message, requiring the response-collecting function to be activated 'N' times, where 'N' equals the number of subordinates. This is where the overhead of the 2-phase commit protocol is most evident. Similarly, during the second phase, the coordinator disseminates a 'commit' message to all the participants, who subsequently relay messages back. The coordinator declares a successful commitment if all participants return the positive “ready-to-commit” response. The additional cost of the second phase commit is roughly equivalent to that of the first phase.

In the context of our use case illustrated in Fig. 5.3, every primary transaction, functioning as a 2PC protocol coordinator, views each child sub-transaction as a 2PC participant. Therefore, S5 assumes the role of a 2PC coordinator for its sub-transactions S1 and S2, while both S1 and S2 embody the 2PC participant functionality. It's pivotal to recognize that a transaction can simultaneously be a parent and a child transaction, thereby necessitating functionalities to act both as a coordinator for its sub-transactions and as a participant when it's a sub-transaction within a parent transaction.

The anticipated cost implications for the transaction mechanism supporting sub-transactions are projected to be high. A more detailed discourse on cost estimations is reserved for Chapter 6.

# CHAPTER 6 TABS+ TOOL FOR PROOF OF CONCEPT AND EVALUATION

We created a proof-of-concept tool, named ***T**ransforming **A**utomatically **B**PMN Models into **S**mart Contracts* and referred to as ***TABS+*** for short, to demonstrate the feasibility and suitability of our approach. At present, a demonstration and restricted-duration online access to the tool is available for academic and research purposes. The tool can be accessed at the following URL: https://blockchain.cs.dal.ca/tabs (Liu et al., 2023d). Additionally, the source code for TABS+ is available upon formal request, which can be directed to Chris.Liu@dal.ca. In the near future, we plan on preparing appropriate documentation and include the code in a repository to facilitate future potential collaborations. In the following sections, we will first provide a description of the tool, and then we follow it by showing the results of the tool evaluation.

## 6.1 Transforming Automatically BPMN Models into Smart Contracts

Transformation starts with a BPMN model specification expressed in XML as per BPMN specifications (BPMN 2.0 Introduction, 2022). The TABS tool invokes Camunda software platform (Camunda Modeler conformance to BPMN and BPEL, 2016) that is used by the developer to create a BPMN model and store it in XML format. Fig. 6.1 shows a partial screen of the tool when creating a BPMN model for the supply-chain application. Once the modeler creates a BPMN model and stores it in an XML format, it is used as input to transformations using tabs appearing under the title. The first two tabs deal with BPMN modeling, one for a BPMN choreography, and one for a model expressed in core BPMN elements. Note that the term choreography here refers to the BPMN concept of choreography that shows “conversations” that eventually need to be elaborated upon by the modeler into a model expressed using BPMN’s core elements. Thus, the term “choreography”, as used in BPMN, is not to be confused with the term choreography as used in previous work on transforming BPMN models into smart contracts, in which choreography refers to the choreography of processes into which a BPMN model is transformed and wherein the choreography is a major part of the job mediator’s job of the system architecture of Fig. 4.2.

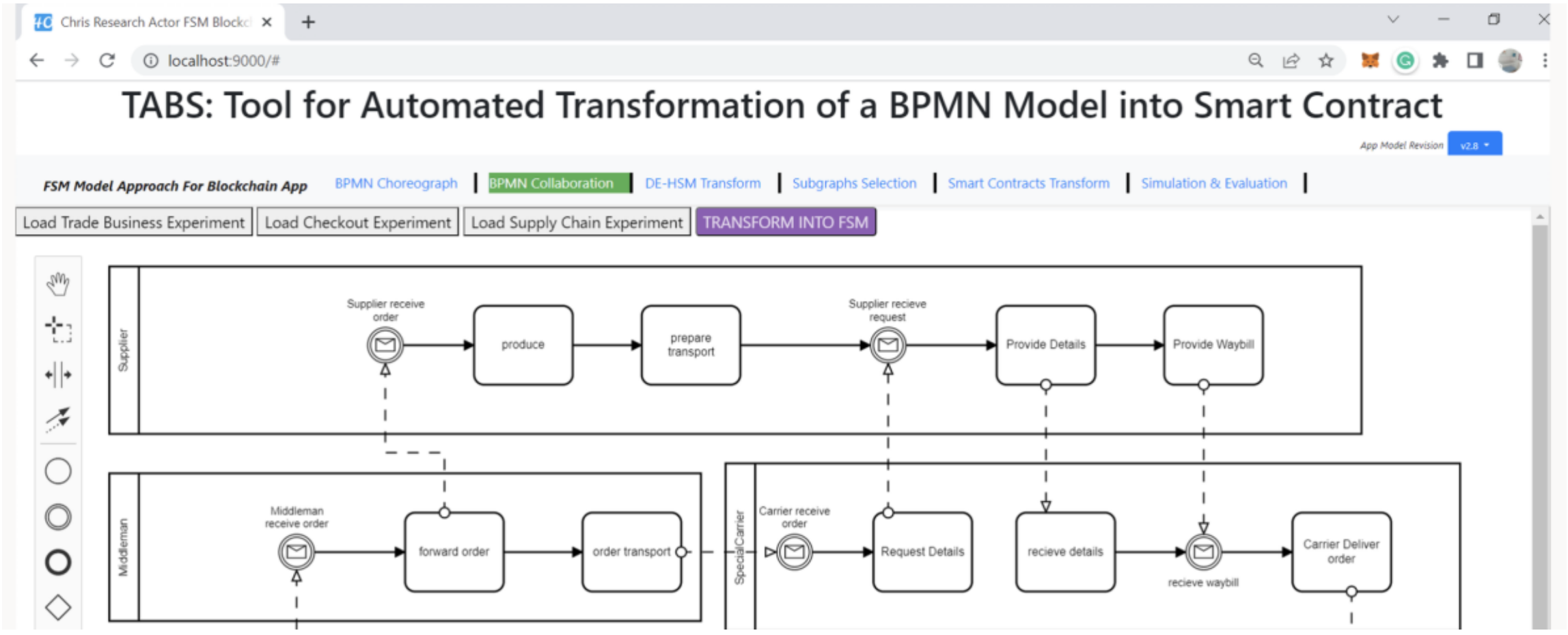

Fig. 6.1 TABS Tool Invoking Camunda Platform to Create a BPMN Model of The Supply-Chain Application (Adopted from (Bodorik et al., 2023)).

Once the BPMN model is developed, the modeler's interaction with the tool consists of (i) guiding the transformation of the BPMN model into a smart contract using the tabs and (ii) supplying the template methods, for BPMN tasks, with code. We facilitate creation of a smart contract and sidechain smart contract(s) either for Ethereum or Hyperledger fabric blockchains However, the tool does not facilitate automated creation of the underlying blockchain fabric itself. In fact, the tool supports using two blockchain types: (i) Hyperledger fabric and (ii) Ethereum or EVM-based blockchain such as Quorum. Either type can be used for the mainchain or for a sidechain(s), wherein the mainchain smart contract invokes methods of the smart contract deployed on a sidechain.

Fig. 6.2(a) shows a screenshot after the supply-chain BPMN model, shown in Fig. 2.1, was transformed into the DE-HSM model having the four LSI subgraphs. It shows the BPMN graph and its LSI subgraphs that have been identified by the TABS tool. Fig. 6.2 (b) was not generated by the tool but was included to show the BPMN representation of the corresponding LSI subgraphs of Fig. 6.2 (a).

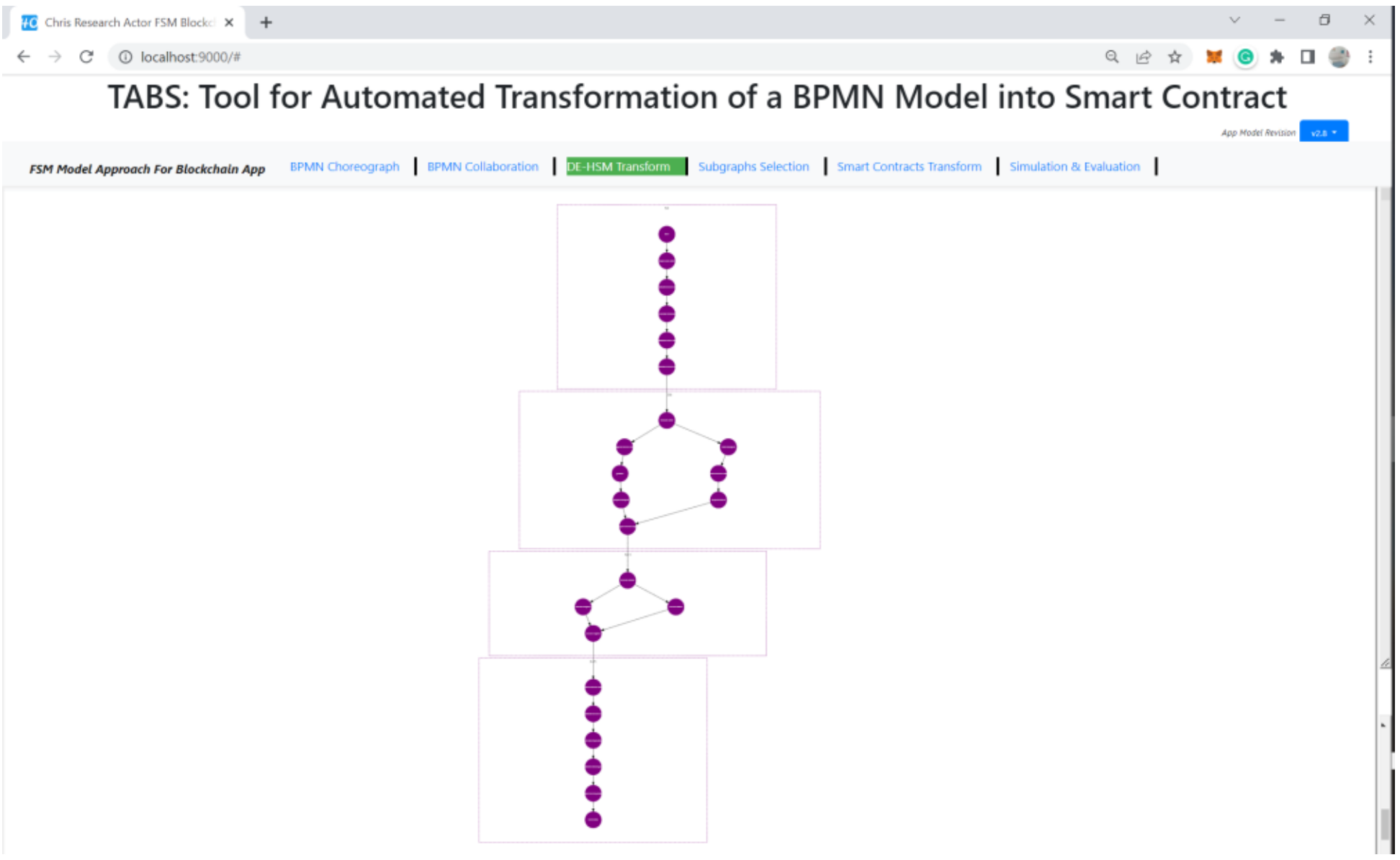

(a) Four found LSI subgraphs

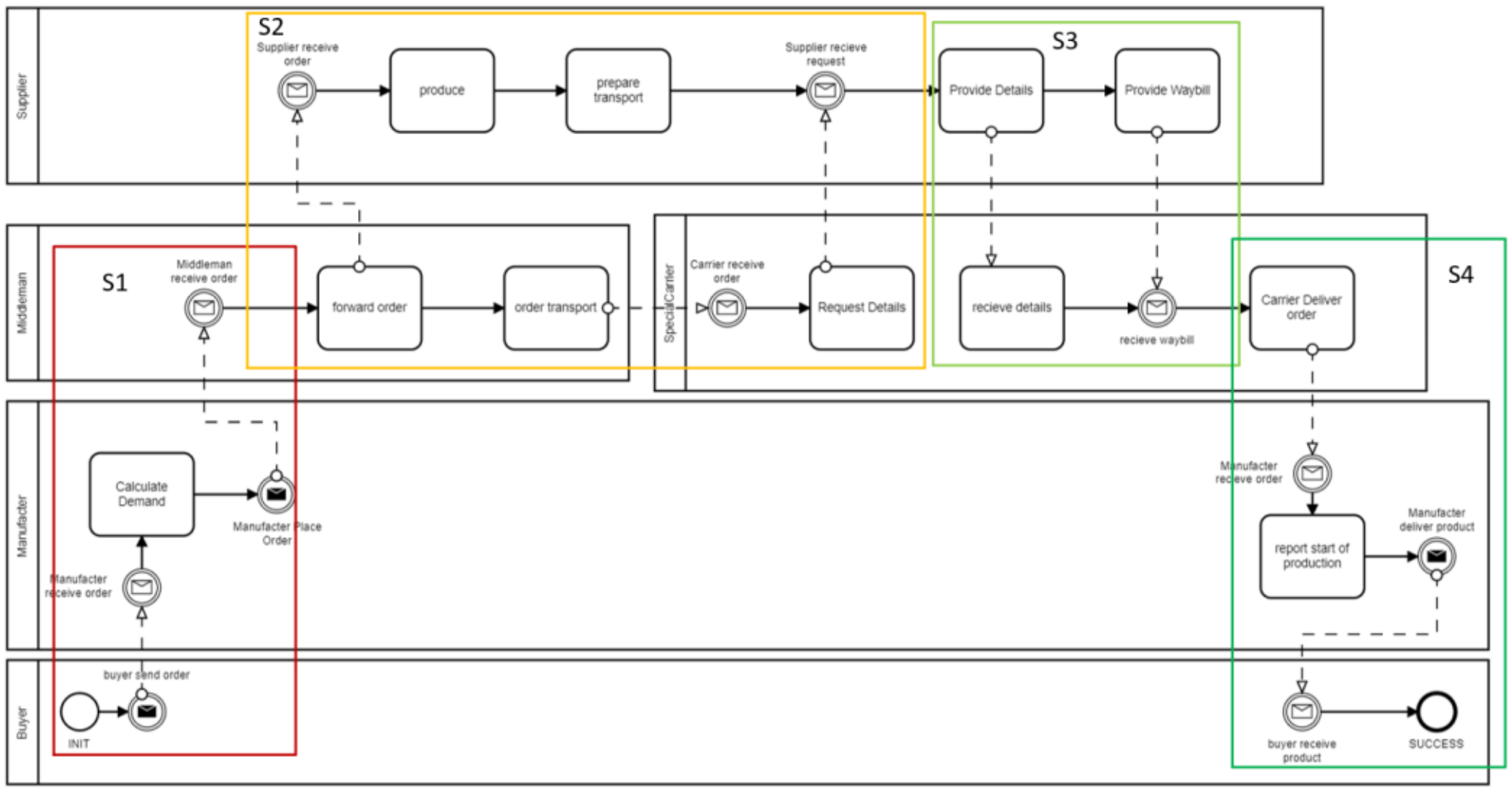

(b) LSI subgraphs in BPMN model

Fig. 6.2 LSI Subgraph after Transformation into A DE-HSM Model (Adopted from (Bodorik et al., 2023))

The modeler/developer is also involved in the decision-making on which of the independent subgraph patterns should be deployed on a sidechain in a form of a separate smart contract that interacts with the main contract deployed and executed on the main blockchain. The modeler than selects which of the Ethereum or Hyperledger fabric is to be used for each of the mainchain and a sidechain – for testing purposes we have local blockchains, one for each of the Ethereum and Hyperledger fabric. On Hyperledger fabric, we also have channels prepared that are used for deployment of smart contracts for both the mainchain and sidechains. For sidechains compatible with Ethereum, we use Quorum sidechain.

Once the selection is made by the developer, the model is transformed into the methods of a smart contract(s) that are deployed and executed. The modeler can examine the generated system by stepping through execution message-by-message, while the tool shows the progress graphically by showing the change to states of the individual FSMs representing the DE-FSM sub-models. This feature is helpful in testing and manual verification. Delays are also shown as execution proceeds step-by-step.

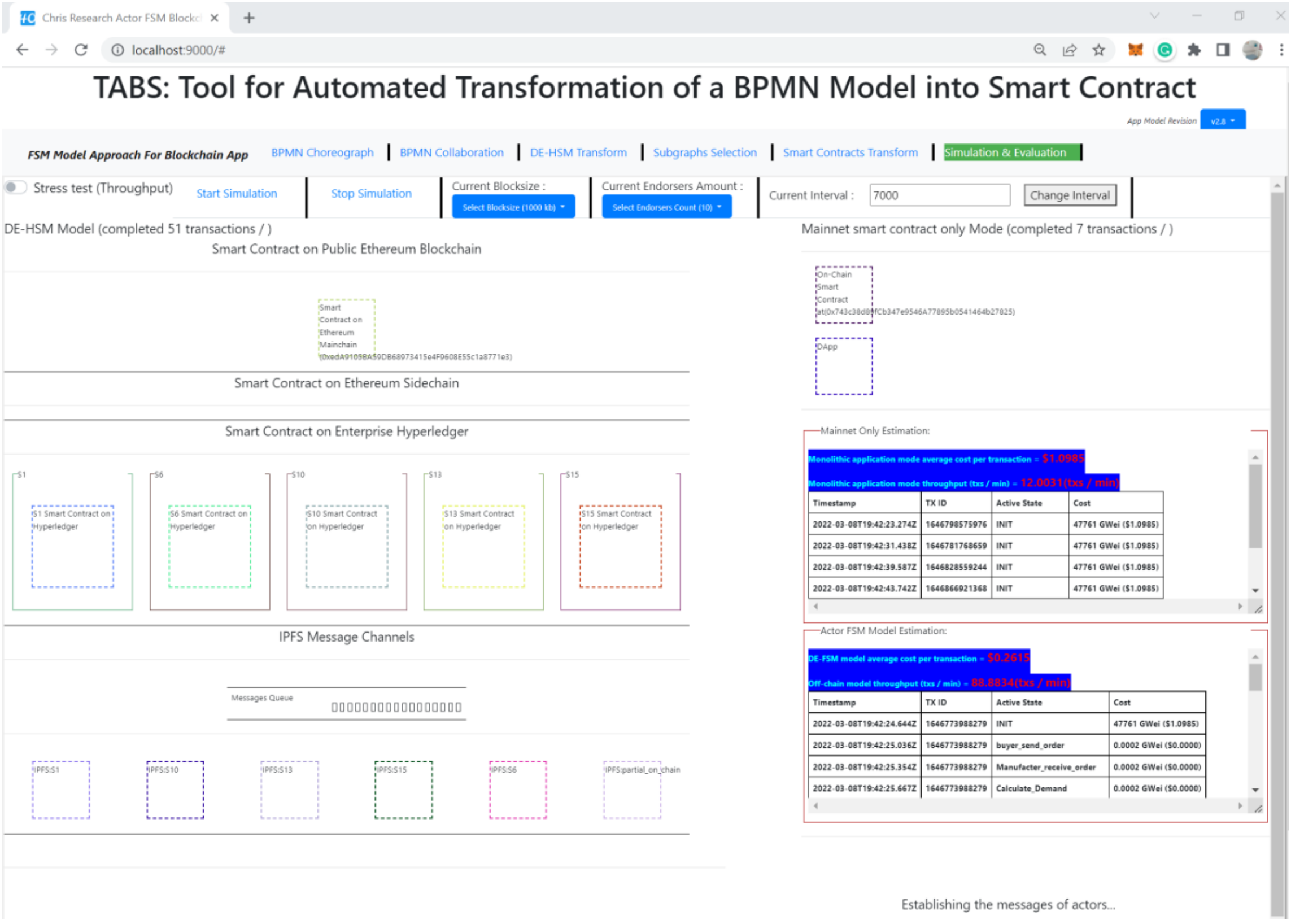

Fig. 6.3 Stress Testing Smart Contract Methods Execution of Bulk Business Processes (Adopted from (Bodorik et al., 2023))

To evaluate the latency and cost of the overall execution, the tool facilitates execution of each method while measuring its latency and cost. It should be noted that we show the cost of execution in Wei (or GWei) for Ethereum or EVM based blockchains/sidechains. Latency and costs are calculated for two cases: (i) When all methods of smart contract are deployed on the mainchain only, and (ii) when execution proceeds on the mainchain with the selected patterns deployed on a sidechain. Thus, the developer can compare the latency and costs when execution is on the mainchain only and when the selected patterns are processed on a sidechain. Fig. 6.3 exhibits a screenshot showing the cost of execution on a local Ethereum mainchain and local Microfab sidechains, where the cost of execution is in GWei and is derived by adding the cost of individual Solidity instructions of the smart contract methods using web3.js (web3.js, 2022). As far as cost is concerned, for Ethereum we estimate the cost in GWei units, using web3.js, which is certainly useful to the modelers. Cost for the case when Hyperledger fabric is used depends directly on the cost of the underlying blockchain infrastructure and we have not tackled as yet how to derive it. As in the case of the pubic Ethereum blockchain, blockchain as a service for Hyperledger fabric is not cheap and it depends on the blockchain configuration in terms of the network nodes, network bandwidth, etc.

## 6.2 Designing and Preparing the Evaluation Environment

In the previous sections, we introduced the TABS+ tool as a Proof of Concept (PoC) to demonstrate the feasibility of our approach and to explore its properties, such as cost and latency. In this subsection, we first outline our evaluation method and then discuss the use cases we tested.

### 6.2.1 Evaluation Method

The TABS tool is a typical two-tier application. The frontend was mainly written in NodeJS, which focuses on the tasks such as:

- Establishing a distributed message channel with backend IPFS nodes (IPFS API, 2023).
- Providing a canvas for users to compose BPMN diagrams, implemented by BPMN.io (bpmn.io, 2023).
- Transforming BPMN diagrams to DE-HSM mode, powered by the GraphViz JS library (GraphViz, 2023).
- Providing Ethereum Solidity and Hyperledger NodeJS abstract interfaces for smart contract transformation, composed of Solidity and NodeJS, respectively (Solidity 0.8.13 documentation, 2022), (Hyperledger Fabric SDK for Node.js, 2023).

- Compiling Smart Contracts with Ethereum using NodeJS Worker Threads (Worker threads | Node.js v17.7.1 Documentation, 2023).
- Compiling Smart Contracts with IBM Blockchain Platform Developer Tools (IBM Cloud Docs, 2023).
- Deploying Ethereum Smart Contracts with EtherJS and Web3.js (docs.ethers.io, 2022)(web3.js, 2023).
- Deploying Hyperledger Chaincode with Microfab API (GitHub - IBM-Blockchain/Microfab, 2023).
- Establishing an IPFS message channel with IPFS API (IPFS cluster consistency model, 2023).
- Implementing evaluation and simulation processes with NodeJS (Node.js, 2023).

According to the analysis conducted using the Count Lines of Code (Cloc) tool, the active code repository of the TABS tool, excluding third-party libraries and focusing solely on our custom-written code, comprises a total of 472,422 lines. Cloc is a renowned lightweight git repository analyzer that offers comprehensive statistics (AlDanial, 2015). A detailed breakdown by programming language can be viewed in Fig. 6.4.

| Language | files | blank | comment | code |
|---|---|---|---|---|
| JavaScript | 572 | 36148 | 88356 | 254905 |
| JSON | 23 | 0 | 0 | 199642 |
| XML | 22 | 1 | 0 | 7349 |
| CSS | 15 | 670 | 440 | 5415 |
| Solidity | 6 | 459 | 1031 | 2207 |
| HTML | 2 | 148 | 298 | 1635 |
| Text | 4 | 12 | 0 | 549 |
| Markdown | 2 | 135 | 2 | 387 |
| INI | 5 | 76 | 557 | 140 |
| Python | 1 | 26 | 32 | 86 |
| Bourne Shell | 12 | 4 | 0 | 79 |
| YAML | 1 | 0 | 0 | 21 |
| Dockerfile | 1 | 4 | 1 | 7 |
| SUM: | 666 | 37683 | 90717 | 472422 |

Fig. 6.4 Code Statistics for The TABS Tool Repository, as Generated by Count Lines of Code (Cloc).

We utilized three DigitalOcean cloud servers (DigitalOcean – The developer cloud, 2020) for hosting their developed TABS tool, blockchains, and blockchain applications, each with 2 CPUs, 4GB RAM, and 80GB disk space, running on Ubuntu OS. Ganache-CLI was installed, configured, and executed on each server with various parameters to simulate realistic working environments. The mainchain blockchain was configured with modified parameters for block time, endorsers, and block size, comparable to the public Ethereum blockchain. Sidechain networks were configured as Quorums for Ethereum smart contracts and Microfab for Hyperledger Chaincode. TABS also allowed testers to deploy smart contracts on actual public Ethereum and Ropsten sidechains if they owned sufficient ETH (etherscan.io, 2023). Each server functioned as an IPFS node within a preconfigured IPFS private cluster, granting each participant exclusive access to the IPFS node and data space.

We improved TABS features to enable blockchain developers and designers to model blockchain applications using BPMN as a starting point (Bodorik et al., 2023). BPMN modeling initiates the evaluation process for a use case. TABS automatically converts the BPMN model to a DE-HSM model at the end of the modeling phase. The modeler then provides the code for the BPMN elements' tasks. The DE-HSM model is presented to the modeler with information about the execution cost on the mainchain alone and on the mainchain with sidechain processing of selected subgraphs. The modeler analyzes the provided data to determine which LSI subgraphs to deploy and process on a sidechain and which Ethereum Solidity and Hyperledger chain-code to use for smart contract creation on the mainchain and sidechain, respectively. The developer can examine the generated code for the smart contract(s).

### 6.2.2 Experimental Applications

We employed three experimental blockchain applications of varying complexity for evaluation purposes:

I. *Supply Chain Management*: Adopted from Weber et al. (2016), this use case has been discussed as a running example throughout this paper (see Fig. 2.1). The model includes ten tasks and four gateways. The supply chain use case begins with the buyer issuing a new order. When the manufacturer receives the order, they calculate demand and place an order with a middleman. The middleman concurrently sends the order to the supplier and requests transport from the carrier. The producer fabricates the product and prepares it for transport. Upon receiving the request from the middleman, the carrier asks for details from the supplier. The supplier provides the necessary information to the carrier and prepares the waybill. The carrier then delivers the order to the manufacturer after receiving the product details and waybill from the supplier. Once the order is received, the manufacturer begins production and delivers the finished product to the buyer, who then accepts the order.

II. *Order Process*: This process, adopted from Fleischmann et al. (2013), is relatively straightforward, with five tasks and two gateways (see Fig. 6.5(a)). The use case starts with the customer preparing an order that is sent to order handling. Order handling checks the order and concurrently (i) confirms the order with the customer and (ii) initiates shipment preparation by the shipper. The shipper prepares the order for shipment and sends it to the customer. Upon receiving the order confirmation and the order from the shipper, the customer accepts the order, completing the workflow. Fig. 6.5(b) shows the BPMN diagram with the identified LSI subgraphs marked by colored rectangles.

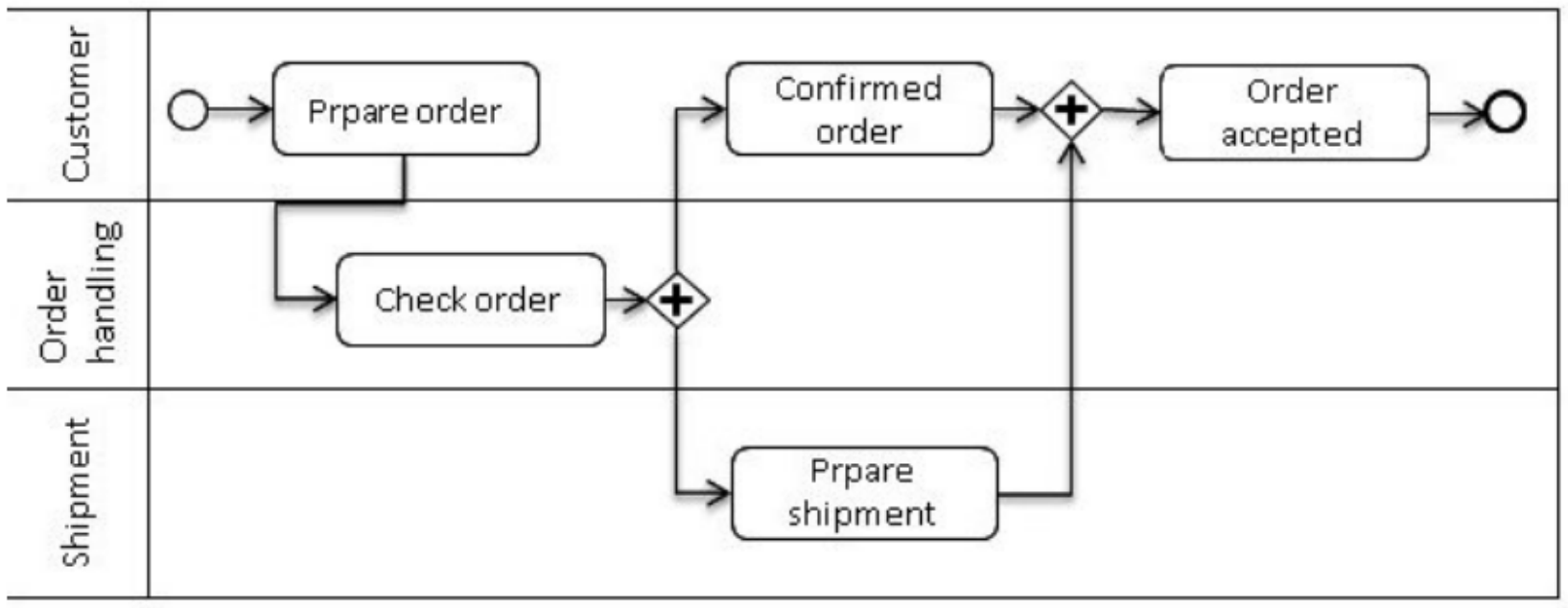

(a) BPMN diagram

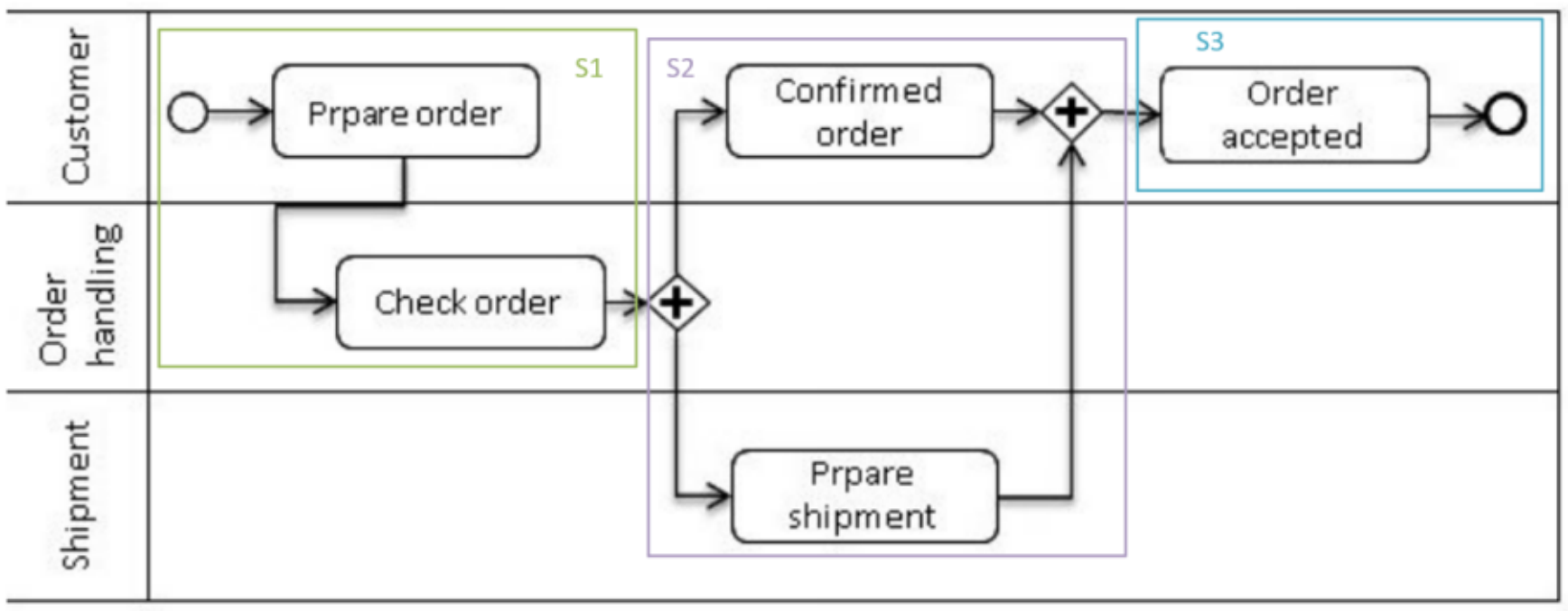

(b) BPMN diagram with found LSI subgraphs

Fig. 6.5 BPMN Diagram for The Order Process Use Case (Adopted from (Fleischmann et al., 2013))

III. *Trade*: Adopted from Asgaonkar & Krishnamachar (2019), this use case contains 19 tasks and five gateways (see Figure 6.6). The model shows that a product is posted for sale, and after an offer is made, the buyer and seller negotiate the price. Once they agree on the price, a contract is prepared and signed, stipulating escrow deposit and delivery terms. When the buyer deposits funds into an escrow account, the shipment proceeds to the port. The shipment involves crossing borders and customs processing. Subsequently, the shipment is loaded onto a ship, delivered to a port, processed at the destination customs, unloaded from the ship to the port, picked up by the buyer, payment terms are executed, and finally, the escrow deposit is returned.

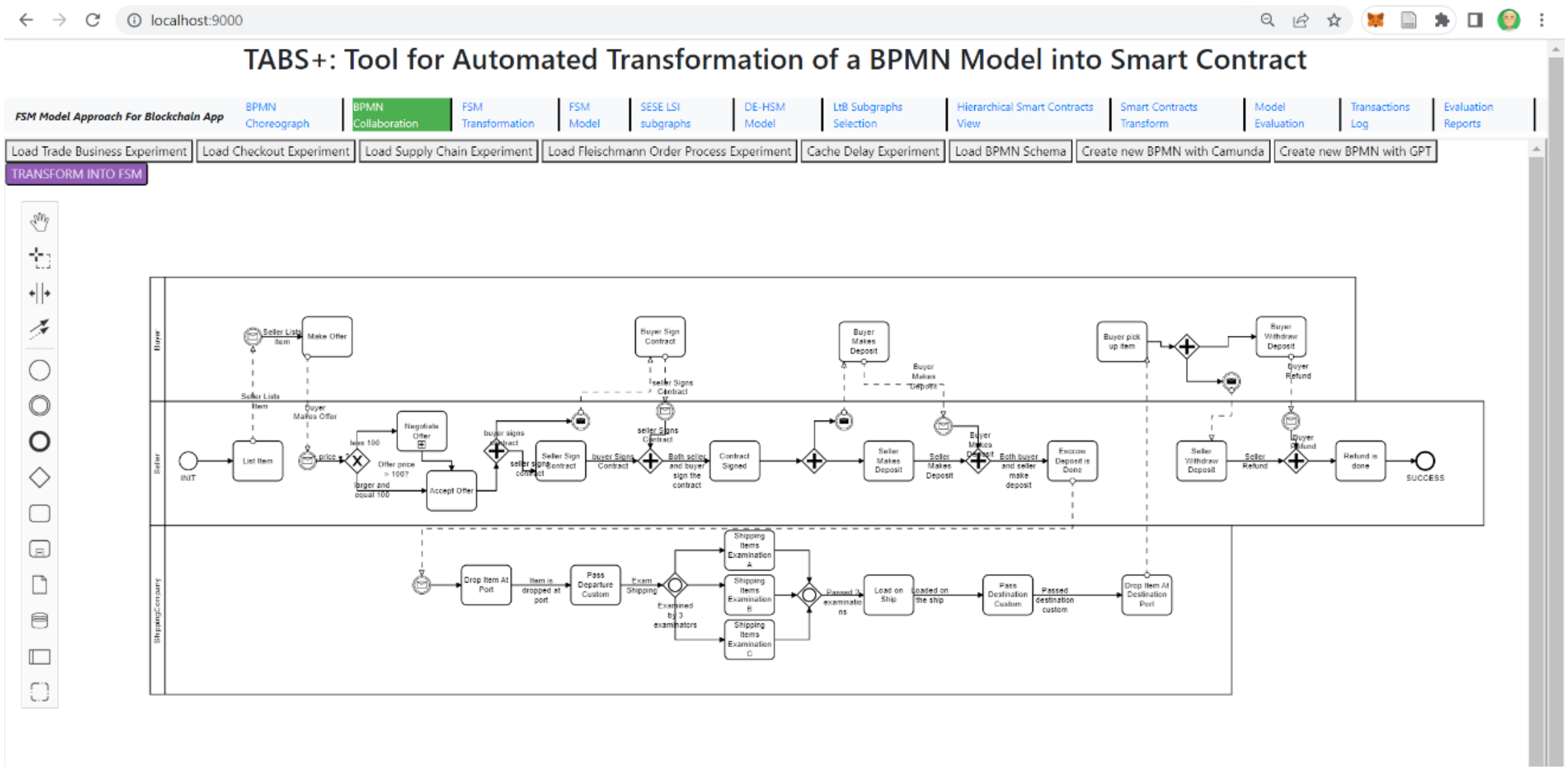

(a) BPMN diagram

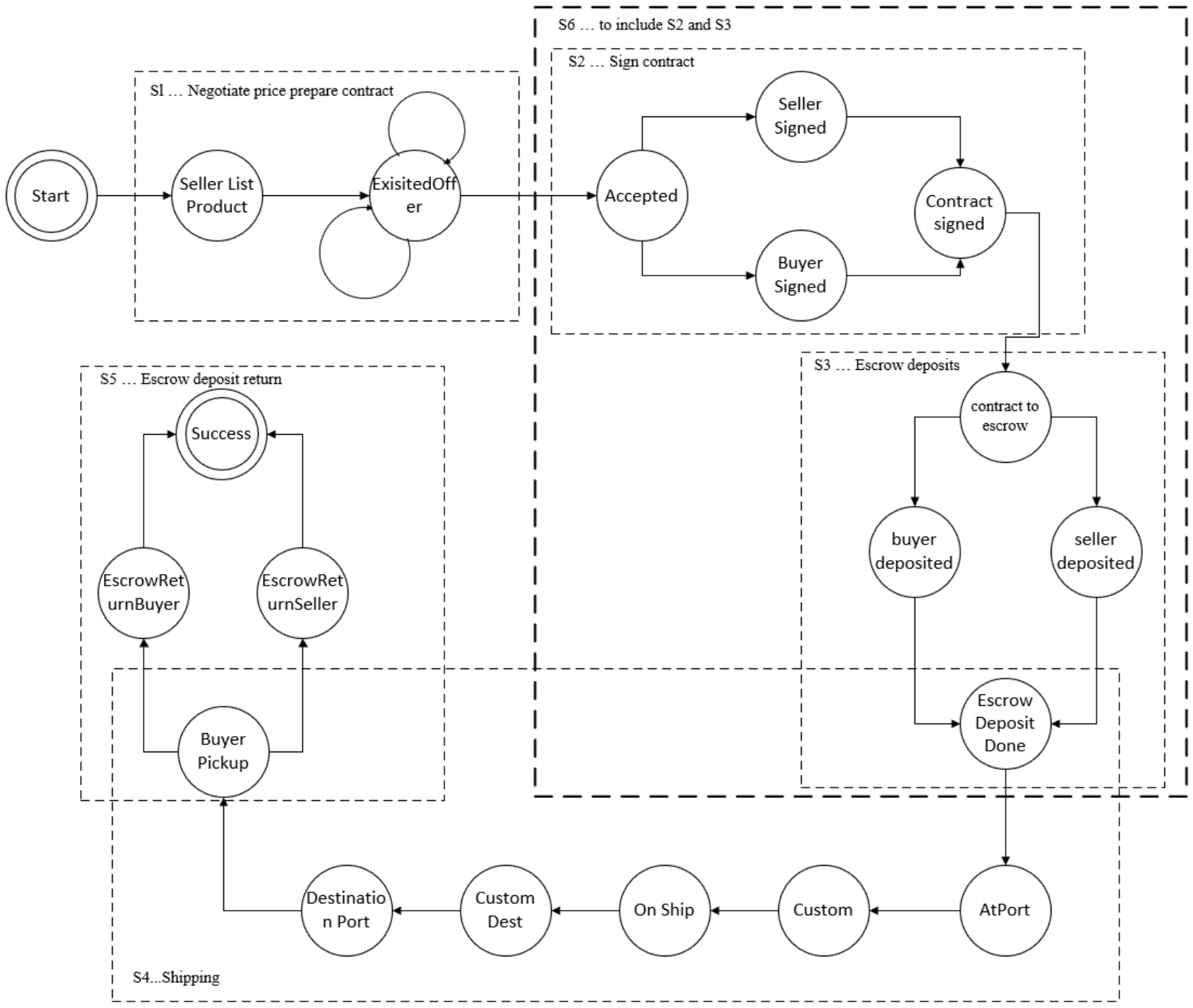

(b) Found LSI subgraphs

Fig. 6.6 Trade Business Use Case (Adopted from (Bodorik et al., 2021))

## 6.3 Mechanisms to Support Nested Long-Term Transactions: Overhead

Before delving into the overall performance evaluation of the application, it is essential to investigate the overhead costs associated with various mechanisms designed to support long-term transactions using multi-method smart contracts. This evaluation aims to compare the overhead costs of different deployment modes of smart contracts, specifically focusing on accessing data in private workspaces.

The evaluation is conducted in two distinct scenarios:

- *Supporting Long-Term Transactions Without Nesting*: In this situation, we will analyze the overhead costs of mechanisms that facilitate long-term transactions without involving nested structures. This will provide insights into the efficiency and cost-effectiveness of the methods when dealing with straightforward long-term transactions.
- *Supporting Long-Term Transactions With Nesting*: This scenario will extend the evaluation to include nested transactions within long-term transactions. The complexity of nested structures may introduce additional overhead, and this part of the evaluation will shed light on how different mechanisms handle this complexity.

By comparing these two scenarios, we aim to understand the trade-offs and benefits of various mechanisms in different contexts. This understanding will be crucial in selecting the most suitable approach for implementing long-term transactions with or without nested structures on the blockchain, balancing efficiency, cost, and complexity.

### 6.3.1 Collaborative Transactions without Nesting

Fig. 6.7 shows several screenshots for the use cases shown in figures 2.1 and 4.6:

- Fig. 6.7(a) shows the BPMN model that was created using the TABS tool (TABS simply invokes the Camunda BPMN software editor (BPMN 2.0 Symbols, 2023) to support the creation of a BPMN model and store it in as an XLM file (Business Process Execution Language, 2023).
- Fig. 6.7(b) shows the LSI SESE subgraphs forming the DE-FSM models, i.e., when each subgraph does not have any proper subgraph that is an LSI SESE subgraph itself, as generated by the TABS+ tool.

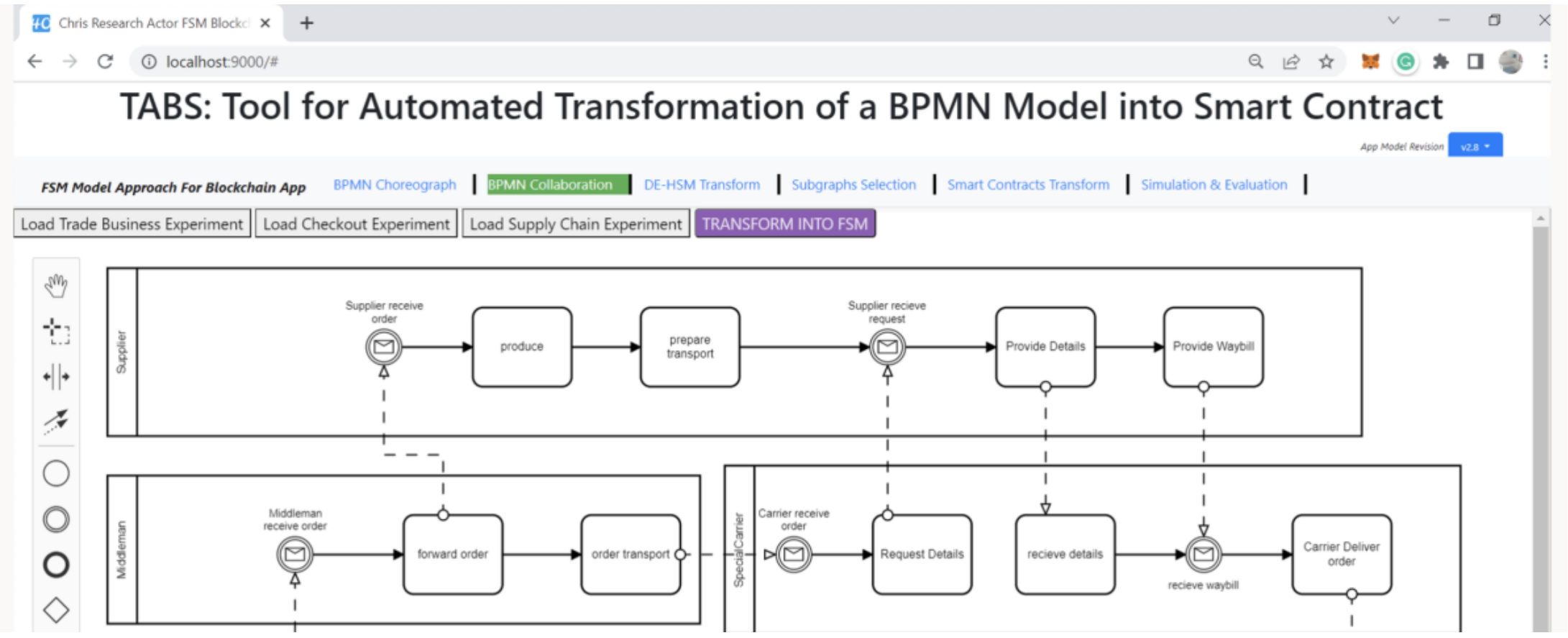

(a) Creation of BPMN model

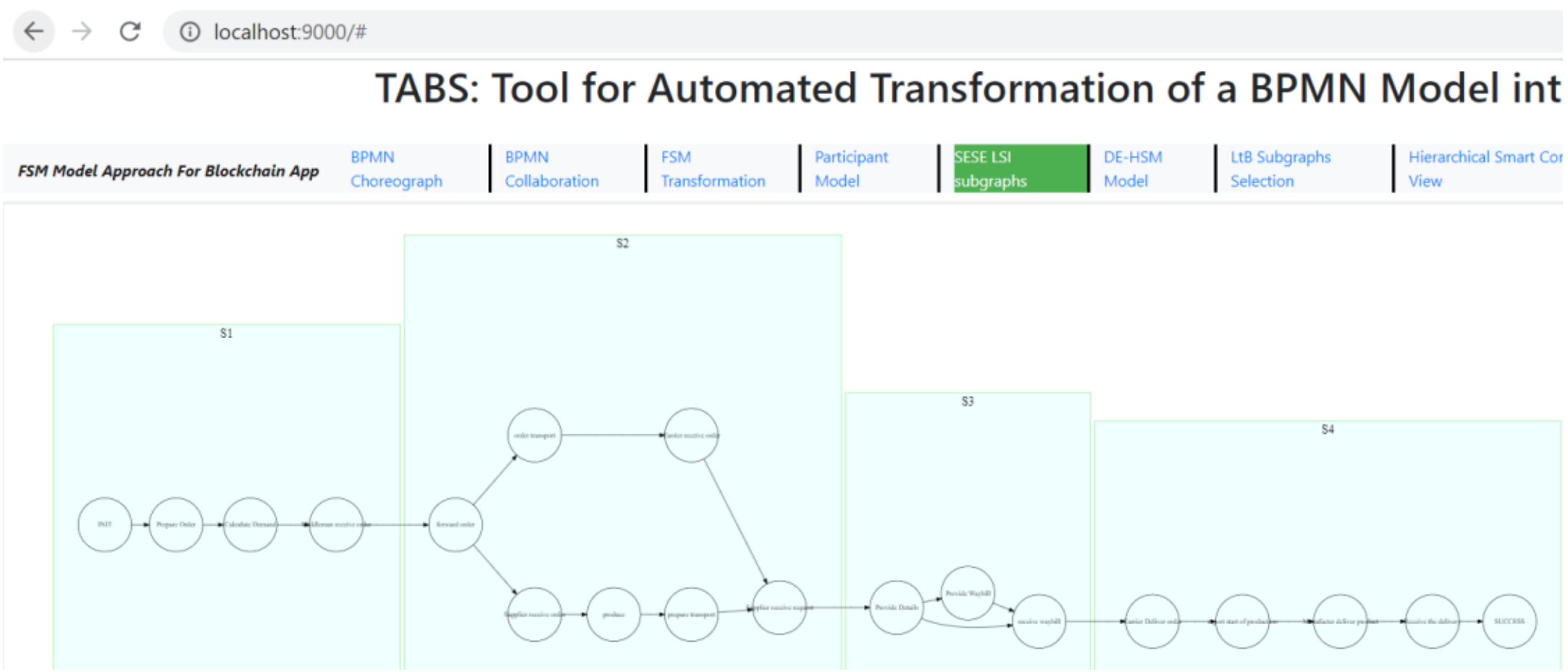

(b) Four found LSI subgraphs

Fig. 6.7 TABS+ Tool Use: Creation of A BPMN Model Finding LSI SESE Subgraphs (Adopted from (Liu et al., 2023b))

Recall that in the BPMN model is represented as an interconnection of the LSI subgraphs, wherein each LSI subgraphs is represented as a DE-FSM model that may concurrent FSMs to represent the concurrent streams of executions within sub-model. The interconnection of the LSI subgraphs is shown in Fig. 6.7(b).

Fig. 6.8 is a partial screen of the tool showing the SESE subgraphs for selection to be treated as collaborative transactions. All SESE subgraphs are listed, while the first four subgraphs, S1, S2, S3, and S4, are the four LSI subgraphs shown in Fig. 6.7(b). The user selects in the right-hand-side pane which of the 10 subgraphs, S1, S2, …, S10, are to be treated as collaborative transactions.

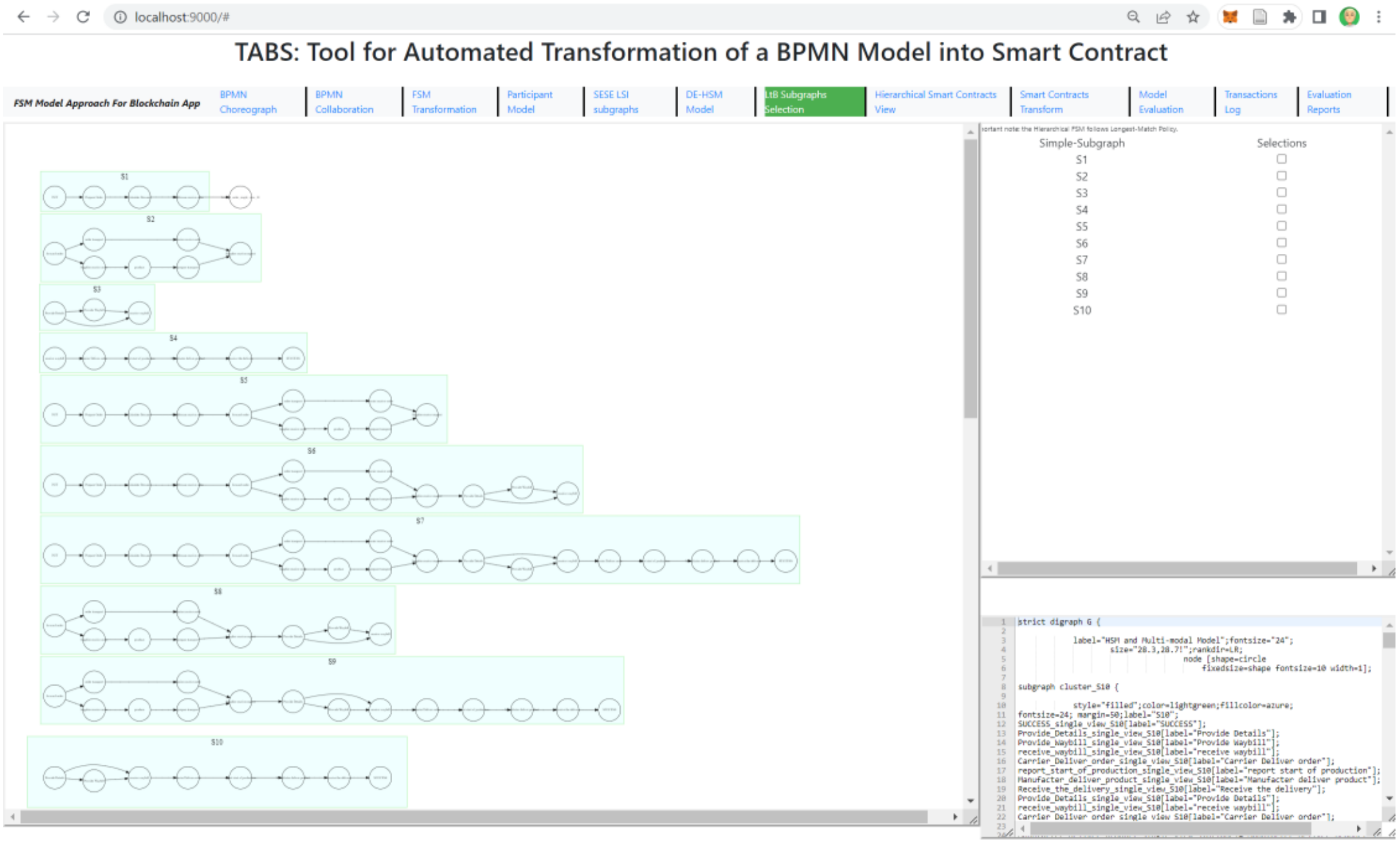

Fig. 6.8 TABS+ Tool: Listing and Selecting SESE Subgraphs (Adopted from (Liu et al., 2023b))

Recall that to support the collaborative transaction mechanism, we described three architectural alternatives in the subsection 3.3 that depend on the selection of the private workspace to store the transaction's state and the ledger access. We now discuss if and how the TABS+ supports each alternative.

- Cache is hosted in the private data structure, if such a data structure is provided by the native blockchain: We do not support this alternative in the TABS+, as it requires specialized treatment, depending on how the private data structure is implemented in the native blockchain. Recall that the private data structure must persist across invocations of the transaction methods and must be sharable by all actors participating in the transaction.
- The cache is hosted on the ledger and all methods are in the same smart contract: However, as the ledger is accessible to any user who has access to the ledger, an attacker, who has access to the ledger, has access to the scripts and hence it feasible for the attacker to deduce form the smart contract methods' scripts where the private workspace is located on the ledger and hence access it. Consequently, the method will provide least privacy and should be supported by cryptographic methods.

- Slave smart contract is used to host the cache and the methods of the collaborative transaction: The methods are invoked by the main smart contract that contains all non-transaction methods that invoke the transaction methods of the slave smart contract. There are two variations:
  - The slave smart contract is deployed on the same chain as the main smart contract.
  - The slave smart contract is deployed on a sidechain, while the main smart contract is deployed on the mainchain.

As we do not support the first case and the last case had two subcases, the TABS+ tool supports the following three alternatives to achieve the support for BPMN collaborative transactions:

- *sc-all*… One smart contract is used to host all methods, whether they are transaction methods or non-transaction methods.
- *sc-2m* … Two smart contracts, one is used to host the transaction methods and the cache, while the second one is used to host all non-transaction methods. Both contracts are deployed on one chain, referred to as the mainchain.
- *sc-2s* … Two smart contracts, one is used to host all non-transaction methods and is deployed on the mainchain, while the second one, deployed on a sidechain, is used to host all transaction methods.

For the last case, when there are two smart contracts and sidechain processing is used, we used the same cost calculations for both the mainchain and the sidechain. Thus, the estimated cost when there is sidechain processing is higher than when both smart contracts are on the mainchain. As sidechain processing incurs overhead relative to processing on the mainchain only, it is cost-wise advantageous only when sidechain processing is cheaper than processing on the mainchain.

However, as we are interested in determining the costs for all the above alternatives, we configured the tool to provide smart contracts for all three options. To determine the cost of executing patterns for each of the alternative mechanisms, we utilize the Ethereum blockchain tools that enable estimation of the cost of execution of smart contract methods written in the Solidity language. We use the Remix compiler that provides for compilation of Solidity smart contract method into executable code and provides estimates of the total cost of executing a smart contract method by relatively detailed cost calculation for each of the method's instruction. We use the Remix compiler to measure the cost of execution for each method of the smart contract for each of the alternative transaction mechanism approaches and for the base case (labeled as no-xa) that does not support collaborative transactions. To calculate the cost of the mechanisms, we performed measurements using a smart contract that contains the following methods:

- Smart contract method *m (inputParameter objectX(x)*. It receives a parameter that is an object of a specified size, and it invokes the methods m1 and m2 in that order.
- method *m1(x1: objectX)* … the method writes to the ledger the object passed as an input parameter.
- method *m2(x1: objectX, x2:objectX)* … the method reads from the ledger the object written by the method m1 and then writes it back to the ledger again as a new object.

We use the Remix compiler to measure the gas cost of execution for each method of the smart contract and thus estimate the relative overhead cost of providing a transaction mechanism when compared to the base case when execution is without a transaction mechanism. The table 6.1 and Fig. 6.9 show the cost for each of the alternative mechanism in the numerical and graphical representation, respectively. It's important to mention that we've standardized the gas price at 20 Gwei to maintain consistency with the default gas price of Ganache, which we used for cost evaluation in our preceding research papers (Liu et al., 2022a; Bodorik et al., 2023). The first column of the table contains the label identifying the transaction mechanism. The subsequent columns show the cost estimates for the alternative mechanisms when the collaborative transaction reads and writes a ledger object of size X = 75, 512, 1024, and 1875 Kb. Note that for current Ethereum, the 1875Kb is the maximum block size that can be used for recording one transaction on the ledger (Ethereum's New 1MB Blocksize Limit, 2021).

Table 6.1  CPU Processing Cost Estimates (in GWei) (Adopted from (Liu et al., 2023b))

| ***Label*** | ***7kb*** | ***512 kb*** | ***1024 kb*** | ***1875 kb*** |
|---|---|---|---|---|
| *no-xa* | 4545000 | 31027200 | 62054400 | 113625000 |
| *sc-all* | 9090582 | 62054982 | 124109382 | 227250582 |
| *sc-2m* | 9319500 | 63621120 | 124113000 | 227254200 |
| *sc-2s* | 9405000 | 64204800 | 127242240 | 232987500 |

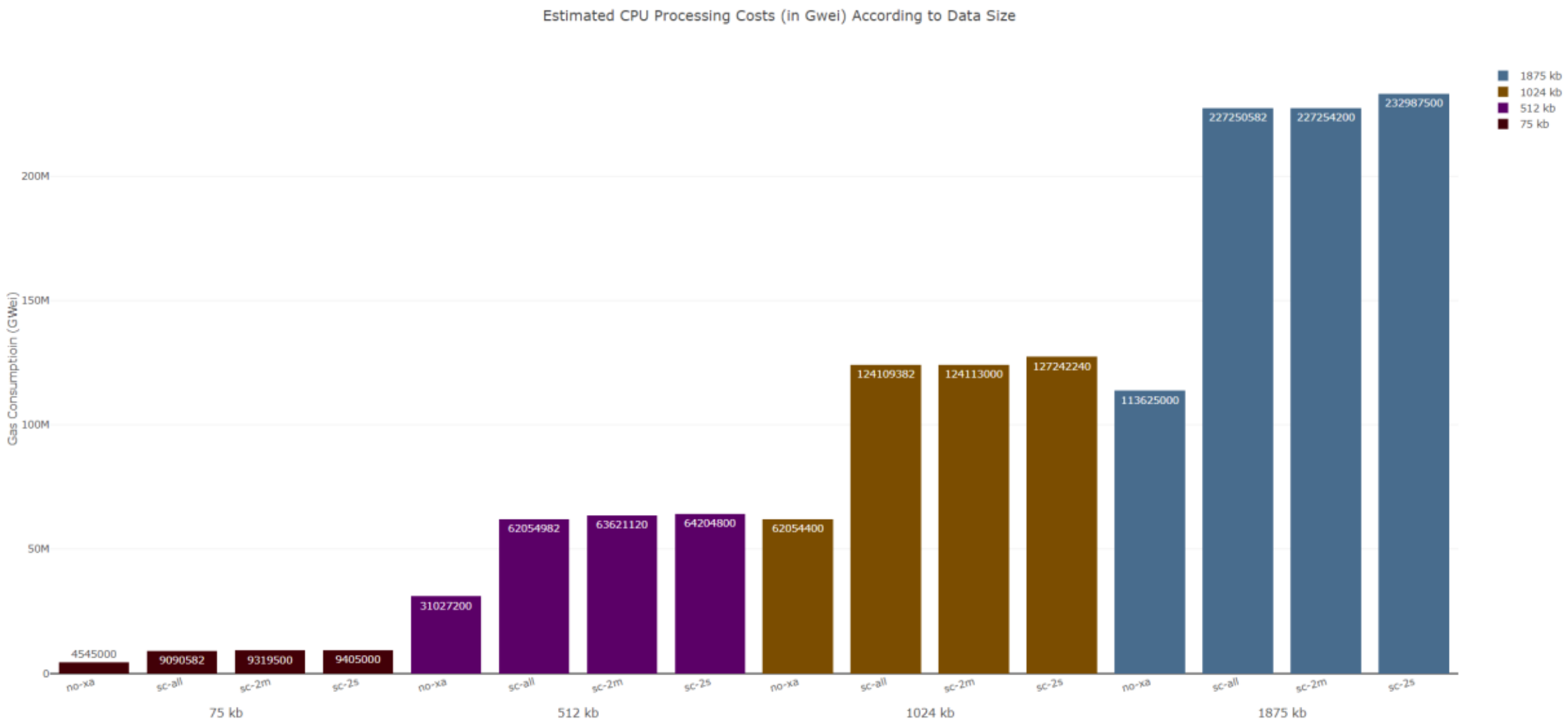

Fig. 6.9  Estimated CPU Processing Costs (in GWei) as A Function of Data Size (Adopted from (Liu et al., 2023b))

The figure and the table show that the cost is indeed directly proportional to the size of the ledger data read/written by each of the approaches. The cost of the transaction mechanism for this particular transaction doubles the cost in comparison to not having transactions. This is not surprising as the data that is read or written is stored in the cache first, and upon the transaction commit the ledger access is replayed from the cache to the ledger itself. Considering that the cache is also on the ledger, the access to the ledger in essence doubles and hence the cost, which is directly proportional to the leger reads and writes, also doubles as the table and the figure indicate.

All of the alternative mechanisms we described ensure the ACID properties of transactions as defined in Section 3.2 and also ensure access control. Although each one also supports privacy, the level of privacy protection differs as different levels of efforts are required on the part of the attacker to subvert privacy. We assume that the attacker has access to the blockchain and hence is able to read any object stored on the ledger as long as the attacker knows where on the ledger it is stored – thus privacy may be subverted for the case when all methods and the private workspace are all hosted in the same smart contract (case labelled as *sc-all*).

For the case labelled as *sc-2m*, when the transaction methods and the private workspace are hosted in a separate smart contract but also hosted on the mainchain, the same approach may be used to subvert the privacy as in the case labelled *sc-all*, except that it would have to be applied twice, once on the main smart contract and once on the smart contract that contains the transaction methods and the private workspace.

The case labelled *sc-2s*, when the transaction methods and the private workspace on hosted in the separate smart contract executed on a sidechain, is similar the case labelled *sc-2m*, except that the attacker also needs access to the sidechain where the smart contract is deployed in addition to having access to the mainchain. Thus, additional information and effort are required in comparison to the previous cases.

To further increase the level of privacy, cryptography may be used to ensure the ledger data written by the transaction is not viewable by the non-transaction actors. To measure the cost of encryption/decryption, the methods *m1* and *m2*, which read and write the ledger, are amended with the public-key cryptography used to encrypt/decrypt any ledger read/write. Thus, the cost of encryption/decryption is expected to be directly proportional to the size of the data that is encrypted/decrypted and thus being expensive, particularly for the code written in the Solidity language for the EVM (Ethereum Virtual Machine (EVM), 2023). Table 6.2 and Fig 6.10 show the cost of cryptography to encrypt/decrypt the written/read ledger data that has the size X = 75, 512, 1024, and 1875 KB. The table and the figure show the cost of the work performed when: (i) there is no transaction mechanism, labeled as *no-xa* in the first column of the table; (ii) the transactional mechanism is supported using sidechain processing for the sub-transactions, labeled as *sc-2s*; and (iii) the transactional mechanism is supported using sidechain processing for the sub-transactions and when the data stored in the private workspace/cache is encrypted/decrypted when writing or reading, labeled as *sc-2s-crypto*. Clearly, the cost of encryption/decryption is significant as it doubles the cost of the transactional mechanism when sidechain processing is used (case labeled *sc-2s-crypto*).

Table 6.2  CPU Processing Cost Estimates (in GWei) (Adopted from (Liu et al., 2023b))

| ***Label*** | ***7kb*** | ***512 kb*** | ***1024 kb*** | ***1875 kb*** |
|---|---|---|---|---|
| *no-xa* | 4545000 | 31027200 | 62054400 | 113625000 |
| *sc-2s* | 9405000 | 64204800 | 127242240 | 232987500 |
| *sc-2s-crypto* | 18268952 | 124706420 | 249411188 | 456684152 |

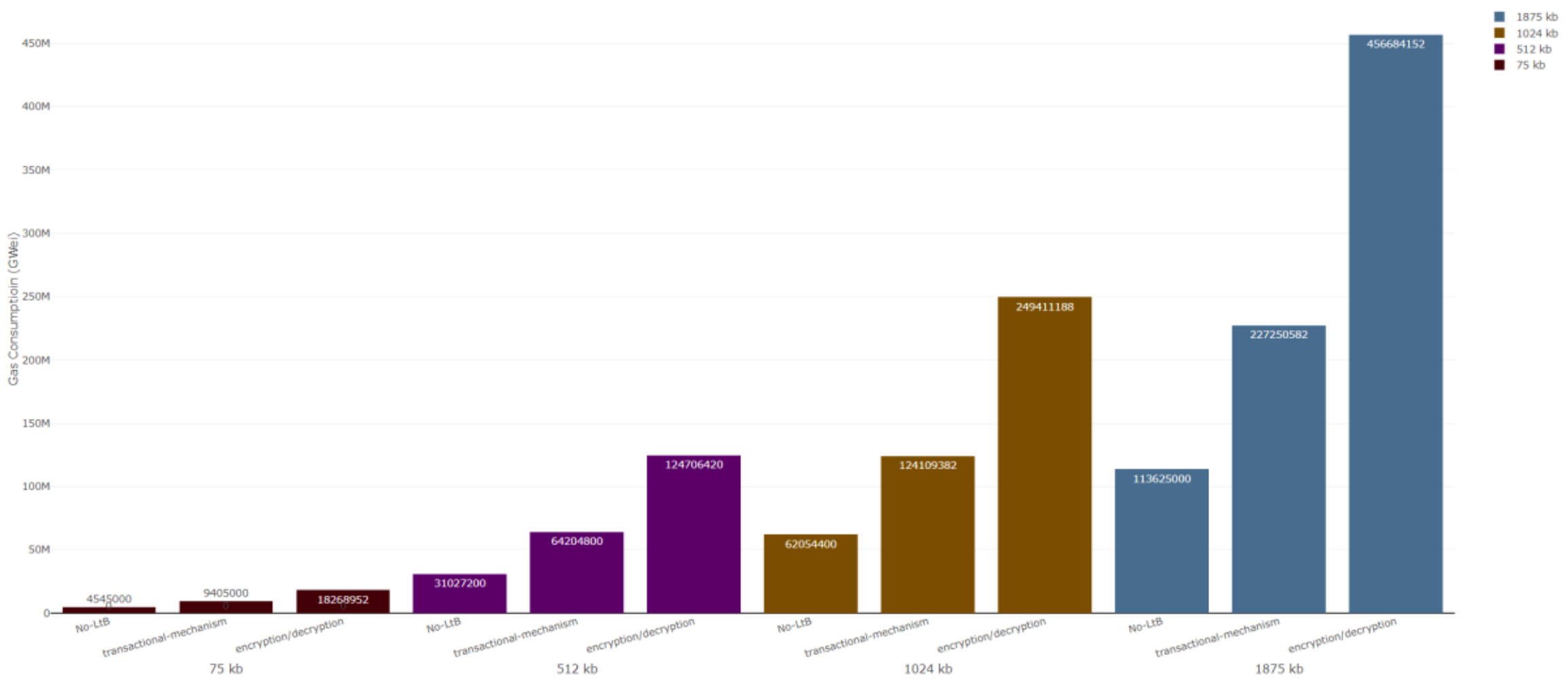

Fig. 6.10  Estimated CPU Processing Costs (in GWei) for Cases Labeled as *no-xa*, *sc-2s*, and *sc-2s-crypto* (Adopted from (Liu et al., 2023b))

### 6.3.2 Supporting Collaborative Transactions with Nesting

The cost estimates for transaction mechanism to support sub-transactions are expected to be high. In comparison to the cost of a transaction mechanism for a collaborative transaction that is not nested, for each nested transaction a Two-Phase Commit (2PC) is implemented. The cost is incurred by a parent transaction coordinator issuing the *get-ready-to commit* command, receiving answers from each of the participant stating readiness for commitment, and then the coordinator issuing the *commit-command* to each of the participants. Fig. 6.10 shows the total cost estimates for performing the 2PC protocol that consists of the coordinator issuing the *Prepare-to-commit* command and receiving a response from each of the participants, and then issuing the *Commit-command* to each of the participants. It also includes the cost for each of the participants to receive the *Prepare-to-commit* command and send the *Ready-to-commit* responses to the coordinator and then receiving the *Commit-command* from the coordinator (the participant does not acknowledge the received *Commit-command* response). The cost of the 2PC-commit protocol coordination needs to be added to the rest of the cost of the transactional mechanism.

To estimate the cost of the 2PC protocol implementation as a part of the mechanism to support the nested transaction, we thoroughly evaluated the cost of each step of the protocol, aligning it with the workflow displayed as a sequence diagram in Fig. 5.5. To gauge the cost of our 2-phase commitment protocol, we employed Remix for interacting with the actual functions of deployed smart contracts. It's noteworthy that the gas price was set at 20 GWei, identical to the default gas price of Ganache. We documented the costs of both phases when the number of participants involved in the 2-phase commit varies from 1 to 5, with the results presented in Table 6.3 for each of the phases of the 2PC protocol. Fig. 6.10 shows the total cost of the 2PC, which is the sum of the costs for the two phases, as a function of the number of participants. As expected, the cost is in a form of a linear function of the number of participants. Furthermore, the cost of a mechanism to support sub-transactions is high due to the 2PC protocol costs. The cost estimates shown in Table 6.3 and in Fig. 6.11 need to be added to the cost of the transactional mechanism when nesting is not used, and when they are added, then the cost of the mechanism is more than doubled when there are two nested sub-transactions and increases linearly with an increase to the number of participants.

Table 6.3 Costs Estimates for The 2-Phase Commit Protocol (in GWei) (Adopted from (Liu et al., 2023b))

| *# of participants* | *Phase 1* | *Phase 2* |
|---|---|---|
| *2* | 627820 | 627400 |
| *3* | 671500 | 671080 |
| *4* | 715180 | 714760 |
| *5* | 758860 | 758440 |
| *6* | 802540 | 802120 |

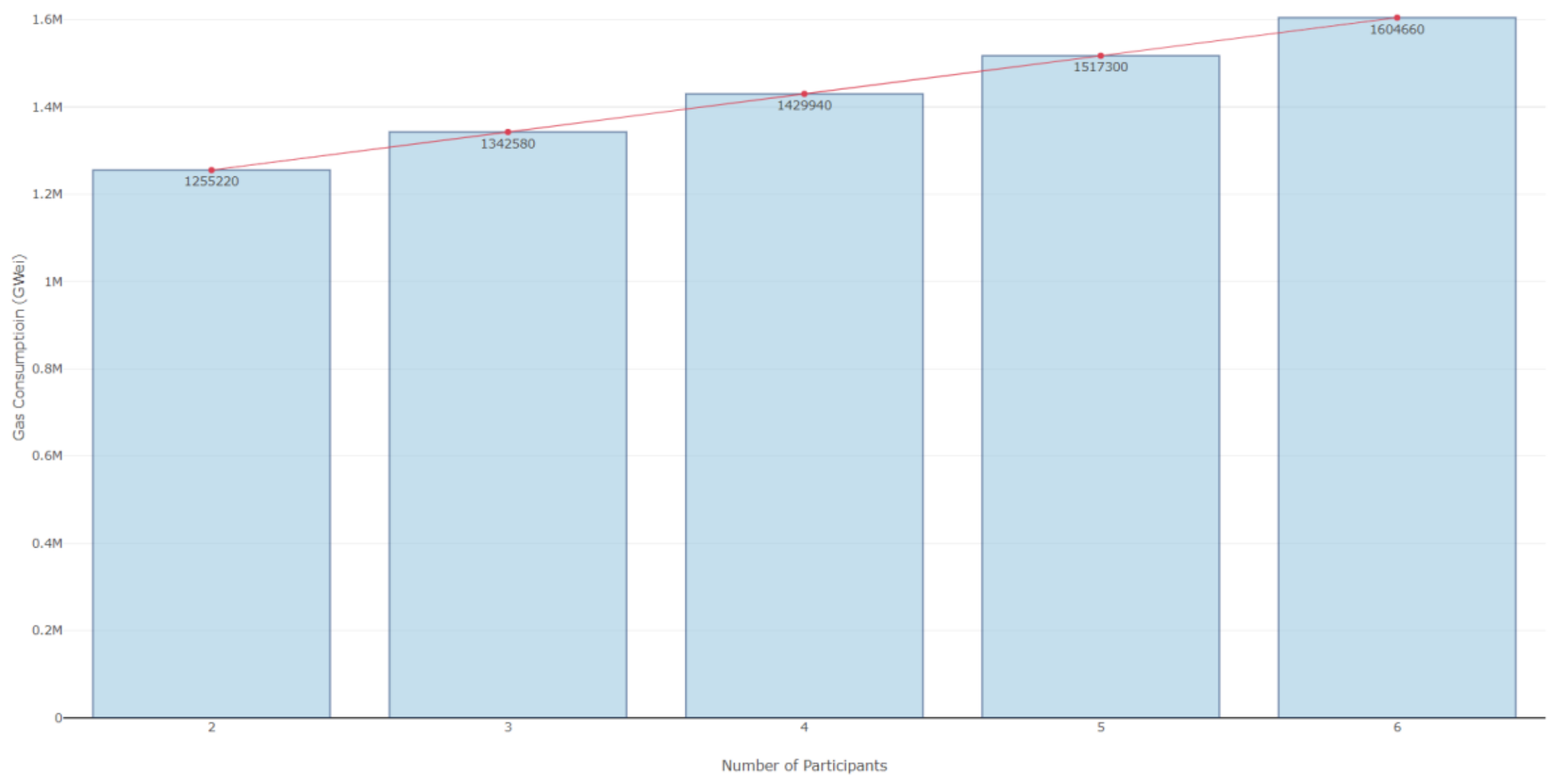

Fig. 6.11 Gas Utilization by The 2-Phase Commit Protocol for Nested Transactions (Adopted from (Liu et al., 2023b)).

## 6.4 Performance Evaluation of Experimental Applications

In the previous section, we examined the overhead cost associated with supporting long-term transactions using multi-method smart contracts. However, as summarized, the overall cost is anticipated to be significantly reduced by leveraging the hierarchical structure of the DE-HSM model and localizing the computations of long-term transactions and their nested transactions on sidechains. Next, we will conduct an overall Performance Evaluation of Experimental Applications by offloading the computations of components of long-term transactions to offline processing. By engaging our proposed methodology, which takes advantage of the concurrent computation capabilities of the DE-HSM multi-modal model, we expect to achieve reductions in both overall cost and latency. In this subsection, we discuss transformation correctness, latency, and cost.

### 6.4.1 Latency and Cost

We now report on evaluation of the three use cases in terms of latency and cost, with a focus on the performance and cost-effectiveness of using sidechains for off-chain processing. The cost is derived for Ethereum using web3.js (web3.js, 2022), which estimates the cost of each instruction in a smart contract in Wei. We provide a comprehensive analysis of observed values for both Ethereum and Hyperledger Fabric, and further discuss the benefits and implications of using sidechains in different scenarios.

### *6.4.1.1 Latency Analysis*

The TABS tool offers various latency measurement capabilities, including the ability to measure latency for individual FSM (Finite State Machine) state transitions, the execution of a sub-model or a smart contract method, and the overall latency for the entire dApp (Decentralized Application). We observed latencies for each use case execution, conducted on both the Ethereum and Hyperledger Fabric platforms. The comparative results are visually represented in Figures 6.12 and 6.13, respectively. Ethereum experiences significantly higher delays than Hyperledger Fabric, primarily due to its more complex consensus algorithm. These delays can impact the overall performance of the use cases when implemented on both platforms. The analysis highlights the importance of considering latency when selecting a blockchain platform for a specific use case, as it can influence the user experience and the efficiency of the overall system.

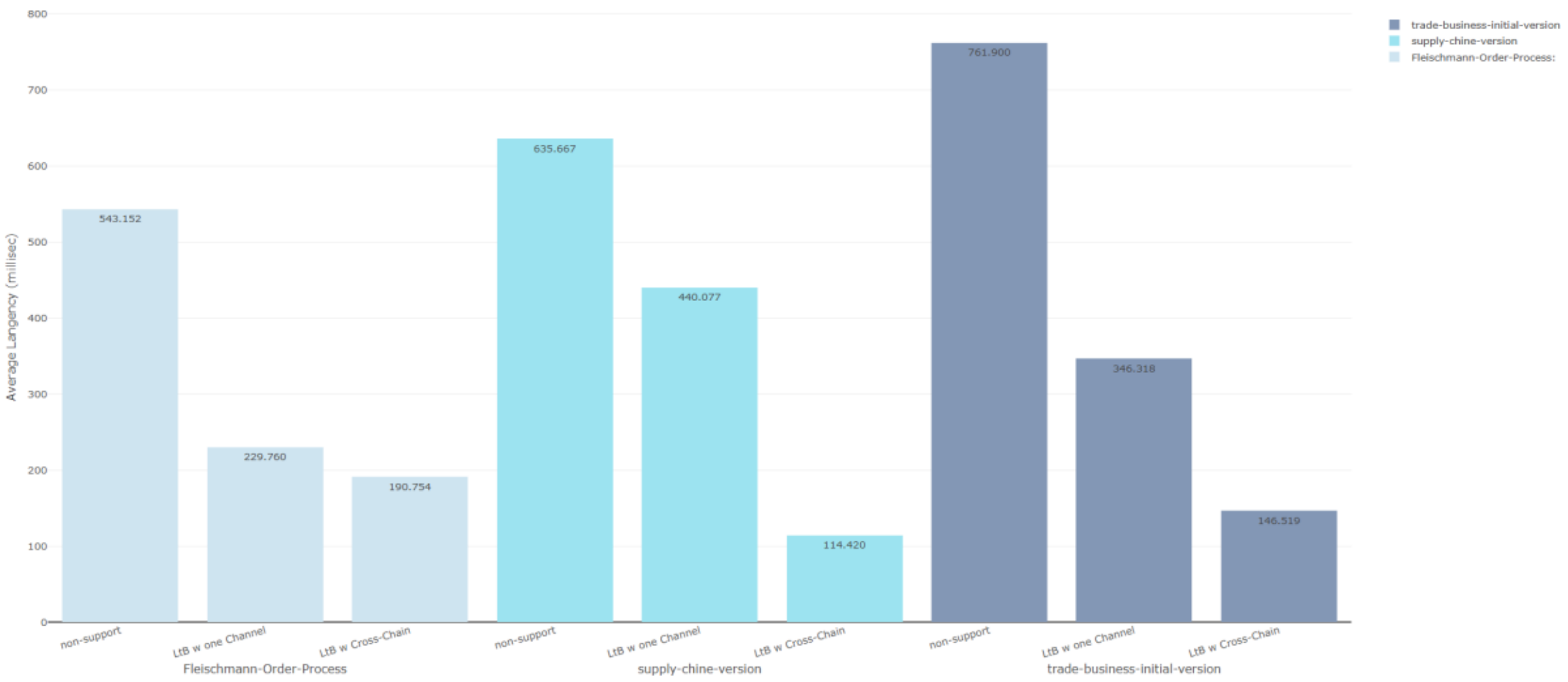

Fig. 6.12 Latency for Each of The Three Experimental Applications for Ethereum

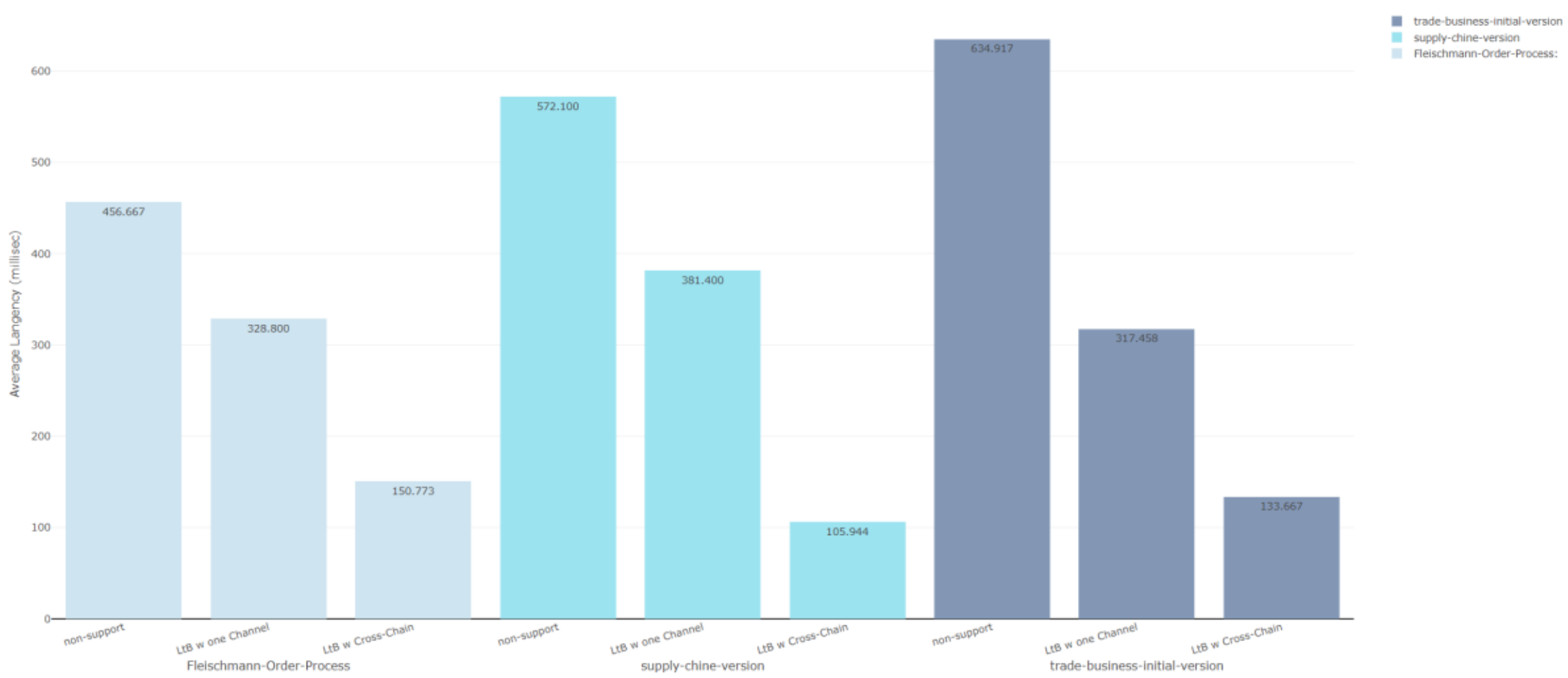

Fig. 6.13 Latency for Each of The Three Experimental Applications for Hyperledger Fabric

#### *6.4.1.2 Cost Analysis*

TABS enables the automatic estimation of costs per method and the overall cost by utilizing web3.js, a popular JavaScript library for interacting with Ethereum blockchain. We analyze the cost in GWei (a unit of Ether) for each of the three use cases and provide a breakdown of the cost in USD for one application workflow of each use case on public Ethereum.

Table 6.4 Cost of Processing for Ethereum Environment (Adopted from (Bodorik et al., 2023))

| | Gas (GWei) | Cost (USD) |
|---|---|---|
| Supply Chain | 95,854 | 0.2615 |
| Order Process | 57,512 | 0.1569 |
| Trade Business | 182,195 | 0.4945 |

It is important to note that Hyperledger Fabric, being a private blockchain, incurs costs primarily associated with hardware and network expenses. These expenses may vary depending on the specific infrastructure setup and are generally lower than those associated with public blockchains like Ethereum.

Supporting LtB (Long-term Blockchain Transaction) transactions and enforcing their properties may introduce overhead costs. However, moving computations from the main public blockchain network to private sidechains can dramatically reduce the cost for overall transactions executed by multi-chain smart contracts. As our experimental results show in the subsection " Sidechain Processing", the three applications almost halve the cost for each long-term transaction by leveraging sidechains. We will explain more details about the cost reduction with multi-chain smart contracts support in the later section.

### 6.4.2 Correctness (Non-conforming vs Conforming Traces)

In the process of transforming BPMN models into smart contracts, a critical aspect is ensuring the correctness of the resulting smart contract by accurately recognizing conforming and non-conforming traces. Conforming traces represent valid inputs from the decentralized application (dApp) in the context of smart contract execution, provided at the appropriate stages. On the other hand, non-conforming traces are incorrect inputs that the smart contract should be able to identify as such.

Our smart contract implementation has undergone thorough testing, including evaluations with both conforming and non-conforming input sequences or traces, resulting in a 100% accuracy in detecting both conformant and non-conformant inputs. This high level of accuracy can be attributed to the DE-HSM modeling process, which transforms the original BPMN model into a DE-HSM model. The DE-HSM model consists of multiple DE-FSM sub-models, each of which corresponds directly to the original BPMN diagram when represented as a directed acyclic graph (DAG).

The DE-HSM modeling process ensures that the smart contract accurately reflects the intended business process as defined in the BPMN model. This, in turn, allows the smart contract to effectively differentiate between conforming and non-conforming traces, maintaining the integrity and correctness of the smart contract execution. By achieving this level of accuracy, developers can have confidence in the reliability of the smart contracts generated through DE-HSM modeling, as they can correctly process and handle both conforming and non-conforming inputs from DApps.

### 6.4.3 Sidechain Processing

A key feature of our approach is the ability to move blockchain patterns, represented as individual DE-HSM (Discrete Event Hierarchical State Machine) sub-models, to sidechains for cost reduction, privacy, or a combination of both. This flexibility allows for more efficient and tailored solutions to specific use case requirements.

#### *6.4.3.1 Sidechains for Cost Reduction*

Using sidechains for cost reduction is only beneficial when sidechain processing is cheaper than mainchain processing. This is crucial for any smart contract targeted for deployment on a blockchain, as they are typically more expensive than centralized systems. For example, processing on the public Ethereum blockchain is costlier than on the Quorum chain, making sidechain processing an attractive option for reducing expenses. For Hyperledger fabric, using channels as sidechains for cost reduction is not practical, as the mainchain is just one of many channels. However, a different, cheaper blockchain can be used as a sidechain in conjunction with Hyperledger fabric as the mainchain.

We calculated the cost of processing one execution of a smart contract for each use-case and compared the cost of processing on the Ethereum mainchain-only and then when a Quorum sidechain is used with the mainchain. We selected all sub-models to be processed on a sidechain for each use-case, as identified by colored or dashed rectangles in figures 4.6, 6.2, 6.5, and 6.6. Fig. 6.14 and Fig. 6.15 demonstrates that due to the lower cost of processing on Quorum, off-chain processing is highly advantageous.

Furthermore, we assessed the latency implications of processing on the Ethereum mainchain compared to the combined processing on the Ethereum mainchain and a Quorum sidechain for each use-case. Latency is reduced by sidechain processing. Similarly, we used sidechain processing for Hyperledger fabric by deploying the main smart contract on one Hyperledger channel and a sidechain on another channel of the same Hyperledger fabric blockchain. As expected, latency with sidechain processing is higher than just processing on the mainchain only, as shown in Fig. 6.16. This is due to the overhead needed to facilitate sidechain processing. However, sidechain processing on a Hyperledger is achieved by using another Hyperledger channel, which incurs the same cost as processing on the mainchain, which is also a Hyperledger channel.

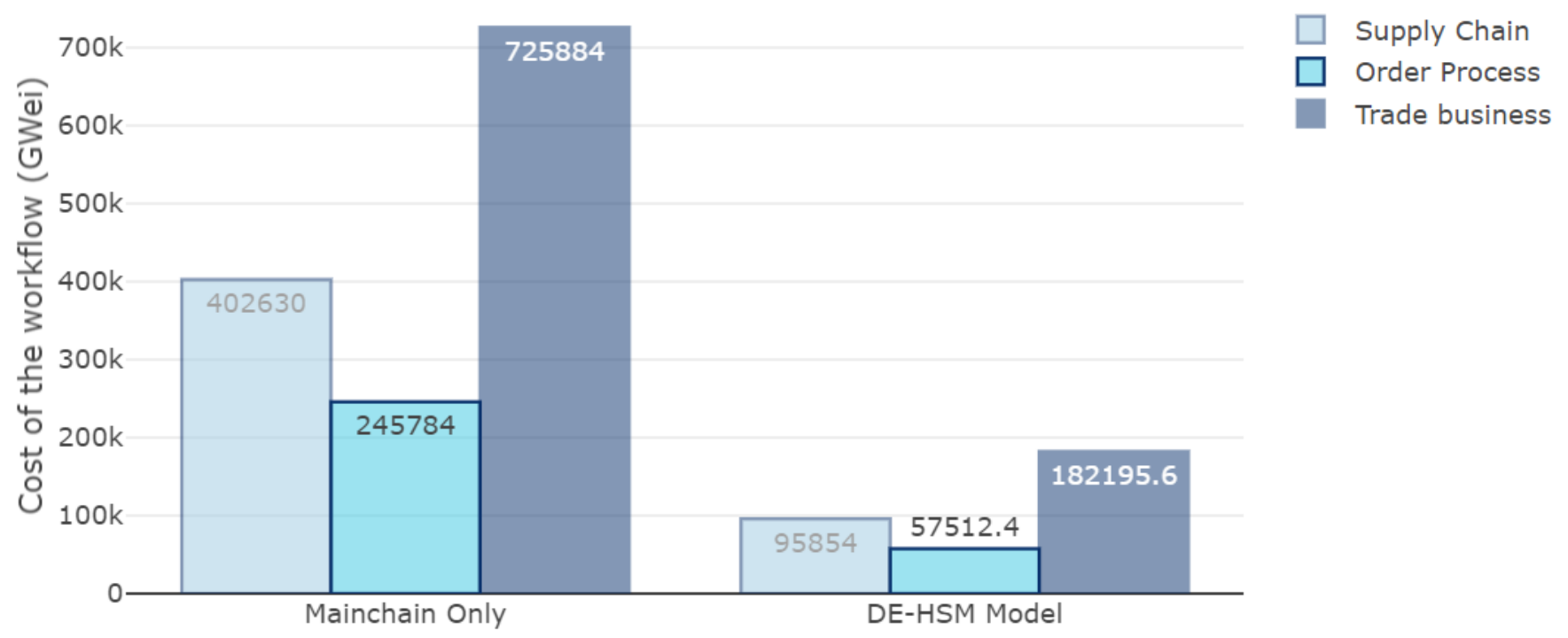

Fig. 6.14 Ethereum Cost of Processing on Mainchain Only vs Processing on Mainchain with Sub-Models in A Sidechain (adopted from (Bodorik et al., 2023)).

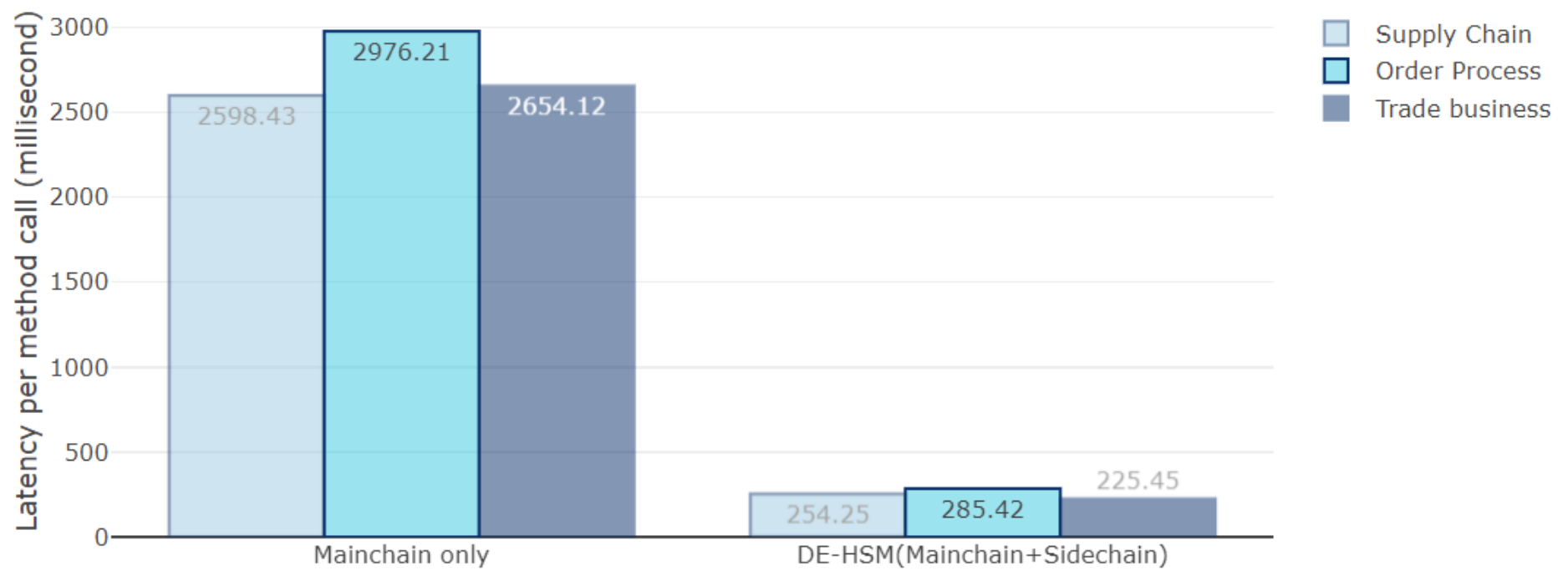

Fig. 6.15 Ethereum Latency of Processing on mainchain vs Sub-Models on Quorum Sidechain (adopted from (Bodorik et al., 2023)).

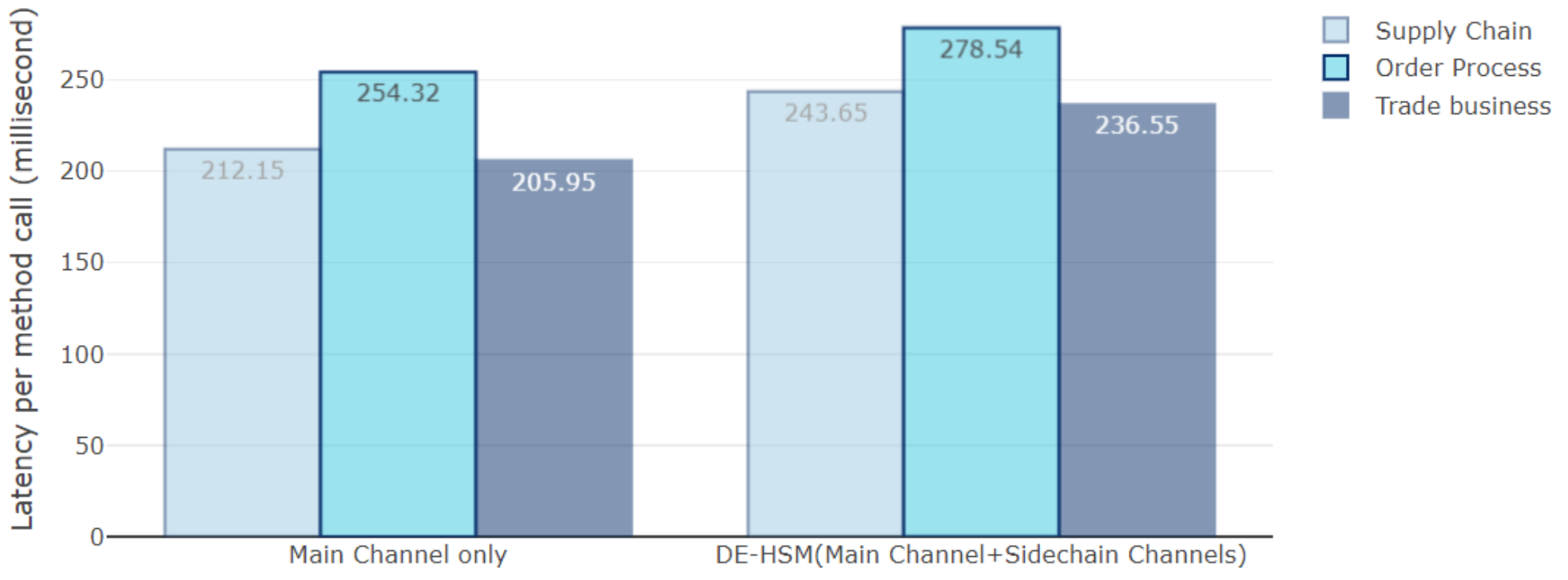

Fig. 6.16 Hyperledger Latency of Processing of All on One Channel vs Processing Sub-Models on Another Channel (adopted from (Bodorik et al., 2023)).

#### *6.4.3.2 Sidechains for Privacy*

In addition to cost reduction, sidechains can also be used for privacy. In more complex smart contracts, such as the trade use case, certain parts of the contract should not be visible to all participants of the smart contract. For instance, after the buyer and seller agree on the price, the price should not be visible to some of the other actors. By choosing to deploy and execute an independent subgraph pattern, i.e., a sub-model, on a sidechain, privacy is provided as only those actors who participate in the sidechain sub-model execution can observe all data pertaining to that sidechain processing. Meanwhile, other actors do not have access to that data. However, all actors have access to the results of computation that is stored on the mainchain. It is important to note that the above statements are subject to the privacy features inherent to specific blockchains. For instance, certain blockchains, including Hyperledger Fabric, facilitate the utilization of private data within a channel, which is not visible to other participants

In conclusion, our use-case evaluations, from both latency and cost perspectives, underscore the merits of sidechain utilization for both cost efficiency and privacy enhancement. The comparative analysis between mainchain-only and sidechain-augmented processing elucidates the context-specific benefits of sidechain processing, contingent on the chosen blockchain platform and the dynamics of the use-case.

# CHAPTER 7 SUMMARY, CONCLUDING REMARKS, AND FUTURE RESEARCH

In the subsequent subsections, we offer a succinct synthesis of the salient accomplishments and academic contributions realized over the course of this research endeavor. We elucidate the core tenets of our discoveries, accentuating the seminal aspects that underscore the import and ramifications of this thesis. Additionally, we proffer a prospective perspective, delineating the envisaged path of our forthcoming research pursuits, with the aspiration that the groundwork established herein perpetuates as a beacon for ensuing scholarly investigations.

## 7.1 MILESTONES AND SIGNIFICANCE

Reflecting on this research journey that commenced in early 2019, we have achieved notable milestones and made substantial contributions to the realms of blockchain technology and smart contract development. Our efforts have culminated in several publications and patent submissions, underscoring the practical relevance and innovative essence of our endeavors. Here, I will provide a concise recap of our primary achievements, their timelines, and references to relevant publications as discussed in previous sections. These scholarly contributions are categorized based on their phases and topics in subsections 7.1.1 and 7.1.2.

### 7.1.1 Master's Studies Contributions:

We introduced an algorithm for the FSM model to pinpoint off-chain computation patterns. Additionally, we developed a tool that automatically detects these patterns in the FSM model, converting them into off-chain executable smart contracts. The publications are listed in Table 7.1.

Table 7.1 Off-chain Computation Contributions

| TITLE | AUTHORSHIP | PUBLICATION VENUE | PUBLISHED DATE, DOI |
|---|---|---|---|
| **Using FSMs to Find Patterns for Off-Chain Computing: Finding Patterns for Off-Chain Computing with FSMs** | **Bodorik, P., Liu, C., Jutla, D.** | **2021 The 3rd ACM International Conference on Blockchain Technology (ICBCT '21)** | 2021-03-26 (https://doi.org/10.1145/3460537.3460565) |
| **A Tool for Moving Blockchain Computations Off-Chain** | **Liu, C., Bodorik, P., Jutla, D.** | **2021 3rd ACM International Symposium on Blockchain and Secure Critical Infrastructure (BSCI'21)** | 2021-06-03 (https://doi.org/10.1145/3457337.3457848) |

### 7.1.2 Doctoral Studies Contributions:

We formulated a multi-modal model to enhance blockchain transaction efficiency and privacy. This method supports blockchain trade attributes and was tested with TABS tools. Our findings were presented at ACM ICBTA and recommended for JAIT journal publication. Relevant publications are listed in Table 7.2.

Table 7.2 DE-HSM Multi-modal Model Contributions

| TITLE | AUTHORSHIP | PUBLICATION VENUE | PUBLISHED DATE, DOI |
|---|---|---|---|
| **Automating Smart Contract Generation on Blockchains Using Multi-Modal Modeling** | **Liu, C., Bodorik, P., Jutla, D.** | **2021 ACM Fourth International Conference on Blockchain Technology and Applications (ICBTA'21); Journal of Advances in Information Technology (JAIT)** | 2021-07-10 (https://doi.org/10.12720/jait.13.3.213-223) |

We started by using BPMN as the base model for blockchain applications and then transformed it into DE-HSM to tackle blockchain transactional issues. Our results were presented at the ICEET conference. Further, we detailed our methods for converting BPMN models into smart contracts and submitted this extended research to the Elsevier Journal BCRA. Relevant publications are listed in Table 7.3.

Table 7.3 BPMN to Smart Contracts Contributions

| TITLE | AUTHORSHIP | PUBLICATION VENUE | PUBLISHED DATE, DOI |
|---|---|---|---|
| **From BPMN to Smart Contracts on Blockchains: Transforming BPMN to DE-FSM Multi-Modal Model** | **Liu, C., Bodorik, P., Jutla, D.** | **2021 IEEE International Conference on Engineering and Emerging Technologies (ICEET'21)** | 2021-09-30 (https://doi.org/10.1109/ICEET53442.2021.9659771) |
| **TABS: Transforming Automatically BPMN Models into Blockchain Smart Contracts** | **Bodorik, P., Liu, C., Jutla, D.** | **Elsevier Blockchain: Research and Applications** | 2023-02-23 (https://doi.org/10.1016/j.bcra.2022.100115) |

We introduced the Long-term Blockchain Transactions (LtB) concept for multi-chain settings. We compared long-term and blockchain transactions, proposed a way to represent long-term transactions in smart contracts, and evaluated its efficiency. We further detailed how our TABS methodology supports BPMN collaborative transactions involving multiple participants and how it identifies patterns suitable for such transactions. Our approach also includes an auto-generated transaction mechanism, showcased by our TABS+ tool, with an overview of associated costs. The associated publications are listed in Table 7.4.

Table 7.4 Long-term Blockchain Transactions Contributions

| TITLE | AUTHORSHIP | PUBLICATION VENUE | PUBLISHED DATE, DOI |
|---|---|---|---|
| **Supporting Long-Term Transactions in Smart Contracts** | **Liu, C., Bodorik, P., Jutla, D.** | **2022 IEEE Fourth Blockchain Computing and Applications (BCCA'22)** | 2022-10-14 (https://doi.org/10.1109/BCCA55292.2022.9922193) |
| **Long-Term Blockchain Transactions Spanning Multiplicity of Smart Contract Methods** | **Liu, C., Bodorik, P., Jutla, D.** | **Springer BlockSys: Blockchain and Trustworthy Systems (BlockSys'23)** | 2023-11-29 (https://link.springer.com/download/epub/10.1007/978-981-99-8101-4.epub) |
| **Nested Blockchain Transaction for Multiple Methods of A Smart Contract** | **Liu, C., Bodorik, P., Jutla, D.** | **ACM Distributed Ledger Technologies (DLT)** | [Manuscript submitted for publication] |
| **Tabs+: Transforming Automatically BPMN Models to Smart Contracts With Nested Collaborative Transactions** | **Liu, C., Bodorik, P., Jutla, D.** | **ACM Distributed Ledger Technologies (DLT)** | [Manuscript submitted for publication] |

We have submitted to the United States Patent and Trademark Office (USPTO) our proposed system and method for the automated generation of smart contracts from Business Process Model and Notation (BPMN) models using Discrete Events Hierarchical State Machines (DE-HSM) multi-modal models, intended for execution by smart contracts within a multi-chain environment. The details are provided in Table 7.5.

Table 7.5 Patent Submissions

| ORGANIZATION | AUTHORSHIP | U.S. PATENT APPLICATION NO. | SUBMITTED DATE |
|---|---|---|---|
| **U.S. Patent and Trademark Office** | **Bodorik, P., Liu, C., Jutla, D.** | 17,968,047 | 2022-10-10 |

Our research has predominantly revolved around the transformation of Business Process Model and Notation (BPMN) into smart contract methods, with a spotlight on managing collaborative transactions. We have introduced algorithms and modeling techniques to pinpoint Single-Entry Single-Exit (SESE) subgraphs within the Deterministic Finite State Machine (DE-FSM) model, which are then transformed back into the BPMN model representation. These SESE subgraphs are selected by the developer to be transformed into transactions that satisfy the ACID properties.

Moreover, we have tackled the challenges of scalability, isolation, and privacy issues associated with blockchain transactions. The research has also proposed methods for supporting blockchain trade transactional properties via multi-modal modeling. The developed TABS tools have been used to test the proposed approach.

## 7.2 Concluding Remarks

The thesis presented herein contributes to the research on the generation of smart contracts from a Business Process Model and Notation (BPMN) model. In alignment with our published and ongoing research work (Liu, 2021; Bodorik et al., 2021; Liu et al., 2021a, 2021b, 2022a, 2022b; Bodorik et al., 2023; Liu et al., 2023a, 2023b), the interaction between the modeler and the system is minimized. Upon inputting the BPMN model, the modeler merely guides the tool in the generation of the smart contract methods. The modeler aids the tool in the transformation process by providing necessary information, such as the code for template methods implementing BPMN task elements or information on which parts of the smart contract should be deployed on a sidechain. Besides, the modeler/developer can utilize the tool to explore execution properties of the methods of the smart contract. At present, our tool facilitates the creation and deployment of contracts for either Ethereum or Hyperledger blockchains.

The BPMN standard defines a transaction on a sub-process, in which case the whole subprocess must complete or any activity of the subprocess must be undone by a compensating transaction if the subprocess is unable to complete successfully. As was already discussed before, a blockchain also has a transaction that is defined as any set of ledger updates made by an execution of a single smart contract method. However, a trade transaction is long-term and may include many activities, activities performed by a subset of the application participants on a subset of data used by the application. And these three different concepts do not align. How to support the long-term trade transactions, which span a number of calls to the smart contract methods, is still an open question that we are currently addressing. There are also some of the BPMN elements, such as error, escalation, and multiple parallel attributes that our TABS system does not support, but we are working on their incorporation into the system.

Our research is marked by several distinctive features that warrant specific emphasis:

- We conduct an analysis of the BPMN diagram and utilize Discrete Event Hierarchical State Machine (DE-HSM) modeling to identify patterns, referred to as LSI independent subgraphs. These patterns are localized, meaning that once the execution of the pattern commences, it remains local to the pattern until there is an exit from that pattern. Each pattern has only one “entry” and one “exit” point, making them naturally suitable for decomposition purposes that our approach exploits. We use such patterns to form the sub-models, wherein the functionality of each sub-model is represented by a Finite State Machine (FSM) or concurrent FSMs. The choreography of processes, that is, the workflow determining which sub-models are executed in which sequence, is guided by the interconnection amongst the sub-models, while the functionality of each sub-model is represented by an FSM, or possibly a number of concurrent FSMs. The interconnection of the sub-models is represented by the DE-HSM model. Consequently, as the DE-HSM is a model that is an equivalent representation of a BPMN model, the correctness of representing the BPMN model using the DE-HSM model is assured.

- We facilitate the transformation of sub-models, selected by the modeler, into a separate contract that is deployed and executed on a sidechain, while the main smart contract interacts with the dApp and invokes these sidechain methods as appropriate. Sidechain processing may be chosen by a modeler to reduce the cost of processing if sidechain processing is cheaper than processing on a mainchain (e.g., processing on a Quorum sidechain in conjunction with the main contract being deployed on the public Ethereum blockchain); or to support privacy by processing a pattern on a sidechain.
- We have established an automated system for managing nested collaborative transactions via the enhanced TABS, known as the TABS+ approach. A collaborative transaction is a long-term process involving multiple participants. We demonstrated how a BPMN model can be scrutinized to identify suitable BPMN patterns for collaborative transactions, ensuring they possess desirable ACID, access control, and privacy properties. We also outlined how a transaction mechanism that enforces these properties can be constructed using pattern augmentation techniques. The developer is equipped with information about suitable BPMN patterns for defining collaborative transactions and the available transaction mechanisms for their support. The developer then selects which of the identified BPMN patterns should be treated as collaborative transactions and which transaction mechanism should be used. The TABS+ approach and tool automatically generate the smart contract(s) that include the collaborative transactions and the supporting transaction mechanism.

Of course, the primary objective is to relieve the developer of as much responsibility as possible in creating the code for smart contract methods. Our approach, in comparison to other methods for generating smart contracts from BPMN models, is unique in its automated support for:

- *Sidechain Processing*: The system provides a list of patterns suitable for sidechain processing, and the developer simply selects which patterns should be deployed and executed on a sidechain.
- *Collaborative Transactions*: Similarly, the developer selects which patterns, from a system-provided list of patterns suitable for collaborative transactions, should be deployed and executed as such. Additionally, the developer can choose which available transactional mechanism should be used to support these collaborative transactions.
- *Nested Collaborative Transactions*: Our system also supports the nesting of collaborative transactions.

One of the main advantages of our approach is the minimization of the developer's responsibilities. Essentially, the developer does not need to concern themselves with determining which methods should form transactions or how to create a transaction mechanism. In fact, our approach circumvents the issue of BPMN's lack of a way to represent a transaction that involves collaboration of actors. The system analyzes the BPMN model and provides information on which BPMN patterns are suitable as collaborative transactions.

Another underlying benefit due to our approach is that it is semi-agnostic to the blockchain. As long as the monitor part of a smart contract is written for a specific blockchain, any smart contract developed using our approach can be deployed on that blockchain. This is because the monitor part of the smart contract is independent of the smart contracts, given that the smart contract is expressed in terms of the DE-HSM model, i.e., in the interconnection of the DE-FSM models which work on any blockchain for which the smart contract monitor is implemented. Of course, there remains the issue of scripting the individual tasks that need to be performed for a specific blockchain. The next subsection on future work describes how this can be avoided.

## 7.3 Future Research Works

While this research has made notable progress in automating the generation of smart contracts from BPMN models, there are several areas of future research that we aim to explore:

*Blockchain-Agnostic Approach*: Our goal is to make our approach blockchain-agnostic, meaning a smart contract developed using our approach should be deployable and executable on any blockchain. Currently, our collaboration model is blockchain-independent, but to deploy and execute a smart contract developed for one blockchain on another, developers need to provide specific scripts for the task elements. We plan to use a two-layer approach, similar to the Plasma project, where task scripts are executed off-chain, and the smart contract simply guides collaborations and obtains certifications about the results of the tasks.

*Enhanced Privacy and Security*: We aim to bolster the privacy and security of our approach by automatically hardening the smart contract methods and ensuring that the security best practices for the creation of the smart contract methods are followed. We are adapting the approach of (Mavridou & Laszka, 2018), where a set of security patterns is added to the smart contract methods. Besides, the research will leverage the capabilities of Non-Fungible Tokens (NFTs) and Fungible Tokens (FTs) to facilitate transfers across different blockchains. This approach is anticipated to provide an additional layer of security and privacy.

*Dynamic BPMN Transformation*: In the long term, we plan to investigate ways to augment BPMN models with patterns or replace certain patterns with similar BPMN patterns. The objective is to create a repository of patterns that provide certain functionalities. For instance, a BPMN model can represent a supply chain management process. Different BPMN patterns might represent different activities needed for producing different components of the final product. If a suitable pattern exists in the repository, it can be added to the model instead of creating a new one from scratch.

*Productization of the Developer Tool*: We aim to productize the developer tool and make it publicly available for evaluation. Feedback will be collected to further improve the tool. Currently, a demonstration of the tool, as referenced in Chapter 6, is accessible online for a restricted period. Moreover, interested parties may obtain the source code for TABS+ by submitting a formal request to Chris.Liu@dal.ca.

In summary, this research has made notable contributions to the field of blockchain technology and smart contracts, providing new insights and methods for transforming BPMN models into smart contracts. The findings and contributions of this research have the potential to impact the development and implementation of blockchain applications in the future.

# APPENDIX: PUBLICATIONS

This section showcases the scholarly contributions linked to this project. Each entry highlights the authors' roles and offers direct access to the full papers. The works are divided based on their academic phase: Master's and Doctoral.

Author Contribution: For all publications, Chris Liu and Drs. Bodorik and Jutla contributed to the design, with implementation and experimentation conducted by Chris Liu.

## MASTER'S STUDIES CONTRIBUTIONS:

1. **Proceedings ACM ICBCT 2021: Bodorik, P., Liu, C., Jutla, D. Using FSMs to Find Patterns for Off-Chain Computing: Finding Patterns for Off-Chain Computing with FSMs**

In: Bodorik, P., Liu, C., Jutla, D. (2021). Using FSMs to Find Patterns for Off-Chain Computing: Finding Patterns for Off-Chain Computing with FSMs. In *2021 The 3rd International Conference on Blockchain Technology* (ICBCT '21), March 26-28, 2021. ACM, New York, NY, USA, 7 pages. https://doi.org/10.1145/3460537.3460565)

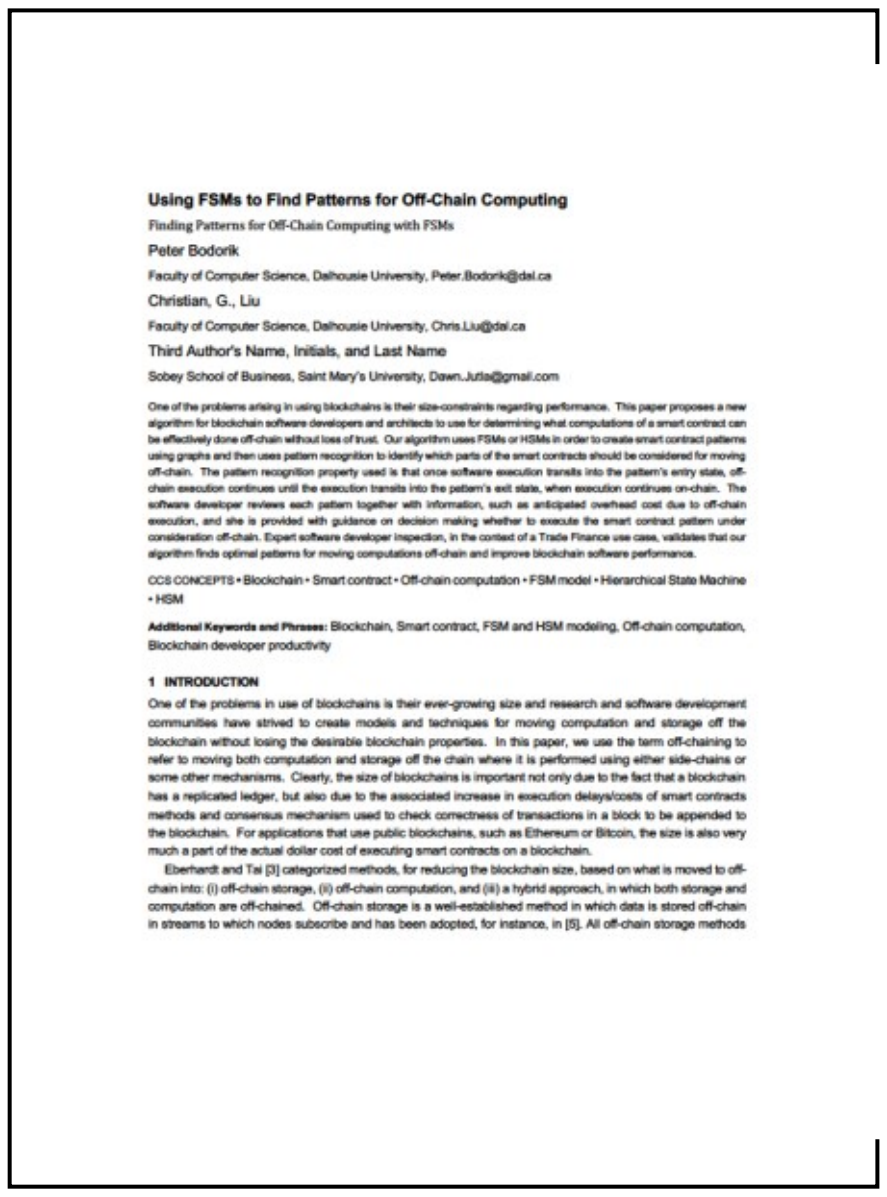

**Using FSMs to Find Patterns for Off-Chain Computing**

Finding Patterns for Off-Chain Computing with FSMs

Peter Bodorik

Faculty of Computer Science, Dalhousie University, Peter.Bodorik@dal.ca

Christian, G., Liu

Faculty of Computer Science, Dalhousie University, Chris.Liu@dal.ca

Third Author's Name, Initials, and Last Name

Sobey School of Business, Saint Mary's University, Dawn.Jutla@gmail.com

One of the problems arising in using blockchains is their size-constraints regarding performance. This paper proposes a new algorithm for blockchain software developers and architects to use for determining what computations of a smart contract can be effectively done off-chain without loss of trust. Our algorithm uses FSMs or HSMs in order to create smart contract patterns using graphs and then uses pattern recognition to identify which parts of the smart contracts should be considered for moving off-chain. The pattern recognition property used is that once software execution transits into the pattern's entry state, off-chain execution continues until the execution transits into the pattern's exit state, when execution continues on-chain. The software developer reviews each pattern together with information, such as anticipated overhead cost due to off-chain execution, and she is provided with guidance on decision making whether to execute the smart contract pattern under consideration off-chain. Expert software developer inspection, in the context of a Trade Finance use case, validates that our algorithm finds optimal patterns for moving computations off-chain and improve blockchain software performance.

**1 INTRODUCTION**

One of the problems in use of blockchains is their ever-growing size and research and software development communities have strived to create models and techniques for moving computation and storage off the blockchain without losing the desirable blockchain properties. In this paper, we use the term off-chaining to refer to moving both computation and storage off the chain where it is performed using either side-chains or some other mechanisms. Clearly, the size of blockchains is important not only due to the fact that a blockchain has a replicated ledger, but also due to the associated increase in execution delays/costs of smart contracts methods and consensus mechanism used to check correctness of transactions in a block to be appended to the blockchain. For applications that use public blockchains, such as Ethereum or Bitcoin, the size is also very much a part of the actual dollar cost of executing smart contracts on a blockchain.

Eberhardt and Tai [3] categorized methods, for reducing the blockchain size, based on what is moved to off-chain into: (i) off-chain storage, (ii) off-chain computation, and (iii) a hybrid approach, in which both storage and computation are off-chained. Off-chain storage is a well-established method in which data is stored off-chain in streams to which nodes subscribe and has been adopted, for instance, in [5]. All off-chain storage methods

2. **Proceedings ACM BSCI 2021: Liu, C., Bodorik, P., Jutla, D. A Tool for Moving Blockchain Computations Off-Chain**

In: Liu, C., Bodorik, P., Jutla, D. (2021). A Tool for Moving Blockchain Computations Off-Chain. In Proceedings of *2021 3rd ACM International Symposium on Blockchain and Secure Critical Infrastructure* (BSCI'21), June 3-7, 2021. ACM, New York, NY, USA, 8 pages. https://doi.org/10.1145/3457337.3457848)

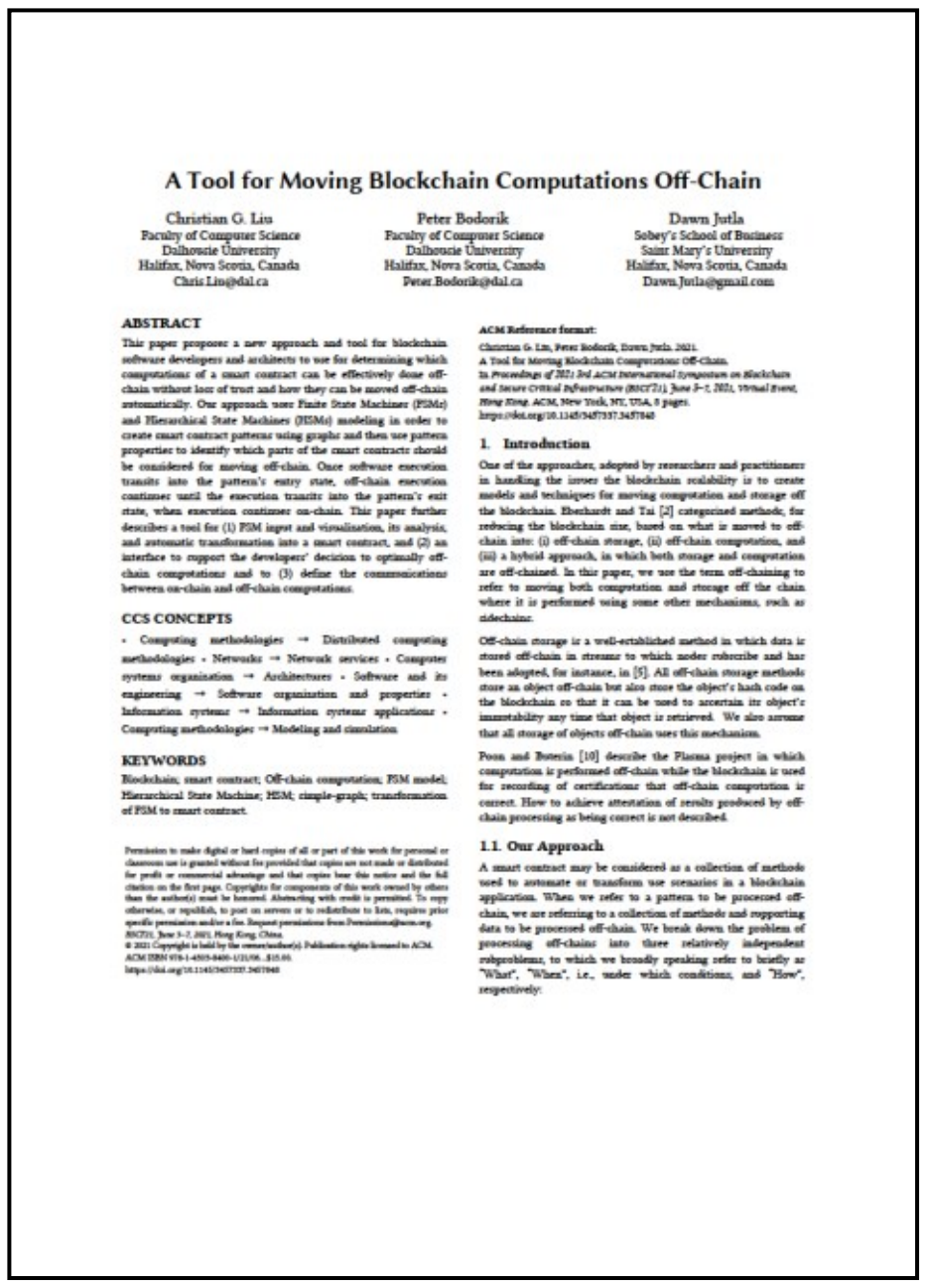

## A Tool for Moving Blockchain Computations Off-Chain

Christian G. Liu, Faculty of Computer Science, Dalhousie University, Halifax, Nova Scotia, Canada, Chris.Liu@dal.ca
Peter Bodorik, Faculty of Computer Science, Dalhousie University, Halifax, Nova Scotia, Canada, Peter.Bodorik@dal.ca
Dawn Jutla, Sobey's School of Business, Saint Mary's University, Halifax, Nova Scotia, Canada, Dawn.Jutla@gmail.com

**ABSTRACT**

This paper proposes a new approach and tool for blockchain software developers and architects to use for determining which computations of a smart contract can be effectively done off-chain without loss of trust and how they can be moved off-chain automatically. Our approach uses Finite State Machines (FSMs) and Hierarchical State Machines (HSMs) modeling in order to create smart contract patterns using graphs and then use pattern properties to identify which parts of the smart contracts should be considered for moving off-chain. Once software execution transits into the pattern's entry state, off-chain execution continues until the execution transits into the pattern's exit state, when execution continues on-chain. This paper further describes a tool for (1) FSM input and visualization, its analysis, and automatic transformation into a smart contract, and (2) an interface to support the developers' decision to optimally off-chain computations and to (3) define the communications between on-chain and off-chain computations.

**CCS CONCEPTS**

• Computing methodologies → Distributed computing methodologies • Networks → Network services • Computer systems organization → Architectures • Software and its engineering → Software organization and properties • Information systems → Information systems applications • Computing methodologies → Modeling and simulation

### 1. Introduction

One of the approaches, adopted by researchers and practitioners in handling the issues the blockchain scalability is to create models and techniques for moving computation and storage off the blockchain. Eberhardt and Tai [2] categorized methods, for reducing the blockchain size, based on what is moved to off-chain into: (i) off-chain storage, (ii) off-chain computation, and (iii) a hybrid approach, in which both storage and computation are off-chained. In this paper, we use the term off-chaining to refer to moving both computation and storage off the chain where it is performed using some other mechanisms, such as sidechains.

Off-chain storage is a well-established method in which data is stored off-chain in streams to which nodes subscribe and has been adopted, for instance, in [5]. All off-chain storage methods store an object off-chain but also store the object's hash code on the blockchain so that it can be used to ascertain its object's immutability any time that object is retrieved. We also assume that all storage of objects off-chain uses this mechanism.

Poon and Buterin [10] describe the Plasma project in which computation is performed off-chain while the blockchain is used for recording of certifications that off-chain computation is correct. How to achieve attestation of results produced by off-chain processing as being correct is not described.

#### 1.1. Our Approach

A smart contract may be considered as a collection of methods used to automate or transform use scenarios in a blockchain application. When we refer to a pattern to be processed off-chain, we are referring to a collection of methods and supporting data to be processed off-chain. We break down the problem of processing off-chains into three relatively independent subproblems, to which we broadly speaking refer to briefly as "What", "When", i.e., under which conditions, and "How", respectively:

## Doctoral Studies Contributions:

3. **Journal of Advances in Information Technology (JAIT) 2021: Liu, C., Bodorik, P., Jutla, D. Automating Smart Contract Generation on Blockchains Using Multi-Modal Modeling**

In: Liu, C., Bodorik, P., Jutla, D. (2021). Automating Smart Contract Generation on Blockchains Using Multi-modal Modeling. *Journal of Advances in Information Technology* (JAIT). 10.12720/Jait, 13, 213–223. https://doi.org/10.12720/jait.13.3.213-223

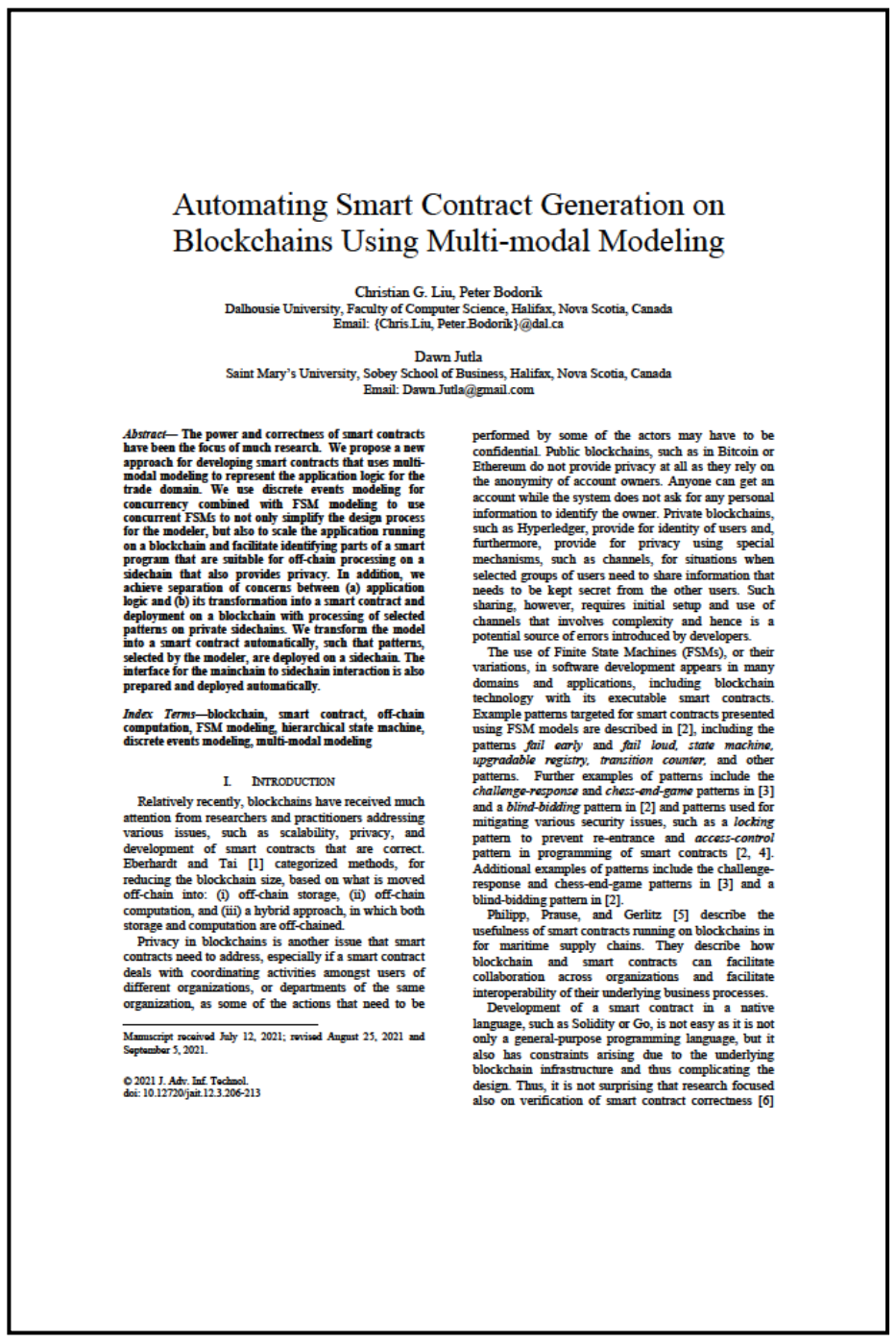

Automating Smart Contract Generation on Blockchains Using Multi-modal Modeling

Christian G. Liu, Peter Bodorik
Dalhousie University, Faculty of Computer Science, Halifax, Nova Scotia, Canada
Email: {Chris.Liu, Peter.Bodorik}@dal.ca

Dawn Jutla
Saint Mary's University, Sobey School of Business, Halifax, Nova Scotia, Canada
Email: Dawn.Jutla@gmail.com

***Abstract*— The power and correctness of smart contracts have been the focus of much research. We propose a new approach for developing smart contracts that uses multi-modal modeling to represent the application logic for the trade domain. We use discrete events modeling for concurrency combined with FSM modeling to use concurrent FSMs to not only simplify the design process for the modeler, but also to scale the application running on a blockchain and facilitate identifying parts of a smart program that are suitable for off-chain processing on a sidechain that also provides privacy. In addition, we achieve separation of concerns between (a) application logic and (b) its transformation into a smart contract and deployment on a blockchain with processing of selected patterns on private sidechains. We transform the model into a smart contract automatically, such that patterns, selected by the modeler, are deployed on a sidechain. The interface for the mainchain to sidechain interaction is also prepared and deployed automatically.**

I. INTRODUCTION

Relatively recently, blockchains have received much attention from researchers and practitioners addressing various issues, such as scalability, privacy, and development of smart contracts that are correct. Eberhardt and Tai [1] categorized methods, for reducing the blockchain size, based on what is moved off-chain into: (i) off-chain storage, (ii) off-chain computation, and (iii) a hybrid approach, in which both storage and computation are off-chained.

Privacy in blockchains is another issue that smart contracts need to address, especially if a smart contract deals with coordinating activities amongst users of different organizations, or departments of the same organization, as some of the actions that need to be performed by some of the actors may have to be confidential. Public blockchains, such as in Bitcoin or Ethereum do not provide privacy at all as they rely on the anonymity of account owners. Anyone can get an account while the system does not ask for any personal information to identify the owner. Private blockchains, such as Hyperledger, provide for identity of users and, furthermore, provide for privacy using special mechanisms, such as channels, for situations when selected groups of users need to share information that needs to be kept secret from the other users. Such sharing, however, requires initial setup and use of channels that involves complexity and hence is a potential source of errors introduced by developers.

The use of Finite State Machines (FSMs), or their variations, in software development appears in many domains and applications, including blockchain technology with its executable smart contracts. Example patterns targeted for smart contracts presented using FSM models are described in [2], including the patterns *fail early* and *fail loud, state machine, upgradable registry, transition counter,* and other patterns. Further examples of patterns include the *challenge-response* and *chess-end-game* patterns in [3] and a *blind-bidding* pattern in [2] and patterns used for mitigating various security issues, such as a *locking* pattern to prevent re-entrance and *access-control* pattern in programming of smart contracts [2, 4]. Additional examples of patterns include the challenge-response and chess-end-game patterns in [3] and a blind-bidding pattern in [2].

Philipp, Prause, and Gerlitz [5] describe the usefulness of smart contracts running on blockchains in for maritime supply chains. They describe how blockchain and smart contracts can facilitate collaboration across organizations and facilitate interoperability of their underlying business processes.

Development of a smart contract in a native language, such as Solidity or Go, is not easy as it is not only a general-purpose programming language, but it also has constraints arising due to the underlying blockchain infrastructure and thus complicating the design. Thus, it is not surprising that research focused also on verification of smart contract correctness [6]

## 4. Proceedings IEEE ICEET 2021: Liu, C., Bodorik, P., Jutla, D. From BPMN to Smart Contracts on Blockchains: Transforming BPMN to DE-FSM Multi-Modal Model

<u>In</u>: Liu, C., Bodorik, P., Jutla, D. (2021). From BPMN to smart contracts on blockchains: Transforming BPMN to DE-HSM multi-modal model. *2021 IEEE International Conference on Engineering and Emerging Technologies* (ICEET), 1–7. https://doi.org/10.1109/ICEET53442.2021.9659771

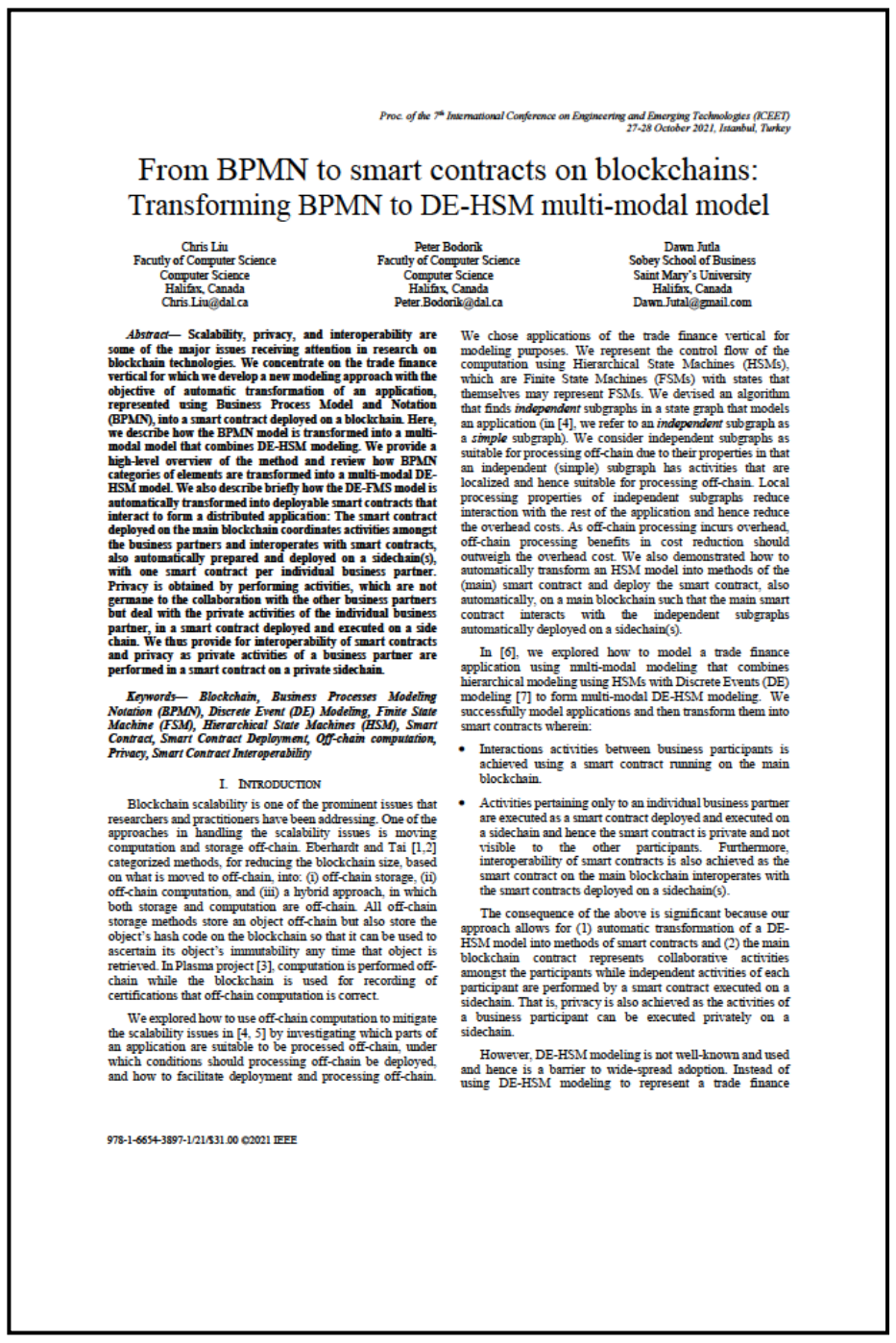

*Proc. of the 7th International Conference on Engineering and Emerging Technologies (ICEET)*
*27-28 October 2021, Istanbul, Turkey*

From BPMN to smart contracts on blockchains: Transforming BPMN to DE-HSM multi-modal model

Chris Liu
Facutly of Computer Science
Computer Science
Halifax, Canada
Chris.Liu@dal.ca

Peter Bodorik
Facutly of Computer Science
Computer Science
Halifax, Canada
Peter.Bodorik@dal.ca

Dawn Jutla
Sobey School of Business
Saint Mary's University
Halifax, Canada
Dawn.Jutal@gmail.com

***Abstract*— Scalability, privacy, and interoperability are some of the major issues receiving attention in research on blockchain technologies. We concentrate on the trade finance vertical for which we develop a new modeling approach with the objective of automatic transformation of an application, represented using Business Process Model and Notation (BPMN), into a smart contract deployed on a blockchain. Here, we describe how the BPMN model is transformed into a multi-modal model that combines DE-HSM modeling. We provide a high-level overview of the method and review how BPMN categories of elements are transformed into a multi-modal DE-HSM model. We also describe briefly how the DE-FMS model is automatically transformed into deployable smart contracts that interact to form a distributed application: The smart contract deployed on the main blockchain coordinates activities amongst the business partners and interoperates with smart contracts, also automatically prepared and deployed on a sidechain(s), with one smart contract per individual business partner. Privacy is obtained by performing activities, which are not germane to the collaboration with the other business partners but deal with the private activities of the individual business partner, in a smart contract deployed and executed on a side chain. We thus provide for interoperability of smart contracts and privacy as private activities of a business partner are performed in a smart contract on a private sidechain.**

***Keywords*— Blockchain, Business Processes Modeling Notation (BPMN), Discrete Event (DE) Modeling, Finite State Machine (FSM), Hierarchical State Machines (HSM), Smart Contract, Smart Contract Deployment, Off-chain computation, Privacy, Smart Contract Interoperability**

I. INTRODUCTION

Blockchain scalability is one of the prominent issues that researchers and practitioners have been addressing. One of the approaches in handling the scalability issues is moving computation and storage off-chain. Eberhardt and Tai [1,2] categorized methods, for reducing the blockchain size, based on what is moved to off-chain, into: (i) off-chain storage, (ii) off-chain computation, and (iii) a hybrid approach, in which both storage and computation are off-chain. All off-chain storage methods store an object off-chain but also store the object's hash code on the blockchain so that it can be used to ascertain its object's immutability any time that object is retrieved. In Plasma project [3], computation is performed off-chain while the blockchain is used for recording of certifications that off-chain computation is correct.

We explored how to use off-chain computation to mitigate the scalability issues in [4, 5] by investigating which parts of an application are suitable to be processed off-chain, under which conditions should processing off-chain be deployed, and how to facilitate deployment and processing off-chain. We chose applications of the trade finance vertical for modeling purposes. We represent the control flow of the computation using Hierarchical State Machines (HSMs), which are Finite State Machines (FSMs) with states that themselves may represent FSMs. We devised an algorithm that finds *independent* subgraphs in a state graph that models an application (in [4], we refer to an *independent* subgraph as a *simple* subgraph). We consider independent subgraphs as suitable for processing off-chain due to their properties in that an independent (simple) subgraph has activities that are localized and hence suitable for processing off-chain. Local processing properties of independent subgraphs reduce interaction with the rest of the application and hence reduce the overhead costs. As off-chain processing incurs overhead, off-chain processing benefits in cost reduction should outweigh the overhead cost. We also demonstrated how to automatically transform an HSM model into methods of the (main) smart contract and deploy the smart contract, also automatically, on a main blockchain such that the main smart contract interacts with the independent subgraphs automatically deployed on a sidechain(s).

In [6], we explored how to model a trade finance application using multi-modal modeling that combines hierarchical modeling using HSMs with Discrete Events (DE) modeling [7] to form multi-modal DE-HSM modeling. We successfully model applications and then transform them into smart contracts wherein:

- Interactions activities between business participants is achieved using a smart contract running on the main blockchain.
- Activities pertaining only to an individual business partner are executed as a smart contract deployed and executed on a sidechain and hence the smart contract is private and not visible to the other participants. Furthermore, interoperability of smart contracts is also achieved as the smart contract on the main blockchain interoperates with the smart contracts deployed on a sidechain(s).

The consequence of the above is significant because our approach allows for (1) automatic transformation of a DE-HSM model into methods of smart contracts and (2) the main blockchain contract represents collaborative activities amongst the participants while independent activities of each participant are performed by a smart contract executed on a sidechain. That is, privacy is also achieved as the activities of a business participant can be executed privately on a sidechain.

However, DE-HSM modeling is not well-known and used and hence is a barrier to wide-spread adoption. Instead of using DE-HSM modeling to represent a trade finance

978-1-6654-3897-1/21/$31.00 ©2021 IEEE

5. Proceedings IEEE BCCA 2022: Liu, C., Bodorik, P., Jutla, D. Supporting Long-Term Transactions in Smart Contracts

In: Liu, C., Bodorik, P., Jutla, D. (2022). Supporting Long-term Transactions in Smart Contracts. *2022 IEEE Fourth International Conference on Blockchain Computing and Applications* (BCCA), 11–19. https://doi.org/10.1109/BCCA55292.2022.9922193

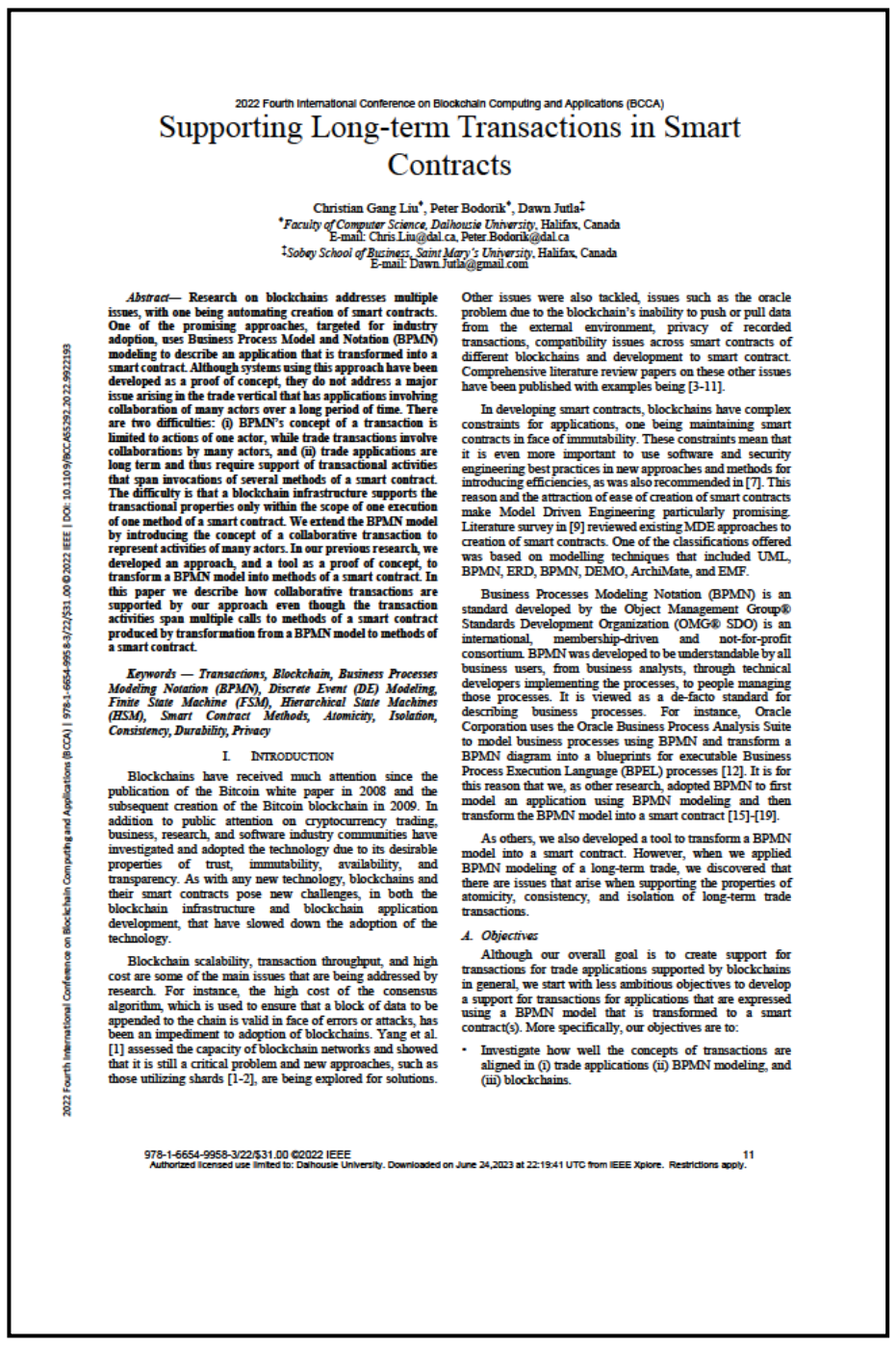

Supporting Long-term Transactions in Smart Contracts

Christian Gang Liu*, Peter Bodorik*, Dawn Jutla‡
*Faculty of Computer Science, Dalhousie University, Halifax, Canada
E-mail: Chris.Liu@dal.ca, Peter.Bodorik@dal.ca
‡Sobey School of Business, Saint Mary's University, Halifax, Canada
E-mail: Dawn.Jutla@gmail.com

*Abstract*— **Research on blockchains addresses multiple issues, with one being automating creation of smart contracts. One of the promising approaches, targeted for industry adoption, uses Business Process Model and Notation (BPMN) modeling to describe an application that is transformed into a smart contract. Although systems using this approach have been developed as a proof of concept, they do not address a major issue arising in the trade vertical that has applications involving collaboration of many actors over a long period of time. There are two difficulties: (i) BPMN's concept of a transaction is limited to actions of one actor, while trade transactions involve collaborations by many actors, and (ii) trade applications are long term and thus require support of transactional activities that span invocations of several methods of a smart contract. The difficulty is that a blockchain infrastructure supports the transactional properties only within the scope of one execution of one method of a smart contract. We extend the BPMN model by introducing the concept of a collaborative transaction to represent activities of many actors. In our previous research, we developed an approach, and a tool as a proof of concept, to transform a BPMN model into methods of a smart contract. In this paper we describe how collaborative transactions are supported by our approach even though the transaction activities span multiple calls to methods of a smart contract produced by transformation from a BPMN model to methods of a smart contract.**

I. INTRODUCTION

Blockchains have received much attention since the publication of the Bitcoin white paper in 2008 and the subsequent creation of the Bitcoin blockchain in 2009. In addition to public attention on cryptocurrency trading, business, research, and software industry communities have investigated and adopted the technology due to its desirable properties of trust, immutability, availability, and transparency. As with any new technology, blockchains and their smart contracts pose new challenges, in both the blockchain infrastructure and blockchain application development, that have slowed down the adoption of the technology.

Blockchain scalability, transaction throughput, and high cost are some of the main issues that are being addressed by research. For instance, the high cost of the consensus algorithm, which is used to ensure that a block of data to be appended to the chain is valid in face of errors or attacks, has been an impediment to adoption of blockchains. Yang et al. [1] assessed the capacity of blockchain networks and showed that it is still a critical problem and new approaches, such as those utilizing shards [1-2], are being explored for solutions. Other issues were also tackled, issues such as the oracle problem due to the blockchain's inability to push or pull data from the external environment, privacy of recorded transactions, compatibility issues across smart contracts of different blockchains and development to smart contract. Comprehensive literature review papers on these other issues have been published with examples being [3-11].

In developing smart contracts, blockchains have complex constraints for applications, one being maintaining smart contracts in face of immutability. These constraints mean that it is even more important to use software and security engineering best practices in new approaches and methods for introducing efficiencies, as was also recommended in [7]. This reason and the attraction of ease of creation of smart contracts make Model Driven Engineering particularly promising. Literature survey in [9] reviewed existing MDE approaches to creation of smart contracts. One of the classifications offered was based on modelling techniques that included UML, BPMN, ERD, BPMN, DEMO, ArchiMate, and EMF.

Business Processes Modeling Notation (BPMN) is an standard developed by the Object Management Group® Standards Development Organization (OMG® SDO) is an international, membership-driven and not-for-profit consortium. BPMN was developed to be understandable by all business users, from business analysts, through technical developers implementing the processes, to people managing those processes. It is viewed as a de-facto standard for describing business processes. For instance, Oracle Corporation uses the Oracle Business Process Analysis Suite to model business processes using BPMN and transform a BPMN diagram into a blueprints for executable Business Process Execution Language (BPEL) processes [12]. It is for this reason that we, as other research, adopted BPMN to first model an application using BPMN modeling and then transform the BPMN model into a smart contract [15]-[19].

As others, we also developed a tool to transform a BPMN model into a smart contract. However, when we applied BPMN modeling of a long-term trade, we discovered that there are issues that arise when supporting the properties of atomicity, consistency, and isolation of long-term trade transactions.

*A. Objectives*

Although our overall goal is to create support for transactions for trade applications supported by blockchains in general, we start with less ambitious objectives to develop a support for transactions for applications that are expressed using a BPMN model that is transformed to a smart contract(s). More specifically, our objectives are to:

- Investigate how well the concepts of transactions are aligned in (i) trade applications (ii) BPMN modeling, and (iii) blockchains.

6. Patent Application Submission USPTO 2022: Bodorik, P., Liu, C., Jutla, D. System and Methods for Automated Generation of Smart Contracts from BPMN Models Using DE-HSM Multi-Modal Models Executed by Smart Contracts with Sidechain Processing

In: Bodorik, P., Liu, C., Jutla, D. (2022). System and Methods for Automated Generation of Smart Contracts from BPMN Models Using DE-HSM Multi-modal Models Executed by Smart Contracts with Sidechain Processing (U.S. Patent Application No. 17,968,047). U.S. Patent and Trademark Office.

7. ELSEVIER Journal Blockchain: Research and Applications (BCRA) 2023: Bodorik, P., Liu, C., Jutla, D. TABS: Transforming Automatically BPMN Models into Blockchain Smart Contracts

In: Bodorik, P., Liu, C., Jutla, D. (2023). TABS: Transforming automatically BPMN models into blockchain smart contracts. *Blockchain: Research and Applications* (Elsevier), 100115. https://doi.org/10.1016/j.bcra.2022.100115

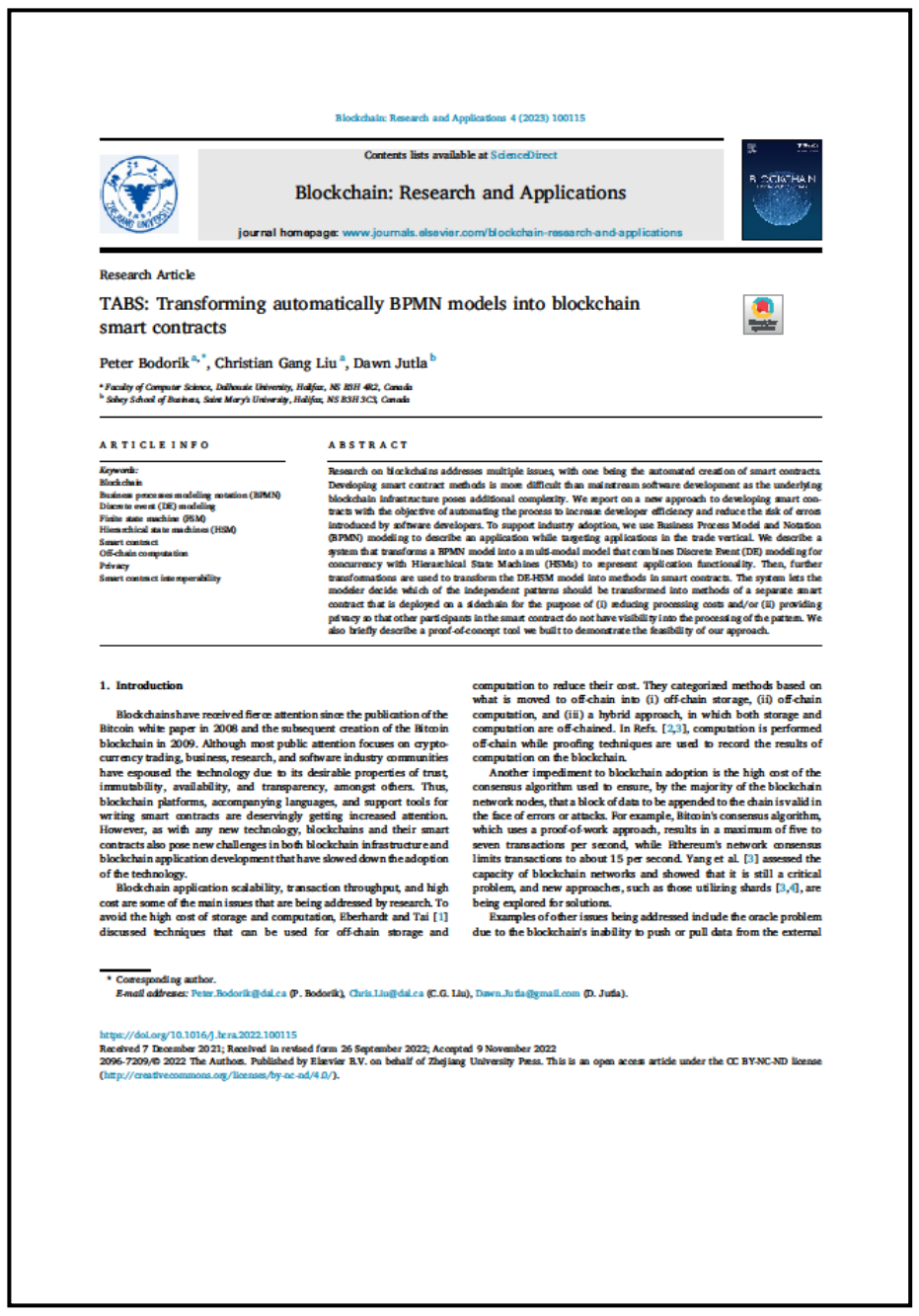
Blockchain: Research and Applications 4 (2023) 100115

Contents lists available at ScienceDirect

Blockchain: Research and Applications

journal homepage: www.journals.elsevier.com/blockchain-research-and-applications

Research Article

TABS: Transforming automatically BPMN models into blockchain smart contracts

Peter Bodorik[a,*], Christian Gang Liu[a], Dawn Jutla[b]

[a] Faculty of Computer Science, Dalhousie University, Halifax, NS B3H 4R2, Canada
[b] Sobey School of Business, Saint Mary's University, Halifax, NS B3H 3C3, Canada

ARTICLE INFO

*Keywords:*
Blockchain
Business processes modeling notation (BPMN)
Discrete event (DE) modeling
Finite state machine (FSM)
Hierarchical state machines (HSM)
Smart contract
Off-chain computation
Privacy
Smart contract interoperability

ABSTRACT

Research on blockchains addresses multiple issues, with one being the automated creation of smart contracts. Developing smart contract methods is more difficult than mainstream software development as the underlying blockchain infrastructure poses additional complexity. We report on a new approach to developing smart contracts with the objective of automating the process to increase developer efficiency and reduce the risk of errors introduced by software developers. To support industry adoption, we use Business Process Model and Notation (BPMN) modeling to describe an application while targeting applications in the trade vertical. We describe a system that transforms a BPMN model into a multi-modal model that combines Discrete Event (DE) modeling for concurrency with Hierarchical State Machines (HSMs) to represent application functionality. Then, further transformations are used to transform the DE-HSM model into methods in smart contracts. The system lets the modeler decide which of the independent patterns should be transformed into methods of a separate smart contract that is deployed on a sidechain for the purpose of (i) reducing processing costs and/or (ii) providing privacy so that other participants in the smart contract do not have visibility into the processing of the pattern. We also briefly describe a proof-of-concept tool we built to demonstrate the feasibility of our approach.

1. Introduction

Blockchains have received fierce attention since the publication of the Bitcoin white paper in 2008 and the subsequent creation of the Bitcoin blockchain in 2009. Although most public attention focuses on cryptocurrency trading, business, research, and software industry communities have espoused the technology due to its desirable properties of trust, immutability, availability, and transparency, amongst others. Thus, blockchain platforms, accompanying languages, and support tools for writing smart contracts are deservingly getting increased attention. However, as with any new technology, blockchains and their smart contracts also pose new challenges in both blockchain infrastructure and blockchain application development that have slowed down the adoption of the technology.

Blockchain application scalability, transaction throughput, and high cost are some of the main issues that are being addressed by research. To avoid the high cost of storage and computation, Eberhardt and Tai [1] discussed techniques that can be used for off-chain storage and computation to reduce their cost. They categorized methods based on what is moved to off-chain into (i) off-chain storage, (ii) off-chain computation, and (iii) a hybrid approach, in which both storage and computation are off-chained. In Refs. [2,3], computation is performed off-chain while proofing techniques are used to record the results of computation on the blockchain.

Another impediment to blockchain adoption is the high cost of the consensus algorithm used to ensure, by the majority of the blockchain network nodes, that a block of data to be appended to the chain is valid in the face of errors or attacks. For example, Bitcoin's consensus algorithm, which uses a proof-of-work approach, results in a maximum of five to seven transactions per second, while Ethereum's network consensus limits transactions to about 15 per second. Yang et al. [3] assessed the capacity of blockchain networks and showed that it is still a critical problem, and new approaches, such as those utilizing shards [3,4], are being explored for solutions.

Examples of other issues being addressed include the oracle problem due to the blockchain's inability to push or pull data from the external

* Corresponding author.
*E-mail addresses:* Peter.Bodorik@dal.ca (P. Bodorik), Chris.Liu@dal.ca (C.G. Liu), Dawn.Jutla@gmail.com (D. Jutla).

https://doi.org/10.1016/j.bcra.2022.100115
Received 7 December 2021; Received in revised form 26 September 2022; Accepted 9 November 2022
2096-7209/© 2022 The Authors. Published by Elsevier B.V. on behalf of Zhejiang University Press. This is an open access article under the CC BY-NC-ND license (http://creativecommons.org/licenses/by-nc-nd/4.0/).

8. Proceedings SPRINGER BLOCKSYS 2023: Liu, C. G., Bodorik, P., Jutla, D. Long-Term Blockchain Transactions Spanning Multiplicity of Smart Contract Methods

<u>In</u>: Liu, C. G., Bodorik, P., Jutla, D. (2023). Long-Term Blockchain Transactions Spanning Multiplicity of Smart Contract Methods. In Springer BlockSys: Blockchain and Trustworthy Systems. https://doi.org/10.1007/978-981-99-8104-5. [Long-Term Blockchain Transactions Spanning Multiplicity of Smart Contract Methods.pdf]

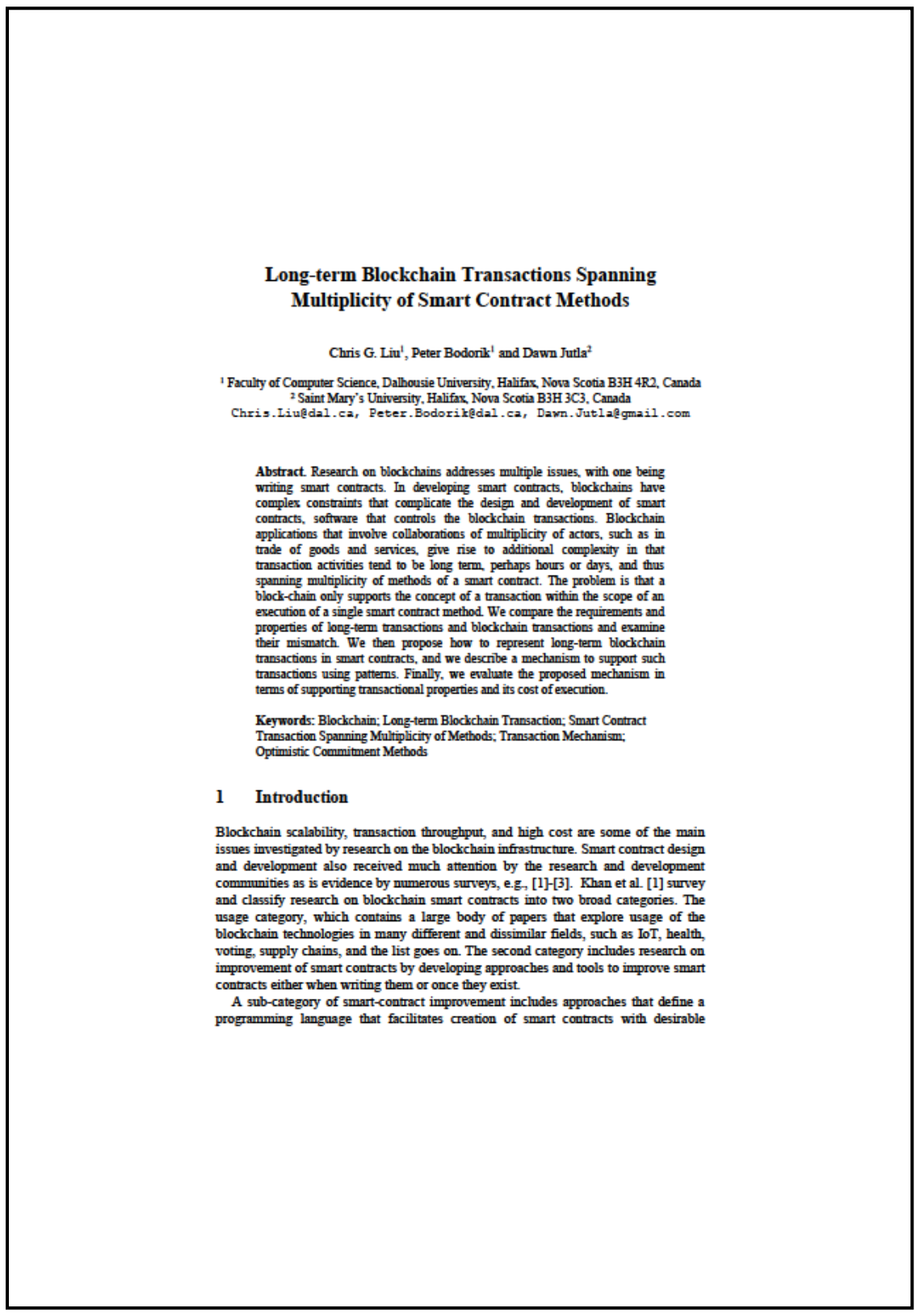

**Long-term Blockchain Transactions Spanning Multiplicity of Smart Contract Methods**

Chris G. Liu[1], Peter Bodorik[1] and Dawn Jutla[2]

[1] Faculty of Computer Science, Dalhousie University, Halifax, Nova Scotia B3H 4R2, Canada
[2] Saint Mary's University, Halifax, Nova Scotia B3H 3C3, Canada
Chris.Liu@dal.ca, Peter.Bodorik@dal.ca, Dawn.Jutla@gmail.com

**Abstract.** Research on blockchains addresses multiple issues, with one being writing smart contracts. In developing smart contracts, blockchains have complex constraints that complicate the design and development of smart contracts, software that controls the blockchain transactions. Blockchain applications that involve collaborations of multiplicity of actors, such as in trade of goods and services, give rise to additional complexity in that transaction activities tend to be long term, perhaps hours or days, and thus spanning multiplicity of methods of a smart contract. The problem is that a block-chain only supports the concept of a transaction within the scope of an execution of a single smart contract method. We compare the requirements and properties of long-term transactions and blockchain transactions and examine their mismatch. We then propose how to represent long-term blockchain transactions in smart contracts, and we describe a mechanism to support such transactions using patterns. Finally, we evaluate the proposed mechanism in terms of supporting transactional properties and its cost of execution.

**1 Introduction**

Blockchain scalability, transaction throughput, and high cost are some of the main issues investigated by research on the blockchain infrastructure. Smart contract design and development also received much attention by the research and development communities as is evidence by numerous surveys, e.g., [1]-[3]. Khan et al. [1] survey and classify research on blockchain smart contracts into two broad categories. The usage category, which contains a large body of papers that explore usage of the blockchain technologies in many different and dissimilar fields, such as IoT, health, voting, supply chains, and the list goes on. The second category includes research on improvement of smart contracts by developing approaches and tools to improve smart contracts either when writing them or once they exist.

A sub-category of smart-contract improvement includes approaches that define a programming language that facilitates creation of smart contracts with desirable

9. Submitted Manuscript (Under Review) ACM DLT 2023: Liu, C., Bodorik, P., Jutla, D. TABS+: Transforming Automatically BPMN Models to Smart Contracts with Nested Collaborative Transactions

<u>In</u>: Liu, C., Bodorik, P., Jutla, D. (2023). Tabs+: Transforming Automatically BPMN Models to Smart Contracts with Nested Collaborative Transactions [Manuscript submitted for publication]. *ACM Distributed Ledger Technologies: Research and Practice* (DLT). [TABS+: Transforming Automatically BPMN Models to Smart Contracts with Nested Collaborative Transactions.pdf]

## TABS+: Transforming Automatically BPMN Models to Smart Contracts with Nested Collaborative Transactions

Transforming BPMN Models to Smart Contracts with Support of Nested Transactions

Christian, G., Liu

Faculty of Computer Science, Dalhousie University, Chris.Liu@dal.ca

Peter Bodorik

Faculty of Computer Science, Dalhousie University, Peter.Bodorik@dal.ca

Dawn Jutla

Sobey School of Business, Saint Mary's University, Dawn.Jutla@gmail.com

Development of blockchain smart contracts is more difficult than mainstream software development because the underlying blockchain infrastructure poses additional complexity. To ease the developer's task of writing smart contract, as other research efforts, we also use Business Process Model and Notation (BPMN) modeling to describe application requirements for trade of goods and services and then transform automatically the BPMN model into the methods of a smart contract. In our previous research we described our *TABS* approach and tool to *Transform Automatically BPMN models into Smart contracts* that also supports sidechain processing. In this paper, we describe how the TABS approach is augmented with the support of a BPMN collaborative transaction by several actors, which is an extension of the BPMN concept of a transaction that can be defined only on activities of a single actor. Our approach analyzes the BPMN model to determine which patterns in the BPMN model are suitable for use as collaborative transactions. The found BPMN patterns that are suitable as transactions are shown to the developer who decides which ones should be deployed as collaborative transactions. We describe how the nested BPMN collaborative transactions are supported by an automatically created transaction mechanism. We also overview the TABS+ tool, built as a proof of concept, to show that our approach is feasible. Finally, we provide estimates on the cost of supporting the nested BPMN collaborative transactions.